# Broadband multi-wavelength properties of M87 during the 2018 EHT campaign including a very high energy flaring episode★

J. C. Algaba[1], M. Baloković[2,3], S. Chandra[4,5], W.-Y. Cheong[6,7], Y.-Z. Cui (崔玉竹)[8], F. D'Ammando[9], A. D. Falcone[10], N. M. Ford[11,12], M. Giroletti[9], C. Goddi[13,14], M. A. Gurwell[15], K. Hada[16,17,18], D. Haggard[11,19], S. Jorstad[20], A. Kaur[21], T. Kawashima[22], S. Kerby[21], J.-Y. Kim[6,23], M. Kino[24,25], E. V. Kravchenko[9], S.-S. Lee[6], R.-S. Lu (路如森)[26,27,23], S. Markoff[28,29], J. Michail[30,31,15], J. Neilsen[32], M. A. Nowak[33], G. Principe[★★34,35,9], V. Ramakrishnan[36,??], B. Ripperda[38,39], M. Sasada[40], S. S. Savchenko, C. Sheridan[32]
(The Event Horizon Telescope - Multi-wavelength science working group)
K. Akiyama[41,24,42], A. Alberdi[43], W. Alef[44], R. Anantua[42,15,45], K. Asada[46], R. Azulay[47,48,44], U. Bach[44], A.-K. Baczko[49,44], D. Ball[50], B. Bandyopadhyay[51], J. Barrett[41], M. Bauböck[52], B. A. Benson[53,54], D. Bintley[55,56], L. Blackburn[42,15], R. Blundell[15], K. L. Bouman[57], G. C. Bower[58,59], H. Boyce[60,61], M. Bremer[62], R. Brissenden[42,15], S. Britzen[44], A. E. Broderick[63,64,65], D. Broguiere[62], T. Bronzwaer[13], S. Bustamante[66], J. E. Carlstrom[67,54,68,69], A. Chael[70], C.-k. Chan[50,71,72], D. O. Chang[42,15], K. Chatterjee[42,15], S. Chatterjee[73], M.-T. Chen[58], Y. Chen (陈永军)[26,27], X. Cheng[6], I. Cho[43,6,74], P. Christian[75], N. S. Conroy[76,15], J. E. Conway[49], T. M. Crawford[54,67], G. B. Crew[41], A. Cruz-Osorio[77,78], R. Dahale[43], J. Davelaar[81,82,13], M. De Laurentis[83,84], R. Deane[85,86,87], J. Dempsey[55,56,88], G. Desvignes[44,89], J. Dexter[90], V. Dhruv[52], I. K. Dihingia[80], S. S. Doeleman[42,15], S. A. Dzib[62,44], R. P. Eatough[91,44], R. Emami[15], H. Falcke[13], J. Farah[92,93], V. L. Fish[41], E. Fomalont[94], H. A. Ford[50], M. Foschi[43], R. Fraga-Encinas[13], W. T. Freeman[95,96], P. Friberg[55,56], C. M. Fromm[97,78,44], A. Fuentes[43], P. Galison[42,98,99], C. F. Gammie[52,76,100], R. García[62], O. Gentaz[62], B. Georgiev[50], R. Gold[102], A. I. Gómez-Ruiz[103,104], J. L. Gómez[43], M. Gu (顾敏峰)[26,105], R. Hesper[106], D. Heumann[50], L. C. Ho (何子山)[107,108], P. Ho[46,56,55], M. Honma[16,17,109], C.-W. L. Huang[46], L. Huang (黄磊)[26,105], D. H. Hughes[103], S. Ikeda[24,110,111,112], C. M. V. Impellizzeri[113,94], M. Inoue[46], S. Issaoun[15], D. J. James[114,115], B. T. Jannuzi[50], M. Janssen[13,44], B. Jeter[46], W. Jiang (江悟)[26], A. Jiménez-Rosales[13], M. D. Johnson[42,15], A. C. Jones[54], A. V. Joshi[52], T. Jung[6,7], R. Karuppusamy[44], G. K. Keating[15], M. Kettenis[116], D.-J. Kim[41], J. Kim[6], J. Kim[117], J. Y. Koay[46], P. Kocherlakota[78], Y. Kofuji[16,109], P. M. Koch[46], S. Koyama[118,46], C. Kramer[62], J. A. Kramer[44], M. Kramer[44], T. P. Krichbaum[44], C.-Y. Kuo[119,46], N. La Bella[13], A. Levis[57], Z. Li (李志远)[120,121], R. Lico[122,43], G. Lindahl[15], M. Lindqvist[49], M. Lisakov[44], J. Liu (刘俊)[44], K. Liu[44], E. Liuzzo[123], W.-P. Lo[46,124], A. P. Lobanov[44], L. Loinard[125], C. J. Lonsdale[41], A. E. Lowitz[50], N. R. MacDonald[44], J. Mao (毛基荣)[126,127,128], N. Marchili[123,44], D. P. Marrone[50], A. P. Marscher[20], I. Martí-Vidal[47,48], S. Matsushita[46], L. D. Matthews[41], L. Medeiros[129], K. M. Menten[44], I. Mizuno[55,56], Y. Mizuno[80,130,78], J. Montgomery[61,54], J. M. Moran[42,15], K. Moriyama[78,16], M. Moscibrodzka[13], W. Mulaudzi[28], C. Müller[44,13], H. Müller[44], A. Mus[47,48], G. Musoke[28,13], I. Myserlis[131], H. Nagai[24,17], N. M. Nagar[51], D. G. Nair[51], M. Nakamura[132,46], G. Narayanan[66], I. Natarajan[15,42], A. Nathanail[133,78], S. Navarro Fuentes[131], C. Ni[64,65,63], J. Oh[116], H. Okino[16,109], H. Olivares[13], T. Oyama[16], F. Özel[134], D. C. M. Palumbo[42,15], G. Filippos Paraschos[44], J. Park[135], H. Parsons[55,56], N. Patel[15], U.-L. Pen[46,63,136,137,138], D. W. Pesce[15,42], V. Piétu[62], A. PopStefanija[66], O. Porth[28,78], B. Prather[52], D. Psaltis[134], H.-Y. Pu[139,140,46], R. Rao[15], M. G. Rawlings[142,55,56], A. W. Raymond[42,15], L. Rezzolla[78,143,144], A. Ricarte[42,15], F. Roelofs[15,42,13], C. Romero-Cañizales[46], E. Ros[44], A. Roshanineshat[50], H. Rottmann[44], A. L. Roy[44], I. Ruiz[131], C. Ruszczyk[41], K. L. J. Rygl[123], S. Sánchez[131], D. Sánchez-Argüelles[103,104], M. Sánchez-Portal[131], K. Satapathy[50], T. Savolainen[146,141,44], F. P. Schloerb[66], J. Schonfeld[15], K.-F. Schuster[62], L. Shao[108,44], Z. Shen (沈志强)[26,27], D. Small[116], B. W. Sohn[6,7,74], J. SooHoo[41], L. D. Sosapanta Salas[28], K. Souccar[66], J. S. Stanway[147], H. Sun (孙赫)[148,149], F. Tazaki[150], A. J. Tetarenko[151], P. Tiede[15,42], R. P. J. Tilanus[50,13,113,152], M. Titus[41], K. Toma[153,154], P. Torne[131,44], T. Toscano[43], E. Traianou[43,44], T. Trent[50], S. Trippe[155], M. Turk[76], I. van Bemmel[116], H. J. van Langevelde[116,113,156], D. R. van Rossum[13], J. Vos[13], J. Wagner[44], D. Ward-Thompson[147], J. Wardle[157], J. E. Washington[50], J. Weintroub[42,15], R. Wharton[44], M. Wielgus[44], K. Wiik[158], G. Witzel[44], M. F. Wondrak[13,159], G. N. Wong[160,70], Q. Wu (吴庆文)[161], N. Yadlapalli[57], P. Yamaguchi[15], A. Yfantis[13], D. Yoon[28], A. Young[13], Z. Younsi[162,78], W. Yu (于威)[15], F. Yuan (袁峰)[163], Y.-F. Yuan (袁业飞)[164], J. A. Zensus[44], S. Zhang[165], G.-Y. Zhao[43,44], S.-S. Zhao (赵杉杉)[26]
(The Event Horizon Telescope Collaboration)
R. Bellazzini[166], B. Berenji[167], E. Bissaldi[168,169], R. D. Blandford[170], R. Bonino[171,172], P. Bruel[173], R. A. Cameron[170], P. A. Caraveo[174], E. Cavazzuti[175], C. C. Cheung[176], S. Ciprini[177,178], P. Cristarella Orestano[179,180], S. Cutini[180], N. Di Lalla[170], A. Dinesh[181], L. Di Venere[169], A. Domínguez[181], S. J. Fegan[173], A. Franckowiak[182], Y. Fukazawa[183], P. Fusco[168,169], F. Gargano[169], C. Gasbarra[177,184], S. Germani[179], M. Giliberti[169,168], I. A. Grenier[185], E. Hays[186], D. Horan[173], M. Kuss[166], S. Larsson[187,188], I. Liodakis[189], F. Longo[34,35], F. Loparco[168,169], M. N. Lovellette[190], S. Maldera[171], M. N. Mazziotta[169], I.Mereu[180,179], P. F. Michelson[170], N. Mirabal[186,191], T. Mizuno[192], M. E. Monzani[170,193], A. Morselli[177], M. Negro[194], N. Omodei[170], E. Orlando[35,34,170], M. Persic[35,195], S. Rainò[168,169], B. Rani[186,191], A. Reimer[196], O. Reimer[196], M. Sánchez-Conde[197,198], P. M. Saz Parkinson[199], C. Sgrò[166], E. J. Siskind[200], P. Spinelli[168,169], D. J. Suson[201], H. Tajima[202,203], D. F. Torres[204,205], G. Zaharijas[206]
(The *Fermi* Large Area Telescope Collaboration)
F. Aharonian[207,208,209], F. Ait Benkhali[211], J. Aschersleben[212], H. Ashkar[213], M. Backes[234,4], V. Barbosa Martins[★★★214,182], R. Batzofin[215], Y. Becherini[216,217], D. Berge[214,218], M. Böttcher[4], C. Boisson[219], J. Bolmont[220], M. de Bony de Lavergne[221], J. Borowska[218], M. Bouyahiaoui[208], F. Bradascio[221], R. Brose[207], A. Brown[222], B. Bruno[223], T. Bulik[224], C. Burger-Scheidlin[207], S. Casanova[225], R. Cecil[226], J. Celic[223], M. Cerruti[216], T. Chand[4], A. Chen[227], J. Chibueze[4], O. Chibueze[4], G. Cotter[222], J. Damascene Mbarubucyeye[214], J. Devin[228], J. Djuvsland[208], A. Dmytriiev[4], S. Einecke[229], J.-P. Ernenwein[230], K. Feijen[216], G. Fontaine[213], S. Funk[223], S. Gabici[216], D. Glawion[223], J.F. Glicenstein[221], P. Goswami[216], G. Grolleron[220], L. Haerer[208], B. Heß[231], T. L. Holch[214], M. Holler[232], D. Horns[226], Zhiqiu Huang[208], M. Jamrozy[233], F. Jankowsky[211], I. Jung-Richardt[223], E. Kasai[234], K. Katarzyński[235], R. Khatoon[4], B. Khélifi[216], W. Kluźniak[236], Nu. Komin[227], K. Kosack[221], A. Kundu[4], R.G. Lang[223], S. Le Stum[230], F. Leitl[223], A. Lemière[216], M. Lemoine-Goumard[237], J.-P. Lenain[220], F. Leuschner[231], A. Luashvili[219], J. Mackey[207], D. Malyshev[231], G. Martí-Devesa[232], R. Marx[211], M. Meyer[238], A. Mitchell[223], R. Moderski[236], M.O. Moghadam[215], L. Mohrmann[208], A. Montanari[211], E. Moulin[221], M. de Naurois[213], J. Niemiec[225], P. O'Brien[239], S. Ohm[214], E. de Ona Wilhelmi[214], M. Ostrowski[233], S. Panny[232], M. Panter[208], U. Pensec[220], S. Pita[216], G. Pühlhofer[★★★231], A. Quirrenbach[211], S. Ravikularaman[216,208], A. Reimer[232], O. Reimer[232], B. Reville[208], I. Reis[221], H. Ren[208], F. Rieger[208], G. Roellinghoff[223], B. Rudak[236], E. Ruiz-Velasco[208], K. Sabri[228], V. Sahakian[210], H. Salzmann[231], A. Santangelo[231], M. Sasaki[223], J. Schäfer[223], F. Schüssler[221], H.M. Schutte[4], J.N.S. Shapopi[234], A. Sharma[216], H. Sol[219], S. Spencer[223], Ł. Stawarz[233], C. Steppa[215], K. Streil[223], H. Suzuki[240], T. Takahashi[241], T. Tanaka[240], A.M. Taylor[214], R. Terrier[216], M. Tluczykont[226], M. Tsirou[214], C. van Eldik[223], M. Vecchi[212], T. Wach[223], S.J. Wagner[211], A. Wierzcholska[225], M. Zacharias[211,4], A.A. Zdziarski[236], A. Zech[219], N. Żywucka[4]
(H.E.S.S. Collaboration)
S. Abe[242], J. Abhir[243], A. Abhishek[244], V. A. Acciari[245], A. Aguasca-Cabot[246], I. Agudo[247], T. Aniello[248], S. Ansoldi[249,282], L. A. Antonelli[248], A. Arbet Engels[250], C. Arcaro[251], M. Artero[245], K. Asano[242], A. Babić[252], U. Barres de Almeida[253], J. A. Barrio[254], I. Batković[251], A. Bautista[250], J. Baxter[242], J. Becerra González[255], W. Bednarek[256], E. Bernardini[251], J. Bernete[257], A. Berti[250], J. Besenrieder[250], C. Bigongiari[248], A. Biland[243], O. Blanch[245], G. Bonnoli[248], Ž. Bošnjak[252], E. Bronzini[248], I. Burelli[249], G. Busetto[251], A. Campoy-Ordaz[258], A. Carosi[248], R. Carosi[259], M. Carretero-Castrillo[246], A. J. Castro-Tirado[247], D. Cerasole[260], G. Ceribella[250], Y. Chai[242], A. Cifuentes[257], E. Colombo[245], J. L. Contreras[254], J. Cortina[257], S. Covino[248], G. D'Amico[261], V. D'Elia[248], P. Da Vela[248], F. Dazzi[248], A. De Angelis[251], B. De Lotto[249], R. de Menezes[262], M. Delfino[245,283], J. Delgado[245,283], C. Delgado Mendez[257], F. Di Pierro[262], R. Di Tria[260], L. Di Venere[260], D. Dominis Prester[263], A. Donini[248], D. Dorner[264], M. Doro[251], D. Elsaesser[265], J. Escudero[247], L. Fariña[245], A. Fattorini[265], L. Foffano[248], L. Font[258], S. Fröse[265], S. Fukami[243], Y. Fukazawa[266], R. J. García López[255], M. Garczarczyk[214], S. Gasparyan[267], M. Gaug[258], J. G. Giesbrecht Paiva[253], N. Giglietto[260], F. Giordano[260], P. Gliwny[256], N. Godinović[268], T. Gradetzke[265], R. Grau[245], D. Green[250], J. G. Green[250], P. Günther[264], D. Hadasch[242], A. Hahn[★★★★250], T. Hassan[257], L. Heckmann[250], J. Herrera Llorente[255], D. Hrupec[269], R. Imazawa[266], K. Ishio[256], I. Jiménez Martínez[250], J. Jormanainen[270], T. Kayanoki[266], D. Kerszberg[245], G. W. Kluge[261,284], Y. Kobayashi[242], P. M. Kouch[270], H. Kubo[242], J. Kushida[271], M. Láinez[254], A. Lamastra[248], F. Leone[248], E. Lindfors[270], S. Lombardi[248], R. López-Coto[247], M. López-Moya[254], A. López-Oramas[255], S. Loporchio[260], A. Lorini[244], E. Lyard[272], B. Machado de Oliveira Fraga[253], P. Majumdar[273], M. Makariev[274], G. Maneva[274], M. Manganaro[263], S. Mangano[257], K. Mannheim[264], M. Mariotti[251], M. Martínez[245], M. Martínez-Chicharro[257], A. Mas-Aguilar[254], D. Mazin[★★★★242,285], S. Menchiari[244], S. Mender[265], D. Miceli[251], T. Miener[254], J. M. Miranda[244], R. Mirzoyan[250], M. Molero González[255], E. Molina[255], H. A. Mondal[273], A. Moralejo[245], D. Morcuende[247], T. Nakamori[275], C. Nanci[248], V. Neustroev[276], L. Nickel[265], M. Nievas Rosillo[255], C. Nigro[245], L. Nikolić[244], K. Nilsson[270], K. Nishijima[271], T. Njoh Ekoume[245], K. Noda[277], S. Nozaki[250], Y. Ohtani[242], A. Okumura[278], J. Otero-Santos[247], S. Paiano[248], D. Paneque[250], R. Paoletti[244], J. M. Paredes[246], M. Peresano[250], M. Persic[249,286], M. Pihet[251], G. Pirola[250], F. Podobnik[244], P. G. Prada Moroni[259], E. Prandini[251], C. Priyadarshi[245], M. Ribó[246], J. Rico[245], C. Righi[248], N. Sahakyan[267], T. Saito[242], F. G. Saturni[248], K. Schmidt[265], F. Schmuckermaier[250], J. L. Schubert[265], T. Schweizer[250], A. Sciaccaluga[248], G. Silvestri[251], J. Sitarek[256], V. Sliusar[272], D. Sobczynska[256], A. Spolon[251], A. Stamerra[248], J. Strišković[269], D. Strom[250], M. Strzys[242], Y. Suda[266], S. Suutarinen[270], H. Tajima[278], M. Takahashi[278], R. Takeishi[242], F. Tavecchio[248], P. Temnikov[274], K. Terauchi[279], T. Terzić[263], M. Teshima[250,287], S. Truzzi[244], A. Tutone[248], S. Ubach[258], J. van Scherpenberg[250], M. Vazquez Acosta[255], S. Ventura[244], G. Verna[244], I. Viale[251], C. F. Vigorito[262], V. Vitale[280], I. Vovk[242], R. Walter[272], M. Will[250], C. Wunderlich[244], T. Yamamoto[281]
(MAGIC Collaboration)
A. Acharyya[288], C. B. Adams[289], P. Bangale[290], J. T. Bartkoske[291], W. Benbow[15], J. L. Christiansen[292], A. Duerr[291], M. Errando[293], Q. Feng[291], G. M. Foote[290], L. Fortson[295], A. Furniss[294], W. Hanlon[15], O. Hervet[296], C. E. Hinrichs[15,301], J. Holder[290], T. B. Humensky[302,316], W. Jin[†306,315], M. N. Johnson[296], P. Kaaret[297], M. Kertzman[299], D. Kieda[291], T. K. Kleiner[214], N. Korzoun[290], F. Krennrich[298], S. Kumar[302], M. J. Lang[303], M. Lundy[304], G. Maier[214], C. E McGrath[300], M. J. Millard[297], C. L. Mooney[290], P. Moriarty[303], R. Mukherjee[314], W. Ning[306], S. O'Brien[304,307], R. A. Ong[306], M. Pohl[308,214], E. Pueschel[309], J. Quinn[310], K. Ragan[304], P. T. Reynolds[311], D. Ribeiro[295], E. Roache[15], J. L. Ryan[306], I. Sadeh[214], L. Saha[15], M. Santander[¶315], G. H. Sembroski[305], R. Shang[314], M. Splettstoesser[296], A. K. Talluri[295], J. V. Tucci[312], J. Valverde[313,316], V. V. Vassiliev[306], D. A. Williams[296], S. L. Wong[304]
(VERITAS Collaboration)
Z. Chen[26,317], L. Cui[318,317], T. Hirota[16,17], B. Li[26,317], G. Li[318], Q. Liu[26,317], X. Liu[318,317], Z. Liu[318], J. Ma[318], K. Niinuma[319,320], H. Ro[6], N. Sakai[322], S. Sawada-Satoh[323], K. Wajima[321,refinst5], J. Wang[26,317], N. Wang[318,317], B. Xia[26,317], H. Yan[318], Y. Yonekura[324], H. Zhang[318], R. Zhao[26,317], and W. Zhong[26,317]
(EAVN Collaboration)

*(Affiliations can be found after the references)*

### ABSTRACT

*Context.* The nearby elliptical galaxy M87 contains one of only two supermassive black holes whose emission surrounding the event horizon has been imaged by the Event Horizon Telescope (EHT). In 2018, more than two dozen multi-wavelength (MWL) facilities (from radio to $\gamma$-ray energies) took part in the second M87 EHT campaign.

*Aims.* The goal of this extensive MWL campaign was to better understand the physics of the accreting black hole M87*, the relationship between the inflow and inner jets, and the high-energy particle acceleration. Understanding the complex astrophysics is also a necessary first step towards performing further tests of general relativity.

*Methods.* The MWL campaign took place in April 2018, overlapping with the EHT M87* observations. We present a new, contemporaneous spectral energy distribution (SED) ranging from radio to very high-energy (VHE) $\gamma$-rays as well as details of the individual observations and light curves. We also conducted phenomenological modelling to investigate the basic source properties.

*Results.* We present the first VHE $\gamma$-ray flare from M87 detected since 2010. The flux above 350 GeV more than doubled within a period of ≈36 hours. We find that the X-ray flux is enhanced by about a factor of two compared to 2017, while the radio and millimetre core fluxes are consistent between 2017 and 2018. We detect evidence for a monotonically increasing jet position angle that corresponds to variations in the bright spot of the EHT image.

*Conclusions.* Our results show the value of continued MWL monitoring together with precision imaging for addressing the origins of high-energy particle acceleration. While we cannot currently pinpoint the precise location where such acceleration takes place, the new VHE $\gamma$-ray flare already presents a challenge to simple one-zone leptonic emission model approaches, and it emphasises the need for combined image and spectral modelling.

**Key words.** Accretion, accretion disks – black hole physics –galaxies: individual: M87

## 1. Introduction

The M87 elliptical galaxy in the Virgo Cluster is one of the nearest active galactic nuclei (AGN), located at a distance of $16.8 \pm 0.8$ Mpc (Blakeslee et al. 2009; Bird et al. 2010; Cantiello et al. 2018, and Event Horizon Telescope Collaboration et al. 2019d). It also harbours one of the largest known supermassive black holes (SMBHs; $\sim 6.5 \times 10^9\,M_\odot$ determined by Gebhardt et al. (2011), but see also, e.g., Walsh et al. (2013); Liepold et al. (2023); Osorno et al. (2023); Simon et al. (2024) finding a slightly smaller mass). It is one of only two SMBHs where a ring of light emitted from near the event horizon has been directly imaged at 230 GHz, via the global Very Long Baseline Interferometry (VLBI) project, the Event Horizon Telescope (EHT). The ring's size and shape are directly predicted by general relativity (GR) and depend on the black hole (BH) mass and spin and the orientation of the spin axis with respect to our line of sight (see, e.g., Bardeen 1973; Johannsen & Psaltis 2010).

In April 2019, the EHT Collaboration (EHTC) presented the first M87* images (we note that we use M87* to refer to the SMBH rather than its host galaxy) from the 2017 campaign along with modelling and interpretation of the broadband emission (Event Horizon Telescope Collaboration et al. 2019a,b,c,d,e,f). A primary result of this sequence of works by the EHTC was an independent determination of M87*'s mass: $(6.5 \pm 0.7) \times 10^9\,M_\odot$, consistent with the previous measurements of Gebhardt et al. (2011).

For its modelling, EHTC uses state-of-the-art techniques (see, e.g., Event Horizon Telescope Collaboration et al. 2019e, Porth et al. 2019, Gold et al. 2020, M87 2018 II) that combine ideal GR magnetohydrodynamics (GRMHD) with GR Ray Tracing (GRRT). GRRT is typically conducted as a post-processing step and is highly dependent on the assumed electron distributions that are not determined from first-principles methods. For example, the electrons emitting synchrotron radiation are accelerated by and change the local electromagnetic fields, and they may also interact with the radiation field. The addition of constraints that can guide microphysical kinetic models of particle acceleration is a clear way to reduce the current degeneracy in interpretation.

At the same time, the presence of strong, ordered magnetic fields near the SMBH is required to explain the detected linearly polarised radio synchrotron radiation as well as the angular momentum transport in the inflow and the launching of powerful jets of plasma (see, e.g., Blandford et al. 2019; Event Horizon Telescope Collaboration et al. 2021c). M87* currently expels vast bipolar jets. However, in most observations, only the one that is relativistically beamed towards Earth at approximately 20° from the line of sight is visible (Mertens et al. 2016), as the jets are accelerated to apparent velocities of up to $6c$ (where $c$ is the speed of light) on sub-arcsecond scales (Snios et al. 2019). Previous studies have found that the width and the outflow velocity of the approaching jet vary as a function of the distance from the SMBH (Asada et al. 2014), and other physical parameters possibly vary in a similar manner (Asada & Nakamura 2012). These jets radiate across the entire electromagnetic spectrum, with contributions to different parts of the spectral energy distribution (SED) arising at different places along their elongated structure. A complete model ultimately needs to include both the inflow and outflow components and be able to not only match the EHT image properties but also reproduce the inner jet dynamical properties as well as the overall multi-wavelength (MWL) spectrum.

For the 2017 EHT papers on M87*, complementary MWL information, such as the estimated minimum jet power ($P_{\rm jet} \geq 10^{42}$ erg s$^{-1}$; e.g. Reynolds et al. 1996; Stawarz et al. 2006; de Gasperin et al. 2012; Prieto et al. 2016), allowed roughly half of the $\sim 45$ initial GRMHD models to be ruled out, including all models with zero spin. However, MWL constraints on the SED such as the X-ray luminosity were only applied as an upper limit and thus did not exclude many additional models. In stark contrast, the inclusion of MWL data to compare against SEDs for the second 2017 EHTC image, that was of Sagittarius A* (Sgr A*; the SMBH in the centre of the Milky Way), provided a significantly augmented constraining power (Event Horizon Telescope Collaboration et al. 2022a,b,c,d,e,f; Wielgus et al. 2022). Specifically, the infrared and X-ray observations on Sgr A*'s emission from the Keck Observatory, the Very Large Telescope (VLT), the *Chandra* X-ray Observatory, and Nuclear Spectroscopic Telescope Array (*NuSTAR*) provided particularly tight constraints and allowed us to downselect from a large number of original synthetic images that were generated from models covering a much wider parameter space range than the original set for M87*.

We have acquired a full set of complementary, contemporaneous MWL observations for M87* to support modelling and interpretation as well as to provide a legacy dataset for the community. Together with the EHT images, these datasets provide rich input to help distinguish between the currently large range of theoretical scenarios for accretion and jet launching as well as for particle acceleration, that powers the emitted radiation. The first paper in this series was published as EHT MWL Science Working Group et al. (2021) (hereafter M87 MWL2017), and it presents the extensive MWL observations carried out alongside the EHT 2017 campaign on M87 by the EHT-MWL Science Working Group. The working group included EHTC members as well members of more than a dozen other partner facilities on the ground and in space. M87 MWL2017 includes VLBI images and spectral index maps at several frequencies lower than the 230 GHz band probed by EHT together with an SED spanning centimetre-band radio through TeV $\gamma$-rays, a range of more than 17 decades in frequency. In 2017, this source was found to be in a historically low state at all frequencies, with the core flux dominating over the HST-1 knot at high energies, making it possible to combine the high energy core flux constraints with the more spatially precise VLBI data.

In M87 MWL2017 and the present work, we merge the VLBI and spectral constraints by 'tagging' the radio flux points with VLBI constraints on the emitting region size, a constraint that is often missed by single-zone SED modelling approaches. In both papers we also explore M87*'s jet properties in an order-of-magnitude approach via two heuristic single-zone models meant to help reveal general trends and allow rough comparisons between basic properties year to year. Single-zone models are a common first approach that often successfully account for most of the optical and higher energy emission in many AGN, particularly blazars. The success of these techniques implies that a

* Tables from Appendix C (from C.1 to C.15) are only available in electronic form at the CDS via anonymous ftp to cdsarc.u-strasbg.fr (130.79.128.5) or via http://cdsweb.u-strasbg.fr/cgi-bin/qcat?J/A+A/

** Project leader and *Fermi*-LAT corresponding author. For questions concerning *Fermi*-LAT results contact giacomo.principe@ts.infn.it

*** H.E.S.S. corresponding author. For questions concerning H.E.S.S. results contact hess@hess-experiment.eu.

**** MAGIC corresponding author. For questions concerning MAGIC results contact contact.magic@mpp.mpg.de.

† VERITAS corresponding author. For questions concerning VERITAS results contact wjin@astro.ucla.edu, jmsantander@ua.edu.

significant fraction of energy in the jets is converted into particle energy in a relatively localised region. How this process occurs is one of the most urgent and outstanding questions in the field today, relating to identification of the source of very high energy (VHE) particles detected on Earth, such as cosmic rays (CRs) and neutrinos, as well as understanding the electromagnetic counterparts of explosive transient events, such as compact object mergers involving a neutron star. Localisation is thus a key step in constraining the particle acceleration mechanism as well as in investigating the hadronic content of the jet.

M87 MWL2017, however, demonstrated that M87* requires a stratified jet model, meaning that a single-zone leptonic model cannot adequately describe the sub-millimetre and high-energy SED simultaneously. In particular, we showed that the $\gamma$-rays cannot be fit by the same single-zone model that can match the EHT flux and size constraints. While this would seem to exclude scenarios where the $\gamma$-rays detected in 2017 in the quiescent (i.e. non-flaring) state are emitted from the same region as the sub-millimetre radiation in the EHT image, we discuss new advances in the modelling that may indicate otherwise.

Meanwhile, a new constraint that has yet to be incorporated into the EHT modelling is the source emission variability. In the EHTC Sgr A* papers, it was shown that the root-mean-square variability is one of the most difficult constraints to satisfy, and in fact, almost all of our GRMHD models are somewhat too variable compared to the data (see Event Horizon Telescope Collaboration et al. 2022g). Based only on the 2017 observations, it was not possible to draw conclusions regarding the variability of M87*. For prograde orbits, the dynamical timescale for light to circle M87*'s innermost stable circular orbit ranges from just under a week to approximately a month, depending on the spin, that is currently unconstrained. Therefore, daily images taken over a single week can be considered to correspond effectively to a single snapshot. However, one year represents a significant jump in time for M87*, roughly ten to 70 dynamical timescales, providing a chance to temporally resolve its uncorrelated states.

The EHTC has recently published the first results from its April 2018 campaign on M87* (M87 2018 I). The 2018 EHT images of M87* reveal an asymmetric ring structure that is brighter in the southeast and with a diameter of $\sim$43 $\mu$as. This is remarkably consistent with the EHT images obtained from the 2017 observations, strongly confirming the presence of a supermassive black hole with a mass $M_{\rm BH} \sim 6.5 \times 10^9\,{\rm M}_\odot$. The 2018 images show a significant shift in the position angle (by $\sim$30°) of the ring brightness asymmetry with respect to that of the 2017 images, suggesting the presence of variations on yearly timescales in the event-horizon-scale structures. However, the compact flux density in the 2018 EHT data is less well constrained than in 2017 due to the more limited coverage of short-to-intermediate baselines. Bayesian image reconstruction methods yield a relatively large range of $\sim$0.5–1.0 Jy within a scale of $\lesssim 100\,\mu$ (see M87 2018 I for a full description of the 2018 EHT observations).

During the accompanying 2018 MWL campaign, all currently operating ground-based imaging atmospheric Cherenkov telescopes (IACTs; H.E.S.S., MAGIC, and VERITAS, see Sect. 2.4.2) as well as *Fermi*-LAT measured the first $\gamma$-ray flare seen in M87* in eight years. In this paper, we present the results of the 2018 MWL campaign as well as a comparison with the 2017 results. In Sect. 2, we describe the new MWL observations with band-specific results and, where relevant, light curves and comparisons to prior observations, while we report in Appendix B further details on the data processing. In Sect. 3, we present the compiled nearly simultaneous SED of M87* for the 2018 MWL campaign. In Sect. 4, we present the best fits using phenomenological single-zone models similar to those used in M87 MWL2017 and discuss indications for structural changes between 2017 and 2018. In Sect. 5, we give our conclusions. All data files and products are available for download, as described in Sect. 3 and in Appendix. Data availability.

## 2. Observations and data reduction

The 2018 EHT observations of M87 were made in April 2018 using a total of eight stations (in six geographical locations) across the globe. While observations were performed over four nights (21, 22, 25, and 28 April), the second and last sessions suffered from various issues (bad weather conditions across the array, poor baseline coverage, etc.). Therefore, reliable EHT images of M87* could only be obtained from the data taken on 21 and 25 April.

In the following subsections we present brief descriptions of the 2018 MWL campaign on M87*— more detailed descriptions including data processing procedures and band-specific analyses appear in Appendix B. To aid readability, tabulated data are collected in Appendix C. Figure 1 shows a schematic overview of the observations carried between March and May 2018 for the second EHT-MWL campaign on M87. In addition, to observations carried during March–May 2018 we considered for our study also global VLBI (carried on 1 February 2018) and HST (carried on 30 July 2018) observations.

### 2.1. Radio observations

Here we outline the radio-millimeter observations obtained during the 2018 campaign with various VLBI facilities and connected interferometers. Following M87 MWL2017, we use the term "radio core" to represent the innermost part of the radio jet imaged by VLBI. A radio core in a VLBI jet image is conventionally defined as the most compact (often unresolved or partially resolved) feature seen at the apparent base of the radio jet (e.g. Lobanov 1998; Marscher 2008). For this reason, different angular resolutions by different VLBI instruments and frequencies, together with the frequency-dependent synchrotron optical depth, can lead to differences in the identification of a radio core in each observation (see also M87 MWL2017, for a related discussion).

#### 2.1.1. VERA 22 GHz

The nucleus of M87 was densely monitored over 2017 and 2018 at 22 GHz with the VLBI Exploration of Radio Astrometry project (VERA; Kobayashi et al. 2003), as part of a regular monitoring program of a sample of $\gamma$-ray bright AGN (Nagai et al. 2013). A total of 17 and 13 epochs were obtained in 2017 and 2018, respectively (see Table C.1). During each session, M87 was observed for 10–30 minutes with an allocated bandwidth of 16 MHz, sufficient to detect the bright core and create its light curves (additional details appear in Appendix B.1.1). Peak flux density light curves obtained with VERA at 22/43 GHz during 2017–2018 are included in Fig. 2.

#### 2.1.2. EAVN/KaVA 22 and 43 GHz

Between January 2017 and June 2018, M87 was regularly monitored at 22 and 43 GHz with the East Asian VLBI Network (EAVN; Cui et al. 2021; Akiyama et al. 2022), a joint VLBI array of radio telescopes in East Asia. A total of 22 (9 at 22 GHz and 13 at 43 GHz) and 20 (9 at 22 GHz and 11 at 43 GHz) epochs were obtained in 2017 and 2018, respectively (details in Appendix B.1.2 and Table C.1). Peak flux density light curves obtained with EAVN at 22 and 43 GHz during 2017–2018 are included in Fig. 2.

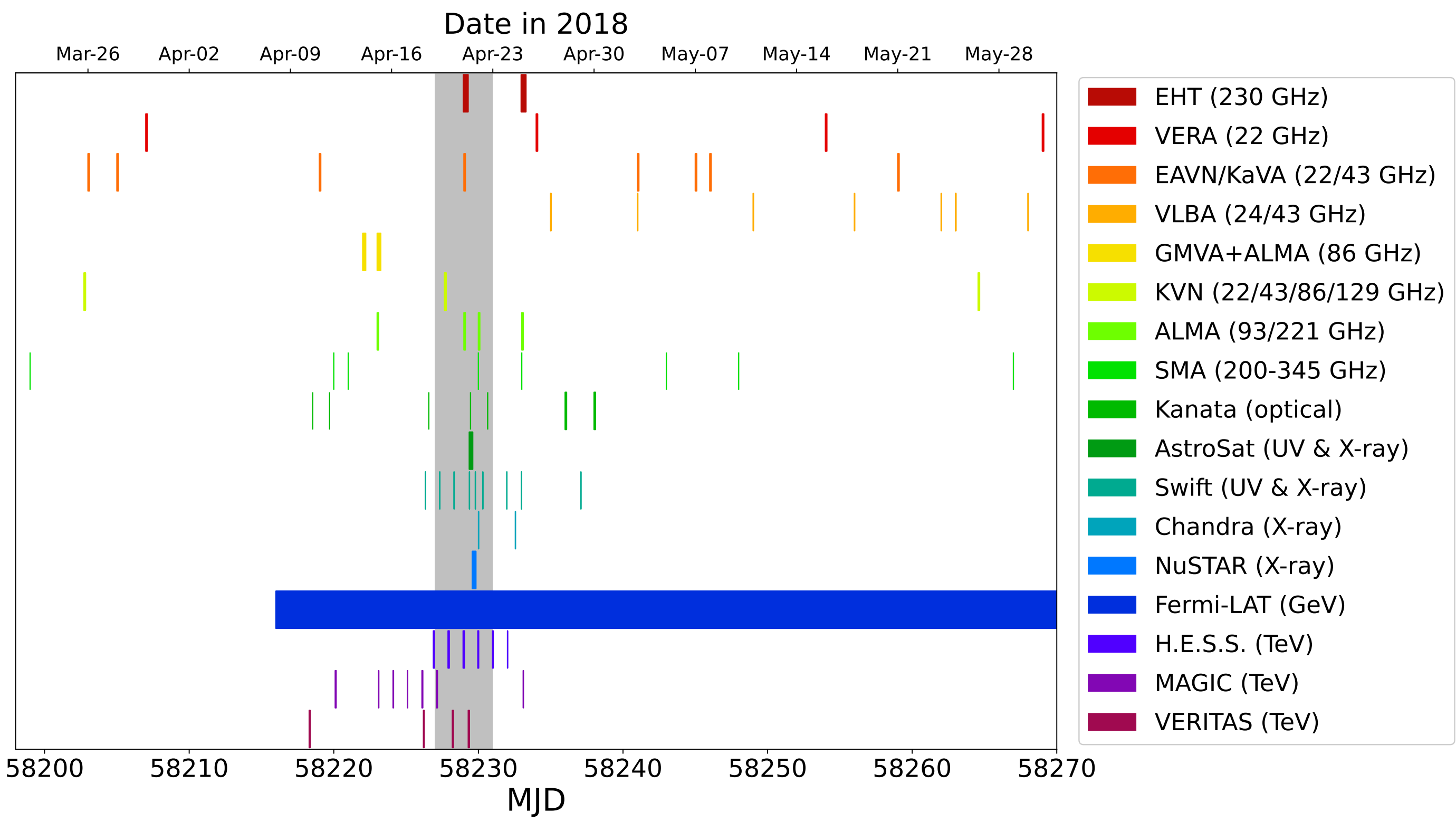

**Fig. 1.** Instrument coverage summary of the 2018 M87 MWL campaign covering MJD range 58198–58270. The grey band indicates the time interval of the VHE $\gamma$-ray flare episode.

### 2.1.3. VLBA 24 and 43 GHz

The program targeting M87 using the National Radio Astronomy Observatory (NRAO) Very Long Baseline Array (VLBA) at central frequencies of 24 and 43 GHz was initiated in 2006 (Walker et al. 2018) and lasted until 2020. Presented here are 10 epochs of observations that were obtained in 2018 and include two imaging (i.e. full-track) and eight short (snapshot) sessions on M87 (see summary in Appendix B.1.3 and Table C.1). Peak flux density light curves obtained with VLBA at 22 and 43 GHz during 2017–2018 are included in Fig. 2.

### 2.1.4. GMVA+ALMA 86 GHz

M87 was observed by the Global Millimetre VLBI Array (GMVA) for the first time in concert with the phased Atacama Large Millimetre/submillimetre Array (ALMA) and the Greenland Telescope (GLT) on 14–15 April 2018 (See details in Appendix B.1.4 and Table C.1).

### 2.1.5. KVN 22, 43, 86, and 129 GHz

M87 was regularly monitored by the Korean VLBI Network (KVN) at 22, 43, 86 and 129 GHz, as part of the Interferometric Monitoring of Gamma-ray Bright Active galactic nuclei (iMOGABA; Lee et al. 2016) program. More information can be found in Appendix B.1.5 and Table C.1. Peak flux density light curves obtained with KVN at 86/129 GHz during 2017–2018 are included in Fig. 2.

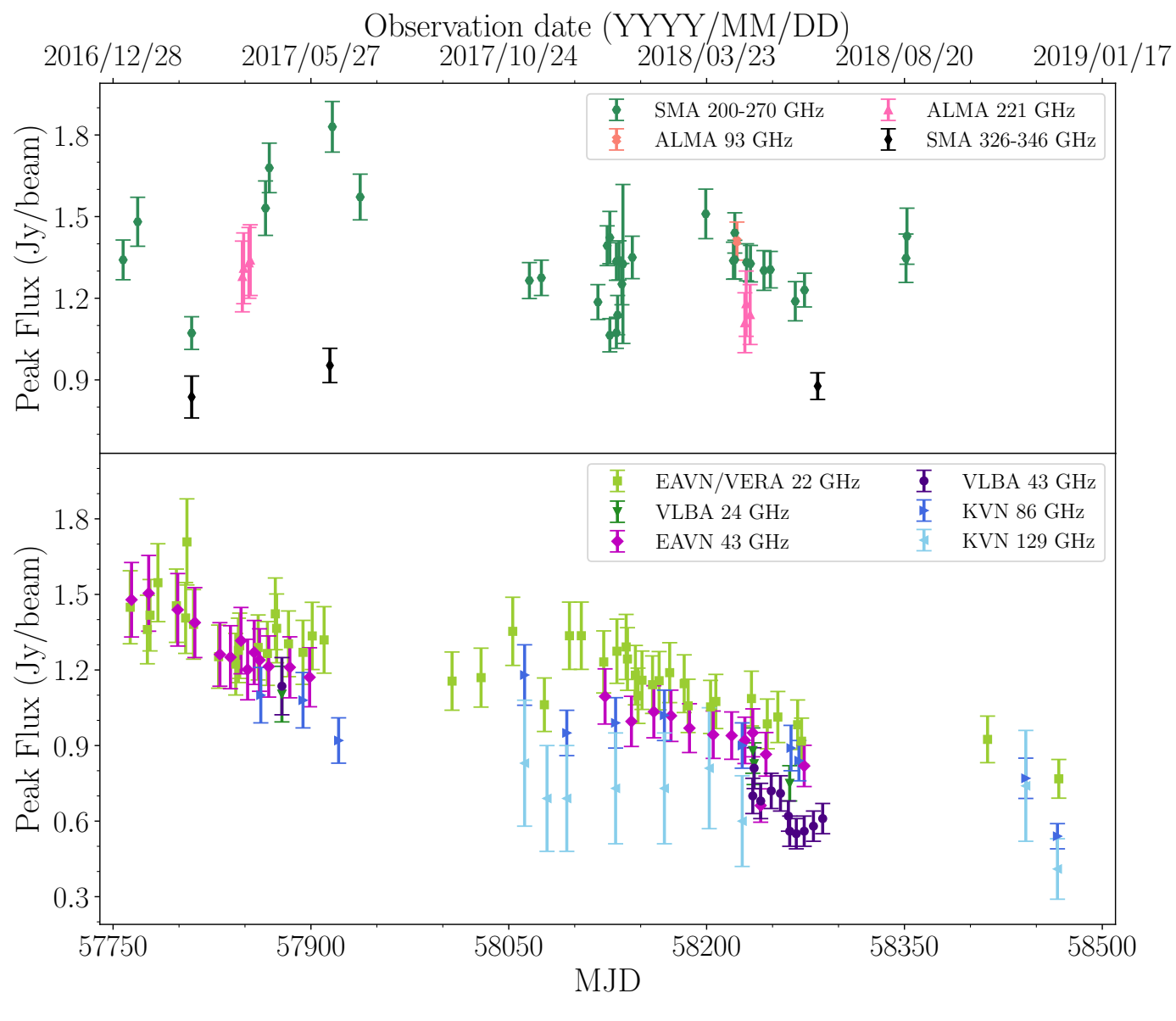

**Fig. 2.** Radio light curves of the M87 core in 2017 and 2018 at multiple bands. The upper and lower panels are for connected interferometers and VLBI, respectively. The corresponding beam sizes are indicated in Table C.1. KVN data at 22 and 43 GHz are not shown here since KVN captures the data from the shortest baselines of EAVN.

#### 2.1.6. Global 43 GHz VLBI

M87 was observed on 1 February 2018 at 43 GHz by a global array of sensitive VLBI antennas in full-track to maximise the imaging sensitivity and resolution. While details of the observation and imaging results will be reported elsewhere (Kim et al., in prep.), we provide a peak flux density of the core and a VLBI-scale total flux density (see Tables C.1 and Table C.1). We estimate these measurements to be uncertain up to ~ 20% considering the absolute flux calibration error and systematic uncertainties in defining the core and the field-of-view (see also Appendix B.1.6).

#### 2.1.7. ALMA 93 GHz and 221 GHz

Observations with phased-ALMA (Matthews et al. 2018; Goddi et al. 2019; Crew et al. 2023) were conducted as part of the 2018 VLBI campaigns with the GMVA at 3 mm / 93 GHz (ALMA Band 3) and the EHT at 1.3 mm / 221 GHz (ALMA Band 6). The main observational and imaging parameters for the 2018 observations are summarised in Table C.2, while those for 2017 are summarised in M87 MWL2017, as well as included in Fig. 2. For each dataset and corresponding image, the table also reports flux-density values both in the central compact core and the extended jet (see Appendix B.1.7 for more detailed descriptions of data acquisition and reduction).

#### 2.1.8. SMA 200–270 and 345 GHz

The long term 1.1–1.4 mm band (200–270 GHz) and 0.87 mm band (345 GHz) flux density light curve data for M87 shown in Fig.2, and later in Fig.14 and 16, were obtained at the Submillimeter Array (SMA) near the summit of Mauna Kea (Hawai'i). M87 is included in an ongoing monitoring program at the SMA to determine flux densities for compact extragalactic radio sources that can be used as calibrators at millimetre wavelengths (Gurwell et al. 2007). A summary of the measurements made from these data for 2017 and 2018 is shown in Table C.2. The reported core flux for M87 is the vector average of baselines longer than 30 k$\lambda$ (see Appendix B.1.8 for more details).

### 2.2. *Optical and UV observations*

#### 2.2.1. Optical observations by the Kanata telescope

Photometric images of M87 were obtained using the 1.5-m Kanata telescope and the Hiroshima Optical and Near-InfraRed Camera (HONIR), that can perform simultaneous two-band imaging (Akitaya et al. 2014). We used optical $R_C$ (634.9 nm) and near-infrared $J$-band (1250 nm) filters for the observations. Images with a field of view of 10″×10″ and a spatial resolution of 0.294″ pixel$^{-1}$ through the band-pass filters were obtained. The central flux of M87 within a radius of 5″ was measured. The measured flux should include three components: radiation from the central region near the BH, radiation from the HST-1 knot (located at a projected distance of ~70 pc from the core), and thermal radiation from the host galaxy. Thus, the measured flux is an upper limit of the flux of the central nuclear region. Additional details appear in Appendix B.2.1.

#### 2.2.2. UV and optical observations from *Swift*-UVOT

During the *Swift* pointings, the UVOT instrument observed M87 in its optical ($v$, $b$ and $u$) and UV ($w1$, $m2$ and $w2$) photometric bands (Poole et al. 2008; Breeveld et al. 2010). An aperture of 5″ radius was used for the flux extraction for all UVOT bands. Compared to the values obtained during the EHT 2017 campaign, we observed with UVOT in 2018 an increase of the flux by a factor of ~3 in optical and a factor of ~ 2 in UV bands. However, the high contribution of the host galaxy flux in the V band and partly in the b band contributes to the large uncertainties on the flux. A detailed description of data, analysis and modelling is given in Appendix B.2.2 (for all the results of the *Swift*-UVOT observations during the EHT 2018 campaign see Table C.5.).

#### 2.2.3. UV observation with HST

We have investigated the HST archive to search for HST UV/optical observations during the 2018 EHT MW campaign. Unfortunately, the HST observation nearest to the campaign was performed more than 3 months later, on 2018 July 30 (MJD 5832.90764). The observations were carried out with the STIS camera (NUV-MAMA detector with F25QTZ aperture) over two HST orbits at the pivot wavelength 2359.7Å (proposal ID 15358). Since none of the other Optical/UV/NUV instruments used in this campaign allow measurements of the core and HST-1 knot fluxes separately at these wavelengths, we have utilised the HST data to obtain fluxes of the core and HST-1 knot.

As in our previous work M87 MWL2017, we have performed photometric decomposition to exclude the host galaxy flux from the image. The host was approximated with two Sersic functions (Sersic 1968) with the core and jet regions masked out. Unfortunately, due to the small FoV of STIS, no stars were available in the images for constructing a PSF. The HST PSF library does not contain PSFs for the STIS instrument, and the HST team advises against using the *TinyTim* package[1] to build PSFs for STIS images. Nevertheless, we have performed a decomposition without taking into account the PSF, rather by masking the central part of the host galaxy and performing the decomposition of outer parts of the galaxy and extrapolating the result to the centre.

Figure 3 shows the residual image (after host model subtraction), where the core and the HST-1 knot are clearly visible. To obtain the fluxes of these features, we performed aperture photometry with an aperture radius of 0.″4 (the same as in M87 MWL2017). Table C.6 gives the fluxes of the core and HST-1 knot before and after host model subtraction. The uncertainties of fluxes before correction for the host galaxy were calculated based on the STIS documentation[2], while we assume uncertainties of the corrected fluxes of 50%, taking into account the method of the decomposition. However, as can be seen in Table C.6, contributions of the host galaxy are small since the band is blue, while the host galaxy is an old elliptical type. The results indicate that the HST-1 knot is significantly fainter than the core, that most likely was also the case during the EHT MW campaign in 2018.

#### 2.2.4. Far-UV observations from AstroSat-UVIT

The UV Imaging Telescope (UVIT) instrument on board the AstroSat mission consists of twin telescopes, each of 38 cm diam-eter, one dedicated to far-UV (FUV: $\lambda$ = 1300 - 1800 Å) and the other to near-UV (NUV: $\lambda$ = 2000 - 3000 Å) & Visible (VIS: $\lambda$ = 3200 - 5500 Å) channels (Kumar et al. 2012). In this work, the far-UV images from UVIT are used to constrain the jet and the HST-1 knot component in the UV emission. The images created for each orbit of the satellite were aligned and merged to create the final image with an effective exposure time of 14.1 ks. The flux contributions of different known components of M87 are extracted by 2D image decomposition modelling. Figure 4 shows the photometric slice made through the galaxy centre in the jet direction obtained in the BaF2 band (F154W: $\lambda_{eff}$ ~1541 Å, with $\Delta\lambda_{eff}$ ~380 Å). In this figure the observed galaxy flux is shown with a thick black line, the total model (a sum of all components) with a red line, the host galaxy model with an orange line, the total jet model, that include the asymmetric Core-Sersic model (Graham & Driver 2005) and a point source for the HST-1 knot, is shown with a green line, and the central source with a blue line. A detailed description of data, analysis and modelling is given in Appendix B.2.3.

[1] www.stsci.edu/files/live/sites/www/files/home/hst/instrumentation/focus-and-pointing/documentation/_documents/tinytim.pdf

[2] https://hst-docs.stsci.edu/stisdhb/chapter-4-stis-error-sources/4-2-summary-of-accuracies

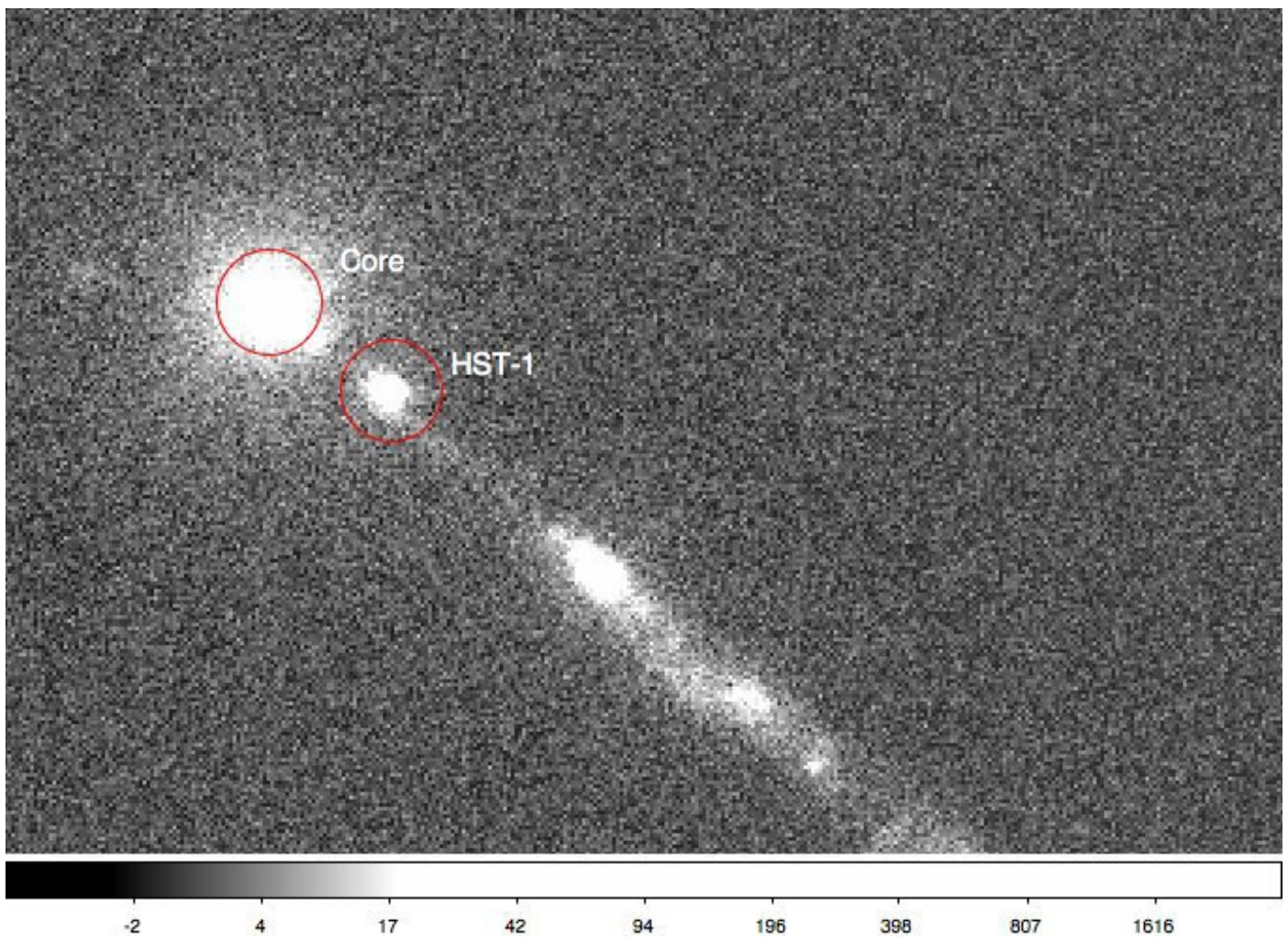

**Fig. 3.** Hubble Space Telescope image of M87 at 2359.7Å with the host galaxy subtracted. The core and HST-1 are indicated; the distance between the features is 1.”00±0.”05, and the pixel size is 0.”0248.

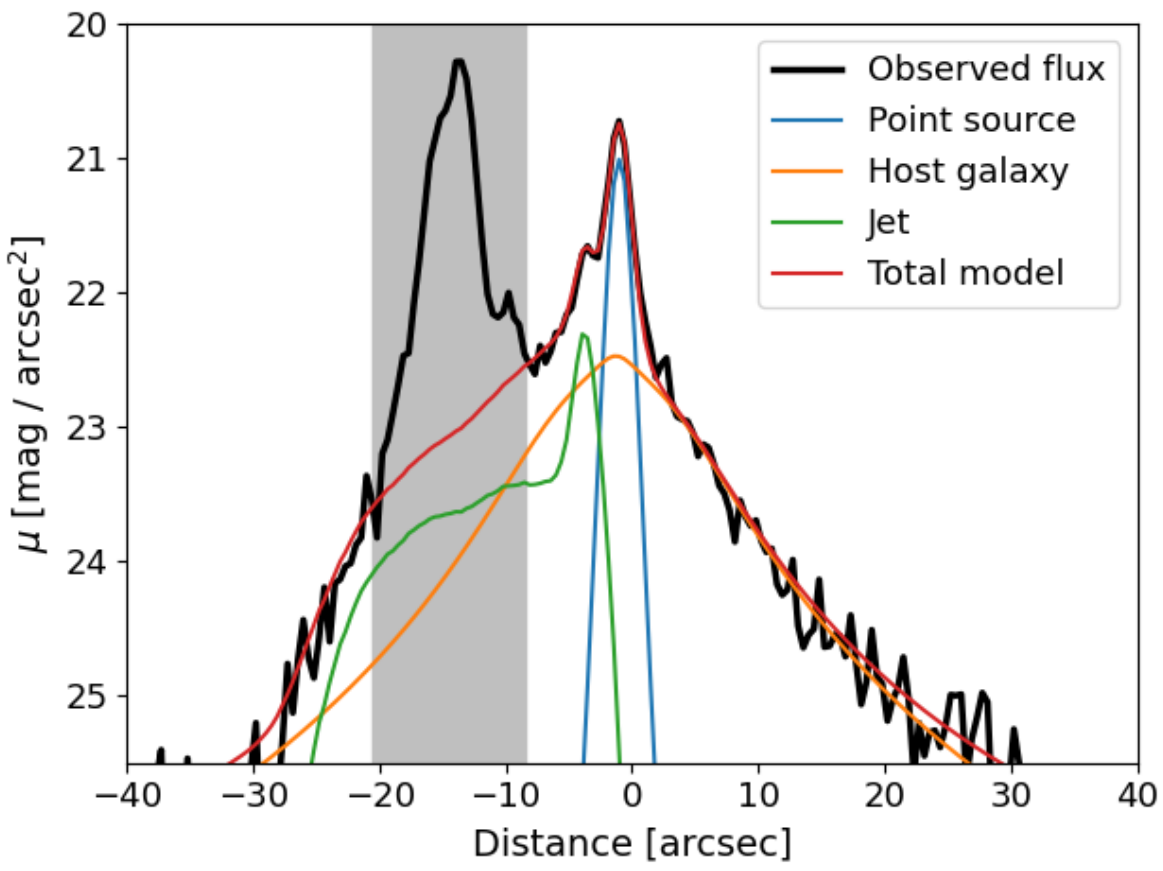

**Fig. 4.** AstroSat-UVIT photometric cut of the decomposition model along the jet direction. The thick black line shows the observed flux along the cut, the blue line is the central point source, the orange line is the host galaxy, the green line is the jet total flux, and the red line is the total model flux. The shaded area shows the masked out region of the jet that has a complex fine structure that cannot be accurately modelled with a simple analytical function and was therefore excluded from the decomposition to reduce the model complexity.

### 2.3. X-ray observations

#### 2.3.1. *Chandra* and *NuSTAR* observations and joint spectral analysis

Observations of M87 with the Advanced CCD Imaging Spectrometer (ACIS) on board the *Chandra*X-ray Observatory were collected via Director’s Discretionary Time (DDT; PI: Wong) on 22 April 2018 (ObsID 21075 for 9.1 ks starting at UT 00:09) and 24 April 2018 (ObsID 21076 for 9.0 ks starting at UT 13:20). An observation with *NuSTAR* (Harrison et al. 2013) was also obtained via DDT on 25 April 2018 (ObsID 60466002002 for 21.1 ks beginning at UT 13:56), sufficiently close in time to the second *Chandra* observation that they can be considered contemporaneous. The data from these observations have been included in previous studies of X-ray flux variability in M87 described in Yang et al. (2019) and Imazawa et al. (2021).

Following the procedure in M87 MWL2017, the two *Chandra* observations are combined in our spectral analysis such that the reported fluxes represent the average for 22–25 April 2018. The variability of the background-subtracted count rates for each of the spatial components is shown in Fig. 5. As is clear from the *Chandra* light curve, the core was significantly brighter in both observations than in April 2017. No significant variability is detected within either of the 2018 observations. However, the core flux did increase between the two observations separated by approximately two days, while the other components remained steady (see Fig. 5). Moreover, an increase in the X-ray brightness of M87 relative to 2017 is apparent directly from the *NuSTAR* spectrum, that exhibits higher S/N out to ≃ 60 keV than the 2017 data (c.f. M87 MWL2017, Fig. 9), despite equal exposure time.

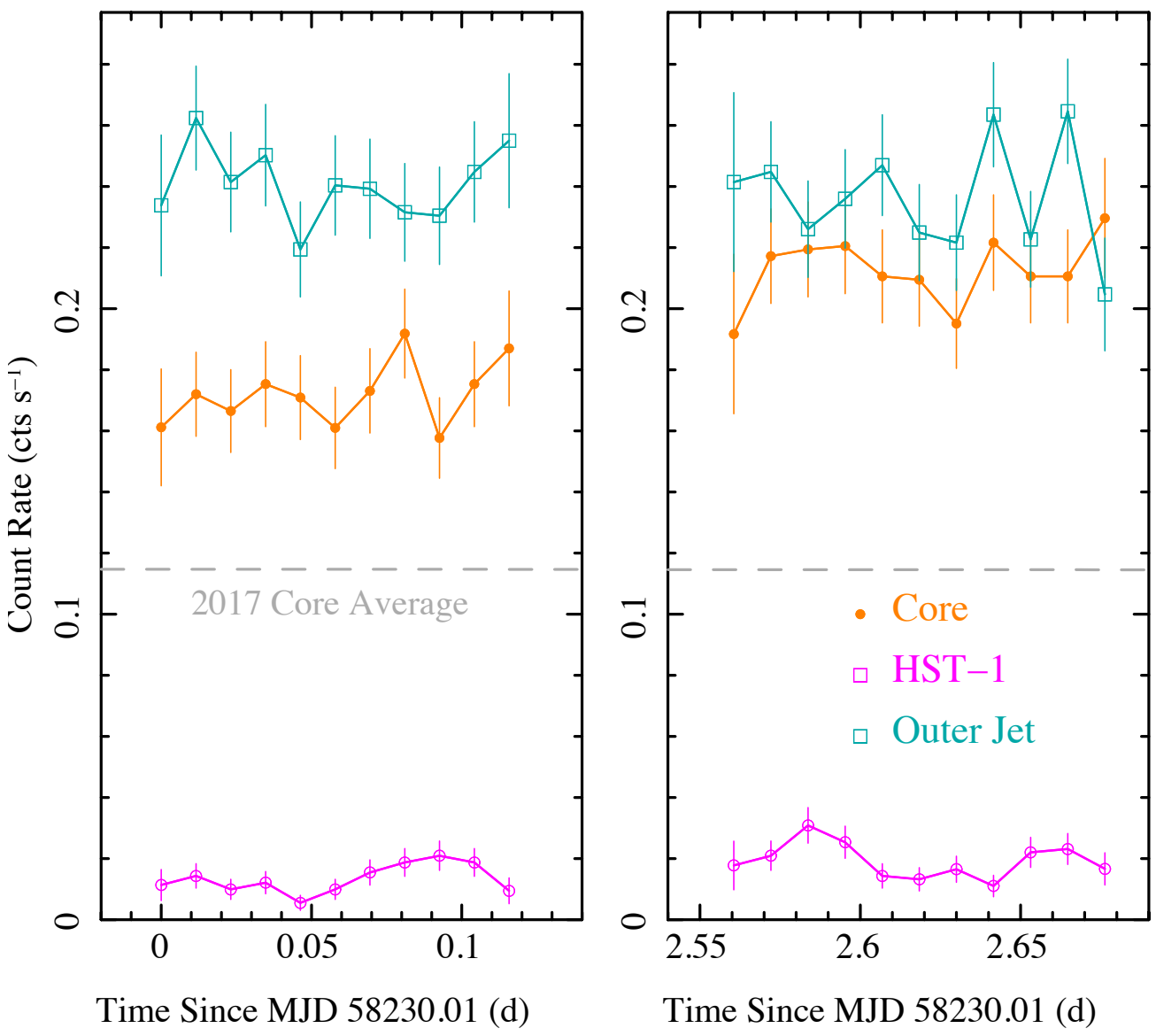

**Fig. 5.** X-ray light curves of the spatially resolved components of M87 from the *Chandra* observations in 2018. The core (orange) shows clear variability between the two different days on which observations have been carried out, while the HST-1 knot and the rest of the jet remain fairly steady.

In particular, we detect a significant increase in the flux of the core component relative to 2017 observations. The unabsorbed core flux in the 2–10 keV band averaged between the two *Chandra* observations in April 2018 was $2.6 \times 10^{-12}$ erg s$^{-1}$ cm$^{-2}$ that represents an ~80% increase on a timescale of one year. Assuming a distance of 16.8 Mpc, we find the average 2–10 keV band luminosity of the core to be $(8.8 \pm 0.4) \times 10^{40}$ erg s$^{-1}$, again a marked increase over $(4.8 \pm 0.2) \times 10^{40}$ erg s$^{-1}$ observed in 2017. The significantly higher luminosity is not seen in the HST-1 knot or the rest of the jet, that both show $\lesssim 10\%$ variability. These results are similar to those obtained by Imazawa et al. (2021). Despite the flux increase, we note that there is no evidence for a change in the shape of the spectrum. Our best fit power-law (PL) photon index for the core is $\Gamma_{\rm core} = 2.03^{+0.12}_{-0.07}$, that is statistically consistent with $\Gamma_{\rm core} = 2.06^{+0.10}_{-0.07}$ from the 2017 EHT observations.

In addition to the observations taken within the EHT MWL campaign, we also examined data taken with *Chandra* earlier in 2018: OBSIDs 20488 (4 January 2018) and 20489 (21 March 2018). These observations were processed and analysed in exactly the same way as described above in order to evaluate the core flux using the same set of assumptions. We find that the core flux in the 2–10 keV band was lower in both epochs, $\simeq 2 \times 10^{-12}$ erg s$^{-1}$ cm$^{-2}$, as shown in Fig. 15.

Although a PL model provides a satisfactory fit to the April 2018 data alone and is suitable given the sensitivity of the spectrum in a typical single observation, the growing archive of *NuSTAR* observations of M87 permits a deeper look at the shape of the X-ray spectrum of the core. While the shape of the spectrum appears to be stable, there is evidence of deviation from a PL. In particular, our analysis of 14 existing *NuSTAR* observations (Sheridan et al., in preparation) reveals statistically significant evidence for a break in the PL around $E_{\rm br} = 10$ keV: compared to PL fits, models with a break improve the Cash fit statistic by $\Delta C > 150$ with the addition of only two free parameters. Relative to the pure PL model, the spectrum is softer below the break ($\Gamma_1$) and harder above the break ($\Gamma_2$), though the exact values depend on how the data are combined. For example, when all observations are stacked, we find $\Gamma_1 = 2.26^{+0.06}_{-0.05}$, $E_{\rm br} = 11 \pm 1$ keV, and $\Gamma_2 = 1.6^{+0.1}_{-0.2}$. When the spectral shape is tied between observations but the flux is allowed to vary, we find $\Gamma_1 = 2.20^{+0.08}_{-0.07}$, $E_{\rm br} = 12.5 \pm 1.2$ keV, and $\Gamma_2 = 0.9 \pm 0.2$. We emphasise that $\Gamma_1$, $E_{\rm br}$, and the existence of a break are weakly sensitive to the stacking method, while $\Gamma_2$ depends more on the details. In our SED modelling (Sect. 4) we therefore consider all three spectral models for the core. Additional details are included in Appendix B.3.1.

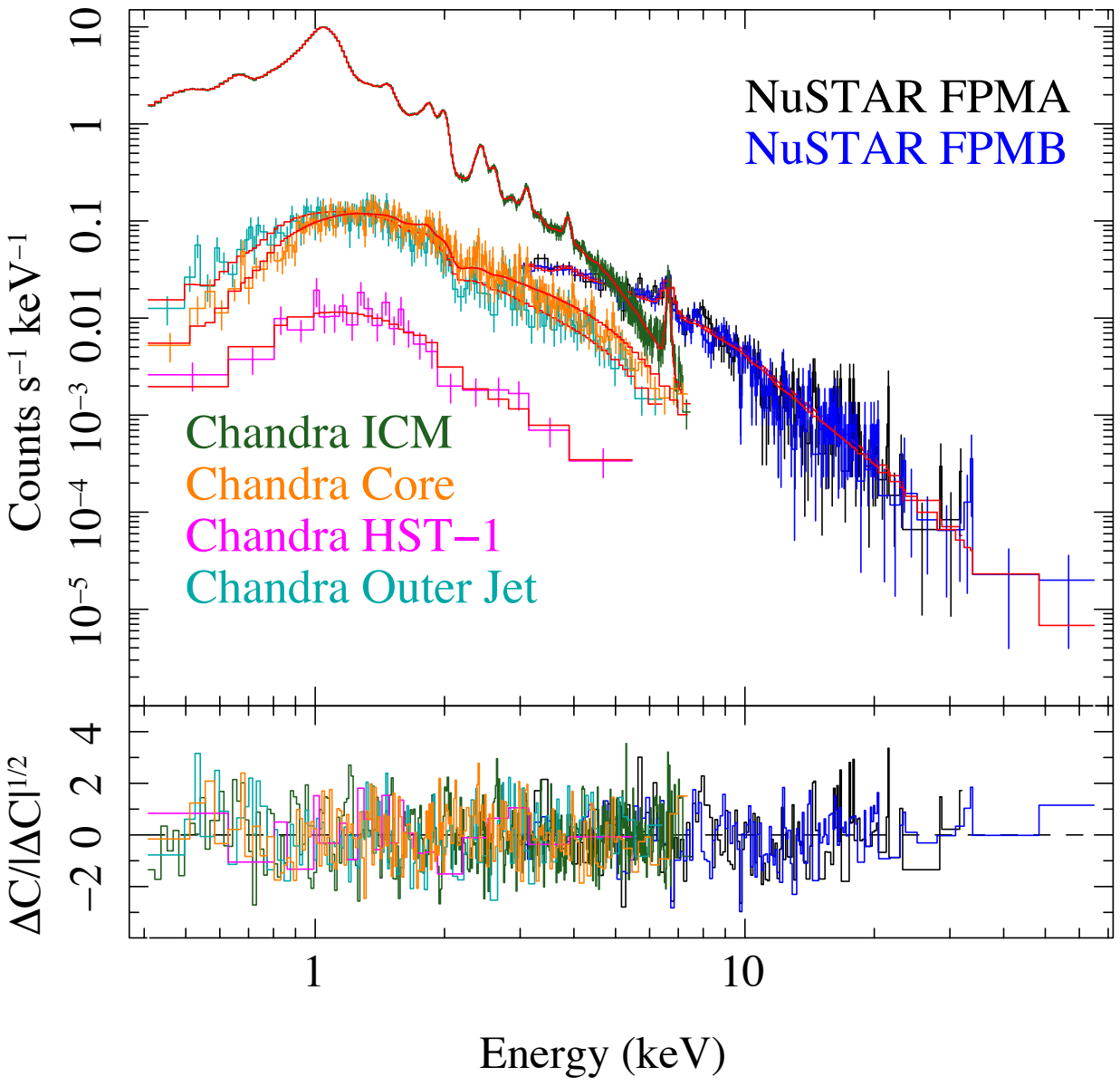

**Fig. 6.** X-ray spectra of M87 from *Chandra* and *NuSTAR* observations in April 2018. Colour-coding for the spatially resolved components is the same as in Fig. 5. The red curves represent the spatially resolved model appropriate for each dataset. The *NuSTAR* model is the sum of the other spatial components (with cross-normalisation constants to account for instrumental differences between *NuSTAR* and *Chandra* as well as between *NuSTAR*'s two focal plane modules).

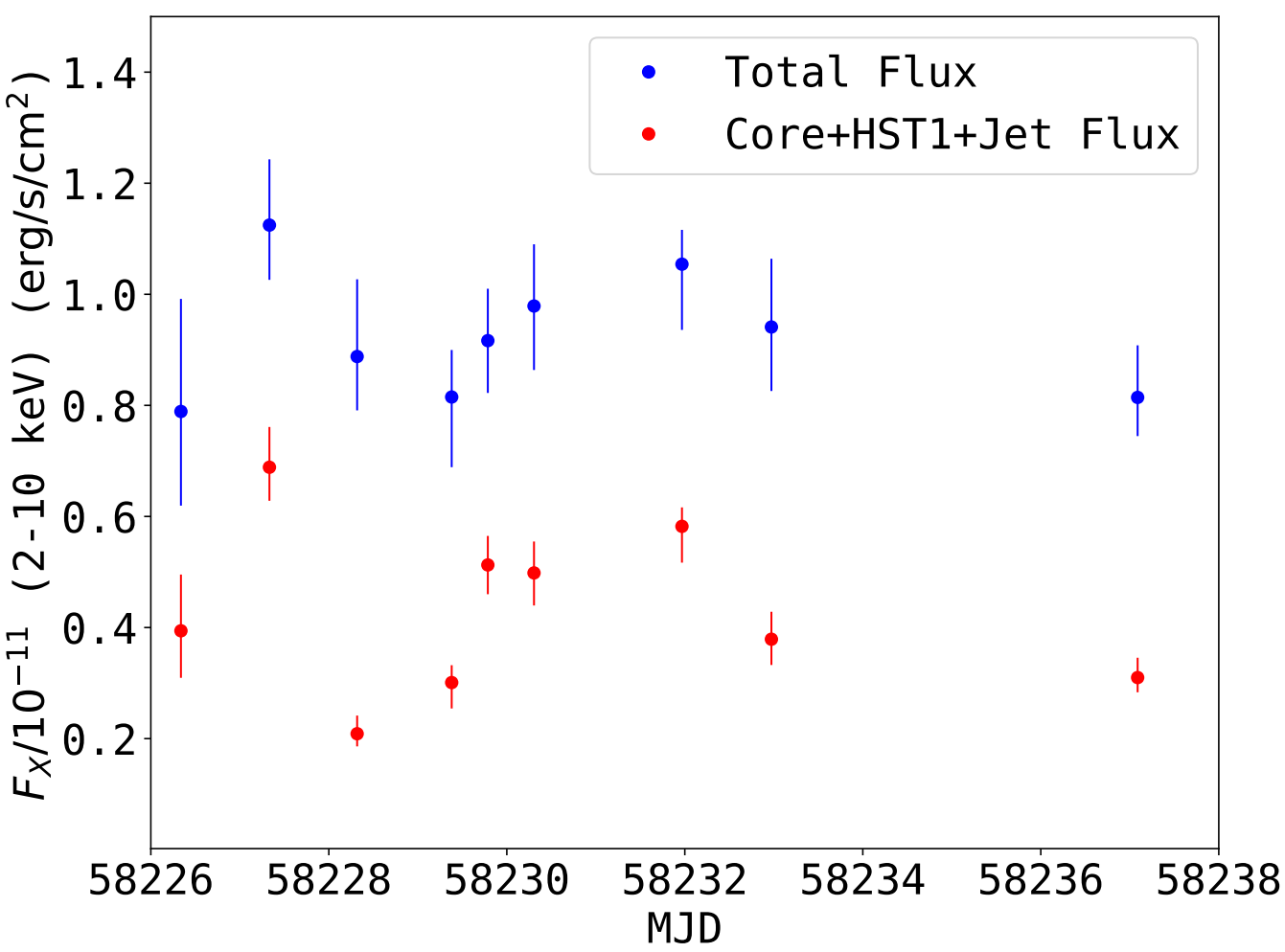

**Fig. 7.** Total flux (blue) and unresolved summed component fluxes (red) of M87 in the 2–10 keV band observed with *Swift*-XRT in 2018.

### 2.3.2. *Swift*-XRT observations and analysis

As a part of the EHT campaign, M87 was observed with *Swift*-XRT from 18 April (MJD 58226) to 29 April 2018 (MJD 58237). A total of 9 observations were conducted in April, with one additional observation on 18 December 2018 (MJD 58240). To put the 2018 observations in context, we also analyse all *Swift*-XRT photon counting mode observations for most of the *Swift*-XRT mission duration, extending back to 2009 and forward to 2021. The details of our data processing and spectral analysis are given in Appendix B.3.2.

Analysing these observations constrains both the long-term and short-term variability of the X-ray emission from M87, even though *Swift*-XRT cannot spatially resolve the inner jet, including the core and the HST-1 knot. The resulting light curve for the 2018 observations is shown in Fig. 7; we include total unabsorbed flux estimates for M87 with background, as well as net unabsorbed flux estimated by removing the contributions from the background model components. The small-scale components' flux is found to be consistently between 40% and 60% of the total. However, the background-subtracted fluxes shown in Fig. 7 should only be interpreted as upper limits on the X-ray emission from the core, the HST-1 knot, and the jet.

### 2.3.3. AstroSat-SXT observations and analysis

AstroSat performed two pointing photon counting (PC) mode observations of M87, the first during 17–18 February 2018 (Ob-

sID: A04_115T02_9000001900; MJD 58166.8) and the second during 21–23 April 2018 (ObsID: A04_115T02_9000002042; MJD 58229.4).

We use a model similar to the one adopted for *Chandra* and *NuSTAR* spectral analysis with the addition of a `vvapec` component to account for the diffuse emission from cluster gas because the SXT samples a larger portion of it. The best-fit normalisation obtained from the SXT spectra is higher than the constraints in M87 MWL2017 by about a factor of three mainly for this reason (additional details appear in Appendix B.3.3).

### 2.4. Gamma ray observations

#### 2.4.1. *Fermi*-LAT observations

We performed a dedicated analysis of the *Fermi* Large Area Telescope (LAT) data (Atwood et al. 2009) of M87 using 12 years of data starting on 4 August 2008, in an energy range between 100 MeV and 1 TeV. We repeated the analysis using a time interval of one month centred on the EHT observation period (i.e. 8 April – 8 May 2018). This choice is also motivated by a problem that affected the solar panel on 16 March 2018, that stopped the observations until 8 April 2018. We summarise these analyses here and include additional details in Appendix B.4.1.

The results from 12 years present a significant excess above the noise level (a test statistic of $TS = 2015$, corresponding to a significance[3] of $> 40\sigma$ with an averaged flux of $F = (1.74 \pm 0.11) \times 10^{-8}$ ph cm$^{-2}$ s$^{-1}$. The obtained spectral behaviour ( $\Gamma = 2.05 \pm 0.03$) is in agreement with the 4FGL-DR2 result for this source (Abdollahi et al. 2020).

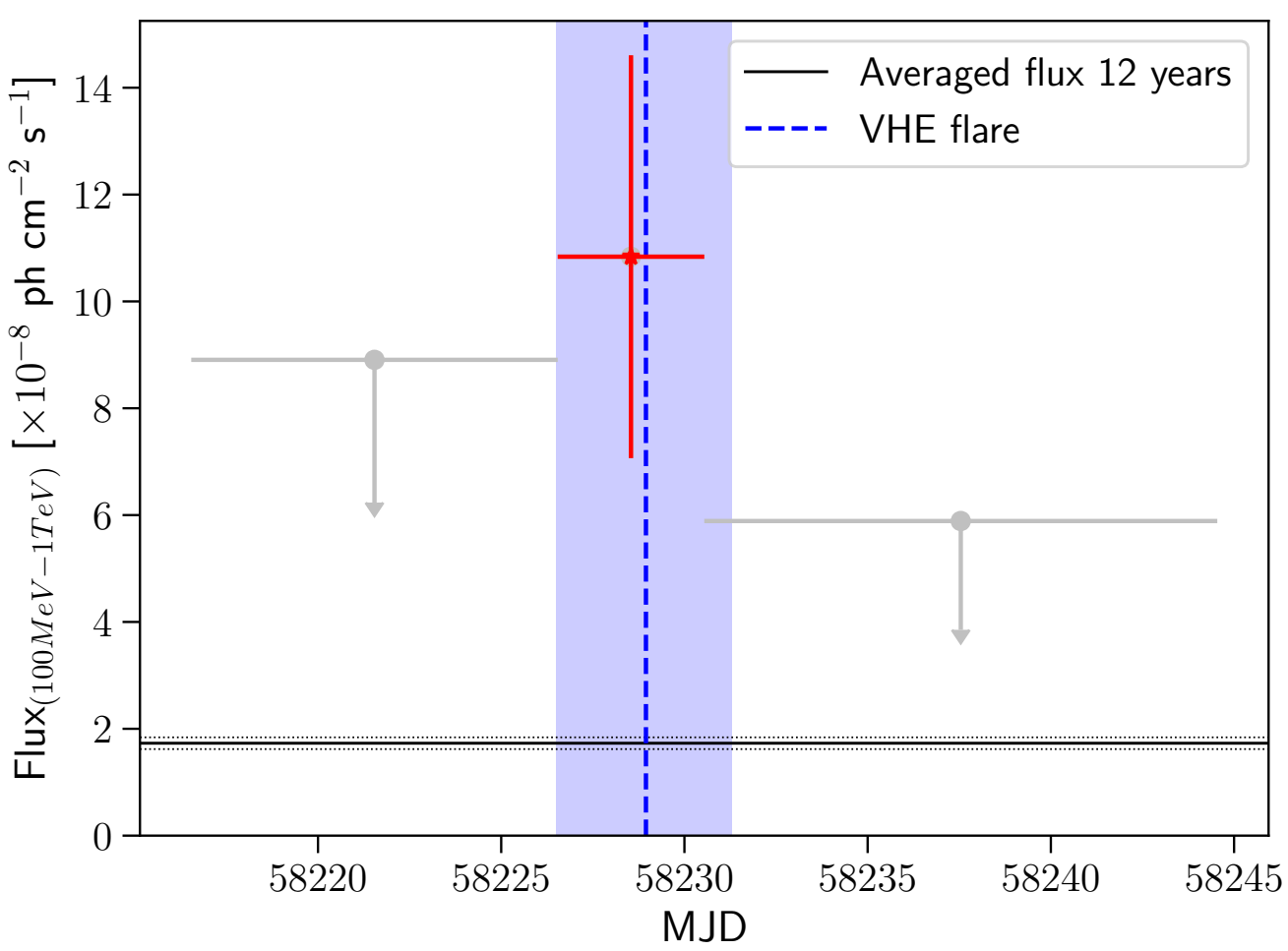

**Fig. 8.** *Fermi*-LAT light curve of M87 on customised time intervals during April 2018. The red point refers to the flux estimate, while grey points indicate upper limits. The $1\sigma$ upper limit is reported when $TS$ <10. The vertical blue line (band) indicates the time of the VHE flaring episode, while the horizontal lines represent the averaged flux over 12 years and its uncertainty $F_{12yr} = (1.74 \pm 0.11) \times 10^{-8}$ ph cm$^{-2}$ s$^{-1}$.

For the period 8 April – 8 May 2018, the source is detected with a significance of $\sim 6\sigma$ ($TS = 38$), $F = (5.0 \pm 1.7) \times 10^{-8}$ ph cm$^{-2}$ s$^{-1}$, three times higher than the average over 12 years. It also presents a soft spectrum with $\Gamma = 2.40 \pm 0.21$. Regarding the variability study, using for the light-curve study either one-day or two-day cadence temporal analyses, the source was detected (with more than three-sigma significance) in both of the analyses only in the one-day (and two-day) bin around the VHE flare episode. Namely, on the one-day bin centred on MJD=58228.0 (20 April 2018), presenting a significance of $3.5\sigma$ and flux $F = (23 \pm 10) \times 10^{-8}$ ph cm$^{-2}$ s$^{-1}$, as well as in the two-day bin centred on MJD=58227.5 (19–20 April 2018), with a significance of $3.4\sigma$ and flux $F = (14 \pm 6) \times 10^{-8}$ ph cm$^{-2}$ s$^{-1}$, respectively. Similarly, on both the one-week time scale and the customised time intervals (see Fig. 8 and Table C.13) the source is significantly detected (with a $TS$=20) only in the week MJD = 58224–58231 (18–24 April 2018), as well as in a four-day period of the VHE flare, namely MJD = 58227–58231, presenting a flux of $(8.1 \pm 2.6) \times 10^{-8}$ ph cm$^{-2}$ s$^{-1}$ and $(10.8 \pm 3.8) \times 10^{-8}$ ph cm$^{-2}$ s$^{-1}$, respectively. The weekly flux represents the highest weekly emission observed for this source so far. The second-highest weekly flux (($(7.2\pm3.1)\times10^{-8}$ ph cm$^{-2}$ s$^{-1}$) was registered in the first week of February 2014, during which the only VHE observations were obtained with MAGIC. The MAGIC data did not show a VHE flare at this time. Finally, for the four-day period of the VHE $\gamma$-ray flare, we performed a spectral analysis of the LAT data finding a PL index of $\Gamma = 2.18 \pm 0.33$, compatible with the one measured in the 12-year analysis.

#### 2.4.2. VHE observations: H.E.S.S., MAGIC, VERITAS

H.E.S.S. observations in 2018 were performed with the four 12 m CT1-4 telescopes (see Table C.14). A total of 13.0 hours of live time of good-quality data were obtained. The average zenith angle of the observations was 42.9°. A total statistical significance of 14.8 $\sigma$ was obtained for the 13.0 hours. For the systematic errors we conservatively estimated an uncertainty of 20% on the flux normalisation and 0.1 on the photon spectral index (H.E.S.S. Collaboration 2006b). Detailed information on the H.E.S.S. sensitivity and angular resolution can be found in Parsons & Hinton (2014) and H.E.S.S. Collaboration (2020).

MAGIC observed M87 for 11.3 hours after quality cuts for this campaign. The observations were carried out at a mean zenith angle of 29.6°. The total statistical significance from MAGIC observations during this campaign is 4.0 $\sigma$. Based on the detailed descriptions in Aleksić et al. (2016) we estimate a systematic uncertainty of $\sim$ 30% on the integral flux that consists of 11% uncertainty of the flux normalisation, ±0.15 on the spectral index and 15% uncertainty of the energy scale. The long-term MWL light curves presented later in Fig. 14 and Fig. 15 also include the MAGIC data collected in 2019. In 2019 MAGIC observed M87 for a total of 29.1 hours after quality selection. The statistical significance obtained from the 2019 observations is 7.0 $\sigma$.

VERITAS (Holder et al. 2006) observed M87 during April 2018 for 10.5 hours after quality cuts with an average zenith angle of about 22.9°. The total statistical significance from VERITAS observations during this campaign is 7.8 $\sigma$. We estimated the systematic uncertainty to be $\sim$ 25% and ± 0.2 on the absolute flux level and spectral index, respectively (Adams et al. 2022).

Further details of the individual analyses can be found in Appendix B.4.2.

The light curves of nightly integrated flux measurements from all VHE instruments above 350 GeV between 10 April 2018 (MJD 58218) and 25 April 2018 (MJD 58233) are shown in Fig. 9 with solid colours. We assumed a PL index of $\Gamma_{\rm VHE} = 2.3$ for the integrated flux calculations. We chose the same analy-sis threshold of 350 GeV as for our 2017 campaign (EHT MWL Science Working Group et al. 2021). Strong moonlight conditions prevented observations after 25 April 2018. In order to investigate possible flux variability, we fitted the flux measurements from all VHE telescopes with a constant flux model. This yields a $\chi^2$/d.o.f. of 42.61/16 corresponding to a probability of p $\approx 3 \times 10^{-4}$. In order to estimate the variability time scale, we fitted the light curve with a Gaussian function:

$$F(t) = F_0 + F_1 \cdot \exp\left(-\frac{(t-t_0)^2}{2\sigma^2}\right), \tag{1}$$

where $F_0$ was added as baseline flux. The Gaussian fit provides a better description of the measured data with $\chi^2$/d.o.f. of 16.46/13 corresponding to a probability of p $\approx$ 0.23. The best fit results are $F_0$=(1.25 ± 0.46) × $10^{-12}$ ph cm$^{-2}$s$^{-1}$, $F_1$=(2.83 ± 0.62) × $10^{-12}$ ph cm$^{-2}$s$^{-1}$, $t_0$=58229.07 ± 0.39 MJD (21 April 2018), $\sigma$=1.32 ± 0.44 days. Defining the variability time scale as the full width at half maximum (FWHM), we obtain FWHM = $2\sqrt{2\ln 2}\,\sigma \approx 2.35\,\sigma$ = 3.10 ± 1.03 days. For comparative purposes, the rising part of light curve was also fitted with an exponential function of the form

$$F(t) = p_0 e^{-|t-58226|/\tau}, \tag{2}$$

where the exponential time scale $\tau$ is 1.88 ± 0.27 days for the rising flux portion. Furthermore, a Gaussian function fit to H.E.S.S. data alone yields a $\chi^2$/d.o.f. of 0.44/2 corresponding to a probability of p $\approx$ 0.80, $F_0$=(0.77 ± 1.06) × $10^{-12}$ ph cm$^{-2}$s$^{-1}$, $F_1$=(4.50 ± 0.95) × $10^{-12}$ ph cm$^{-2}$s$^{-1}$, $t_0$=58229.15 ± 0.11 MJD, and $\sigma$=1.28 ± 0.34 days.

[3] The source significance is determined as the square root of the logarithmic ratio of the likelihood of a source being at a given position in a grid to the likelihood of the model without the source (Mattox et al. 1996), sign~sqrt(2log($\mathcal{L}_{src}/\mathcal{L}_{null}$)).

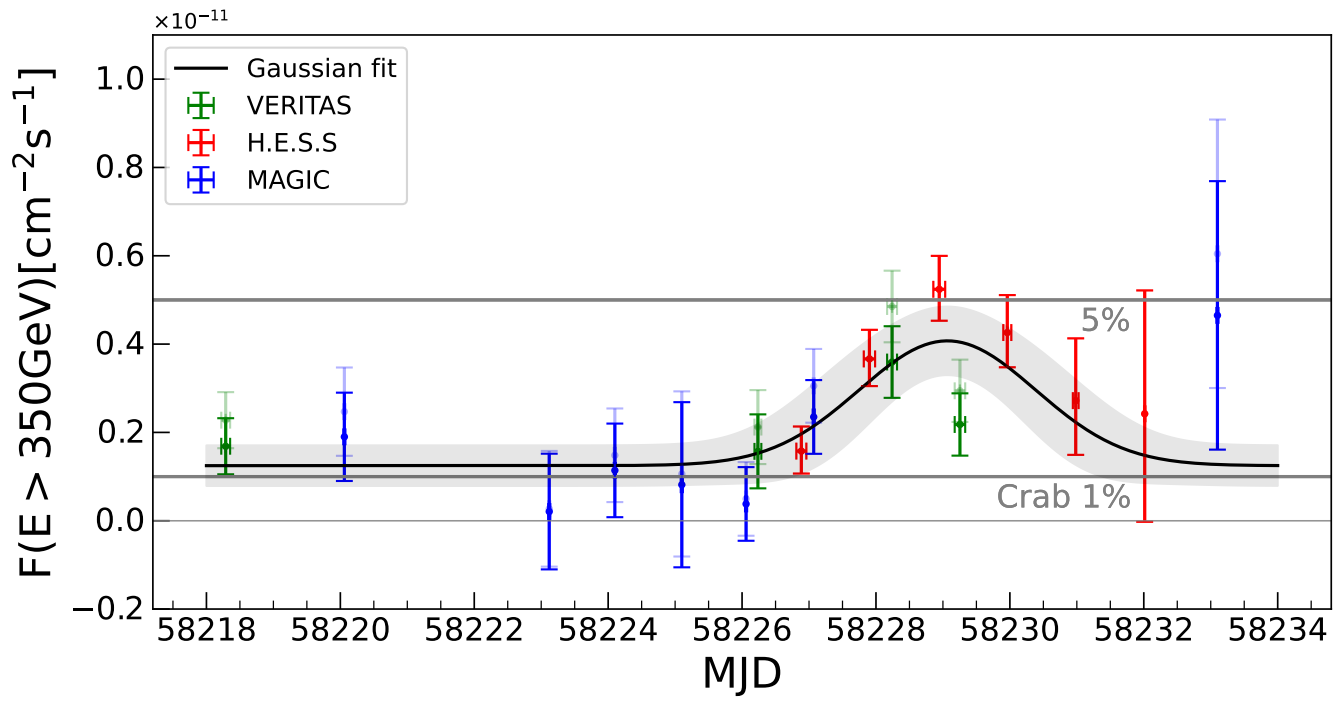

**Fig. 9.** Flux measurements of M87 above 350 GeV with 1 $\sigma$ uncertainties obtained with H.E.S.S., MAGIC, and VERITAS during the coordinated MWL campaign in 2018. The solid and semi-transparent colours represent the original data and scaled (with cross-normalisation factors) data, respectively. The black curve and shaded region represent the best Gaussian fit and 1 $\sigma$ error to the original flux points. The grey lines indicate the percentage of the Crab Nebula's flux level from integrating its SED in the same energy range. The SED of the Crab Nebula is obtained by fitting VERITAS's six years of observations (Meagher & VERITAS Collaboration 2015). For flux points with significance less than 2 $\sigma$, upper limits are given in Table C.14.

Light curves with shorter time binning of less than one night were calculated, but no significant intra-night variations were found. Further discussion of possible systematic cross-normalisation factors between IACTs and how the factors affect the fitting results can be found in B.4.2. To conclude, the statement that a model with a Gaussian peak describes the light curve significantly better than a constant holds even taking potential systematic cross-calibration offsets between the VHE instruments into account. The cross-normalisation factors provide a possible range of fitting parameter values and reduce the tension of the flux difference between VERITAS and H.E.S.S. around MJD 58229 (semi-transparent colours in Fig. 9). However, for this research, there is no compelling reason for a rescaling of the original data, reaffirming the fact that the fit to the flare profile without scaling is acceptable.

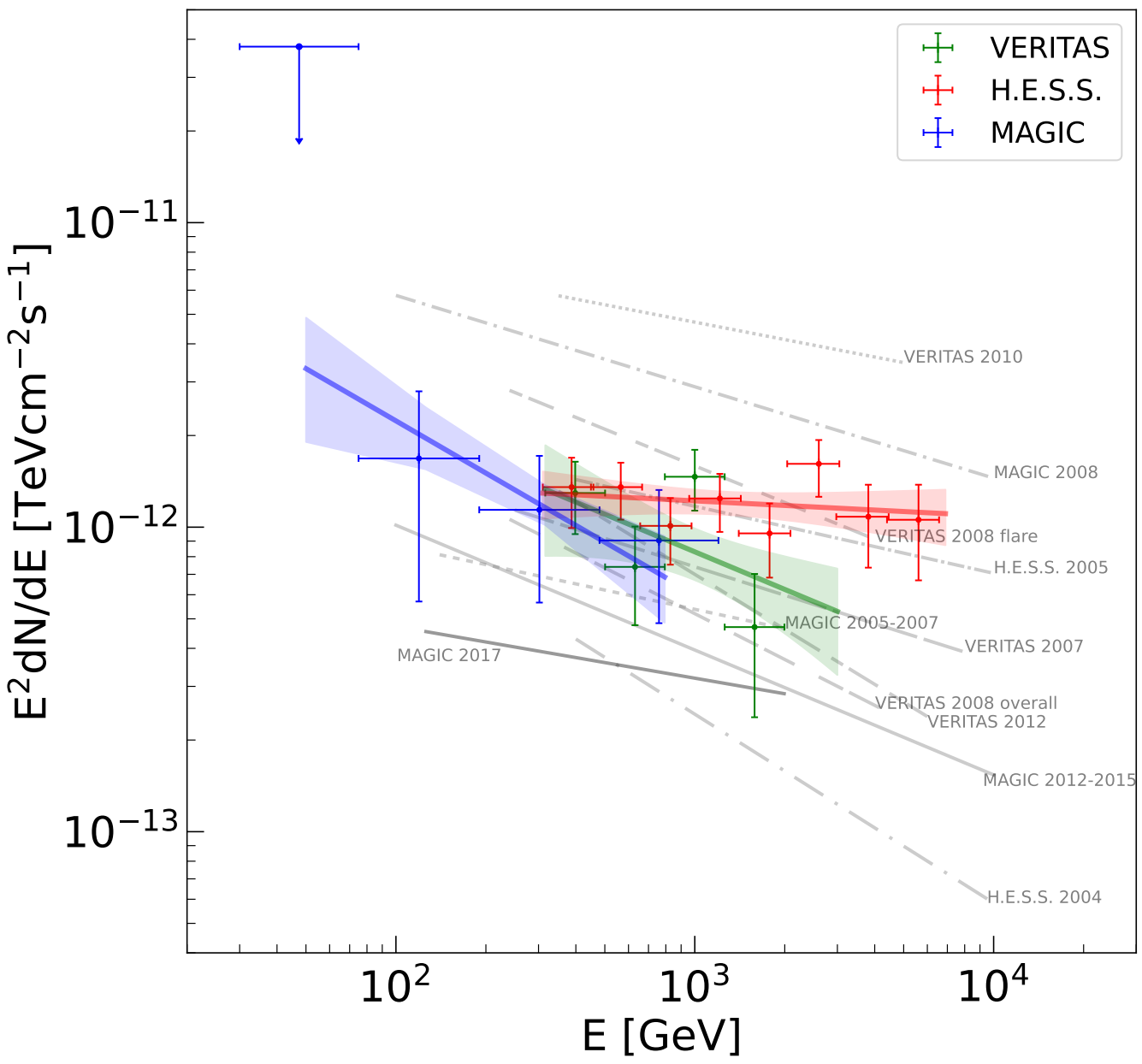

**Fig. 10.** Very high-energy SEDs measured with H.E.S.S., MAGIC, and VERITAS during the coordinated MWL campaign in 2018. Error bars are equivalent to the 1 $\sigma$ confidence level. An upper limit is drawn where the statistical significance is less than 2 $\sigma$ (i.e. 95% confidence level). The thick solid lines represent the best spectral fitting results by using a simple PL function. The shaded regions represent the 1 $\sigma$ error region of the fitting results. Historical spectral fits are taken from H.E.S.S. Collaboration (2006a,b); Aleksić et al. (2012); Acciari et al. (2008, 2010); Albert et al. (2008); Aliu et al. (2012); Beilicke & VERITAS Collaboration (2012); MAGIC Collaboration et al. (2020) and the thin black line indicates the SED obtained with the MAGIC measurements presented in M87 MWL2017.

Individual SEDs obtained with all VHE instruments are shown in Fig. 10 and results of the spectral fits are given in Table C.15. The photon spectra are described with a simple PL function:

$$dN/dE = f_0 \cdot \left(\frac{E}{E_0}\right)^{-\Gamma_{\rm VHE}}, \tag{3}$$

where $f_0$ is the normalisation at the energy $E_0$ and $\Gamma_{\rm VHE}$ is the photon spectral index. It is important to note that the three VHE SEDs are results of different samplings during the observational campaign. MAGIC observed primarily during nights before and after the VHE flux rise discussed below. About half of the VERITAS data were taken before and the other half during the flare. The H.E.S.S. observations cover the whole time span of the flare and lead to the hardest spectral index amongst the three instruments (Fig. 10). This result points to a harder-when-brighter behaviour in VHE, as previously hinted in H.E.S.S. Collaboration (2006a); Albert et al. (2008); Acciari et al. (2010); Aliu et al. (2012); H.E.S.S. Collaboration (2023). We acknowledge that directly comparing the spectral slopes of H.E.S.S., MAGIC, and VERITAS is also limited by the difference in the energy ranges covered by their respective spectra. Differently than the results from H.E.S.S. Collaboration (2024), the spectrum derived from the H.E.S.S. data in this study is limited to a maximum energy of 6.9 TeV. This limitation ensures that each energy bin in Fig. 10 is detected with a significance of at least $2\,\sigma$.

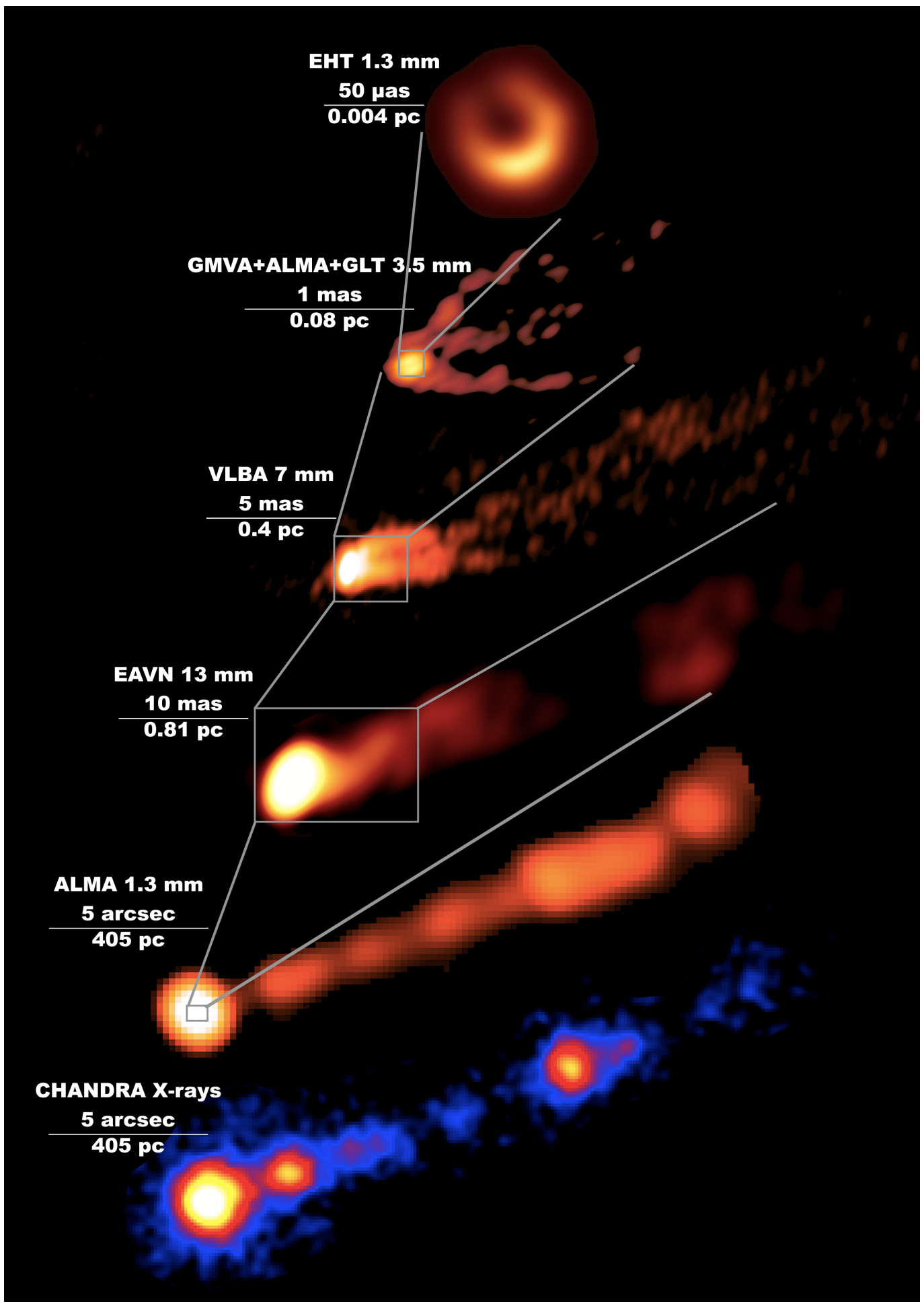

**Fig. 11.** Composite of the M87 MWL images at various scales obtained in radio and X-rays during the 2018 campaign. The instrument, observing wavelength, and scale are shown at the top-left side of each image. We note that the colour-scale has been chosen to highlight the observed features for each scale and should not be used for noise levels, dynamic range, or flux density calculation purposes.

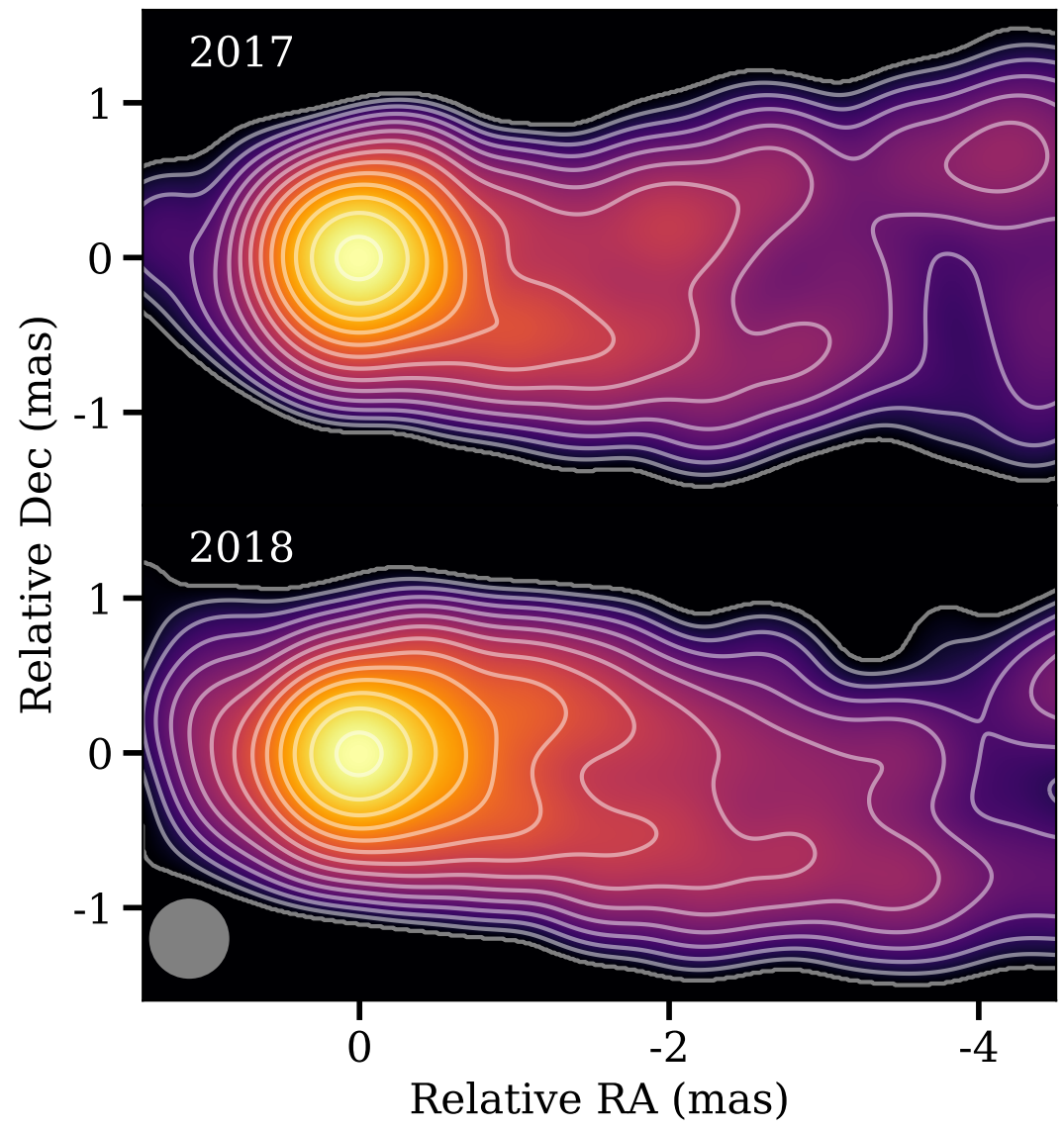

**Fig. 12.** M87 jet structure comparison based on yearly stacked EAVN images at 43 GHz in 2017 and 2018 (see also Cui et al. 2023). The restoring beam size is 0.5 mas circular Gaussian, as indicated in the bottom-left corner. The image has been rotated by $-18°$. Contour levels are scaled as $(1, 2, 4...)\times C_{1\mathrm{st}}$, where $C_{1\mathrm{st}}$ is 1.8 mJy beam$^{-1}$ in 2017 and 1.3 mJy beam$^{-1}$ in 2018. For M87* mass $M_{\mathrm{BH}} = 6.5 \times 10^9 M_\odot$ (Event Horizon Telescope Collaboration et al. 2019a), 1 mas $\approx$ 125 $\mathrm{r_s}$ $\approx$ 0.08 pc, where $r_\mathrm{s}$ is the Schwarzschild radius.

## 3. Results

### *3.1. Multi-wavelength images in 2018*

In Fig. 11, we show a composite of M87 MWL images obtained by radio and X-ray instruments during the 2018 campaign, including the EHT image. The M87 jet is imaged at multiple scales from ∼1 kpc down to a few Schwarzschild radii (defined as $r_\mathrm{s} = 2GM_{\mathrm{BH}}/c^2 = 2r_\mathrm{g}$ where $M_{\mathrm{BH}}$ is the mass of M87*, $G$ the gravitational constant, $c$ the speed of light, and $r_\mathrm{g}$ is the gravitational radius) summarising a contemporaneous MWL view of M87 during the 2018 EHT campaign.

To check the structural variations, we compared the data obtained in 2017 and 2018. While the appearance of the kpc-scale jet remained relatively consistent during this period, our high resolution VLBI observations unveiled some morphological evolution in the pc-scale jet between the two years (see Fig. 12). Notably, these changes encompassed a counterclockwise shift in the jet position angle (PA), occurring within a few milli-arcsecondfrom the core (PA∼281°/283° in 2017/2018; see Cui et al. (2023) for details). Furthermore, the data from 2018 reveal a more symmetric brightness ratio of the north-south jet limbs. All of these variations suggest the existence of year-scale structural evolution transverse to the jet. Moreover, regular VLBI monitoring observations reveal a gradual decrease in core flux (by 10–15%) at 22–129 GHz from 2017 to 2018 as shown in Fig. 2. Closer to the SMBH, GMVA 86 GHz observations in concert with ALMA and GLT spatially resolve the radio core, revealing a ring-like structure connecting to an edge-brightened jet (Lu et al. 2023).

Contemporaneous observations obtained with the VLBA at 24 and 43 GHz were used for the spectral analysis (Sect. 4). Figure 13 compares maps of the spectral index ($\alpha$; defined as $F_\nu \propto \nu^\alpha$) between these two frequencies from observations on 5 May 2017 (panel (a)) and 28 April 2018 (panel (b)). The M87 jet has an edge-brightened structure and does not have distinct components at each frequency to anchor an alignment of different images. We model-fitted the jet structure by a number of two-dimensional Gaussian components, aligned images at the position of the 43 GHz core, and shifted 24 GHz images by the value of the core shift. Finally we measured the shifts from the phase referencing observations, that are 0.036 and 0.012 mas in right ascension and declination, according to Hada et al. (2011) and Ro et al. (2023). Even though the amplitude of the core shift may change with time (Plavin et al. 2019), multiple studies of M87 indicate the magnitude of the core-shift variation in 22–43 GHz is within 10 $\mu$as and negligible (Hada et al. 2012, 2014; Jiang et al. 2021). The core regions in both 2017 and 2018 show relatively flatter spectral index values ($\alpha \approx 0$) than jet regions. However, compared to 2017 observations (Fig. 13(a); M87 MWL2017), the value of the spectral index in 2018 steepens faster downstream from the core region to the jet region. It indicates that the radio emission in the jet close to the core becomes relatively stronger at lower frequencies compared to the emission in the core in 2017, that might be related to changes in the physical conditions of the particle accelerator at the core. In the extended jet regions, the spectral index distribution appears to be more complex: it shows flatter values ($\alpha > -0.5$) at the location of total intensity maxima with steeper values ($\alpha < -1$) that are localised in between. The M87 jet structure and spectral index distribution based on the 8–15 GHz VLBA observations with the highest-to-date dynamic range (Nikonov et al. 2023) are consistent with the development of Kelvin–Helmholtz instabilities in the jet in regions of $10^2 - 10^4\, r_{\rm g}$ from the central engine. Their propagation can produce higher pressure regions near the jet boundary and in its interior, causing variations of the spectral flattening towards the jet edges, as shown in Fig. 13. The observed variability in flat- and steep-spectrum regions could also be influenced by the jet precession, that modulates the observed jet structure over time. However, for the variability results obtained here, rescaling of the original data is not crucial, reaffirming the fact that such a complicated spectral profile in the edge-brightened jet region could be (at least) partly affected by insufficient interferometric (u,v) coverage (e.g. Pashchenko et al. 2023).

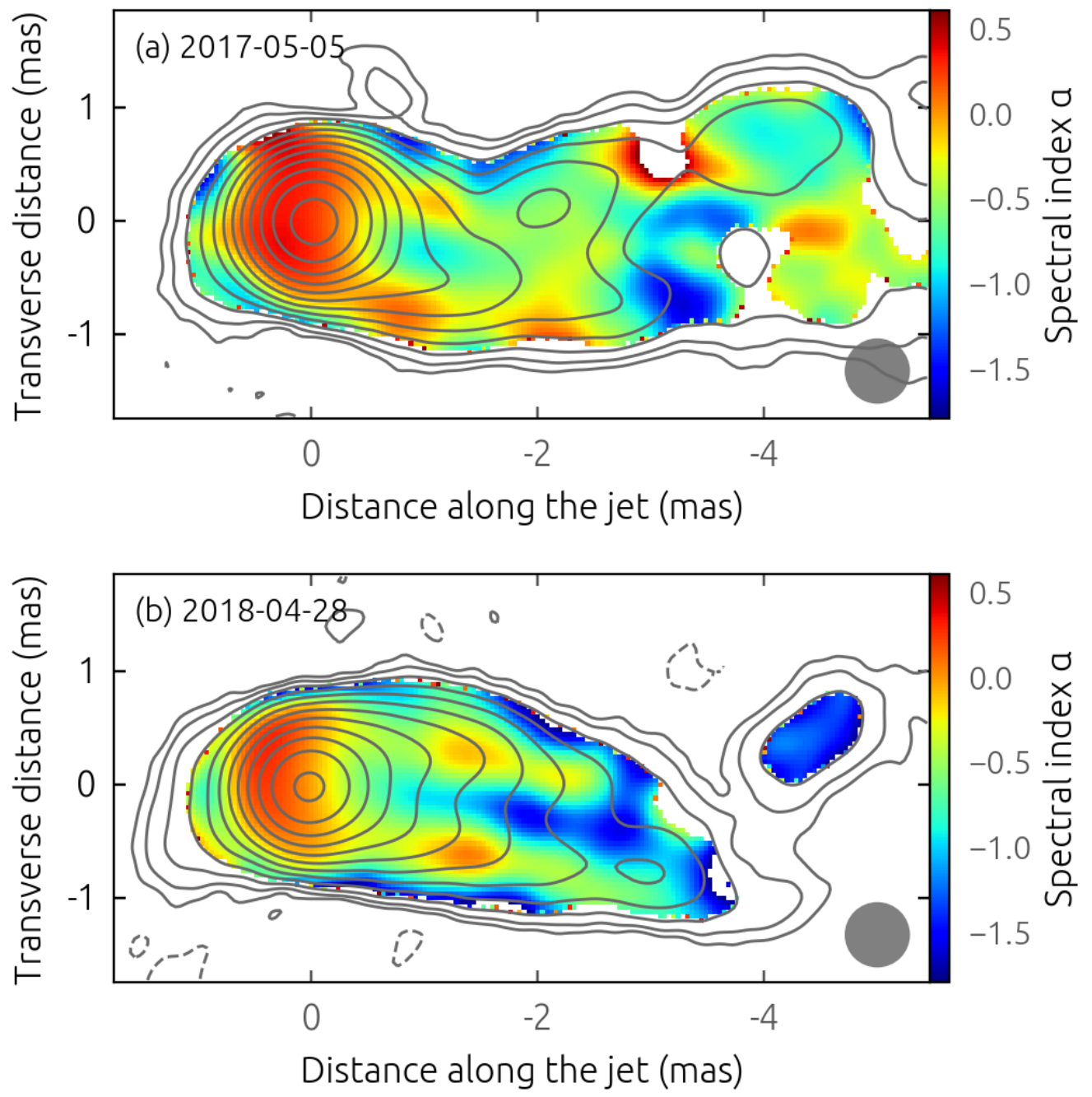

**Fig. 13.** Spectral index maps obtained between 24 and 43 GHz from VLBA 5 May 2017 (a) and 28 April 2018 (b) observations. The restoring beam size is 0.55 mas ×0.55 mas, that is the circular equivalent-area synthesised beam of VLBA at 24 GHz, drawn as a grey circle in the bottom-right corner. The image has been rotated by $-18°$. The contours denote the total intensity at 43 GHz and start at $3.5\sigma$ rms, increasing in steps of two. The $(u, v)$-ranges match between all images.

### 3.2. Change in the mas-scale jet and $\mu$as-scale ring asymmetry position angles

As reported in Sect. 3.1, a slight and counterclockwise variation in the PA of the mas-scale jet was detected with 43 GHz VLBI observations in 2017 and 2018. The observed PA evolution from 2017 to 2018 appears to follow the long-term (quasi-)periodic variation (on time scales of ~10 years) found by the recent VLBI jet monitoring studies (Walker et al. 2018; Cui et al. 2023). Possible scenarios causing such year-scale variations of the jet PA include MHD/HD instabilities (Mizuno et al. 2007; Walker et al. 2018) and Lense-Thirring precession in a tilted accretion disk system (Fragile et al. 2007; Liska et al. 2018; Cui et al. 2023). The observed variation in the north-south limb brightness ratio could also be related to the jet PA variation and/or transient turbulence in the jet flow, which requires further dedicated investigations.

Similarly, the EHT-2018 images (M87 2018 I) reveal a significant shift of the PA of the brightness asymmetry of the ring with respect to that of EHT-2017 images (presenting a counterclockwise variation ~ 30° from 2017 to 2018). The shift direction is as expected if the large-scale jet is aligned with the disk rotation axis (M87 2017 V). Morphological variations of the asymmetric ring at EHT scales could be related to the week-scale variation caused by magnetorotational instabilities (M87 2017 V, M87 2018 I). On the other hand, the observed PA change in the EHT images between 2017 and 2018 may imply the presence of a longer-term (year-scale) variation at this scale as originally suggested by Wielgus & Event Horizon Telescope Collaboration (2020).

Nevertheless, it is worth noting that M87 shows counterclockwise PA variations at both milli-arcsecond and microarcsecond scales between the two years, that sheds light on the possible physical connection between the mas-scale jet and $\mu$as-scale ring. The recent detection of both the ring-like structure and extended jet in a single image using global VLBI at 86 GHz offers a new potential to explore this challenge by bridging the spatial gap between $\mu$as- and mas-scale structures (Lu et al. 2023). Future multi-year comparisons of multi-frequency (230/86/43 GHz) VLBI images will be the key to constraining the origin of variable structures at different scales and their physical connection.

### 3.3. Multi-wavelength light curve and fractional variability

In Figs. 14, 15, and 16 we show the long-term (2000-2022), mid-term (2017-2019) and short-term (April 2018) MWL light curve behaviour of M87 with the observations collected by the instruments participating in the coordinated MWL campaign in 2018.

Comparing the 2018 EHT-MWL campaign (Fig. 16) to 2017 (M87 MWL2017), the source presents brighter emission at high energies, seen both in the X-ray as well as in the $\gamma$-ray energy bands, while in optical and at radio frequencies the core flux remained the same or even a bit dimmer. In particular, *Chandra* observations during April 2018 revealed a core flux increase of ~ 80% with respect to the previous 2017 campaign. While the duration of the VHE $\gamma$-ray flaring episode is well constrained (see Sect. 2.4.2 and further discussion in Sect. 3.3.1), the lack of monitoring X-ray observations of the source does not allow us to fully investigate the duration of the X-ray variability, as we further elaborate in Sect. 3.3.2.

The fractional variability (Vaughan et al. 2003) of the VHE $\gamma$-ray band in 2017 is compatible with zero within uncertainties. The same applies to the 2018 VHE $\gamma$-ray variability. This is caused by the large statistical uncertainties of the VHE data. We note that these large statistical uncertainties are also limiting the calculation of a discrete correlation function (Edelson & Krolik 1988). The fractional variability of the *Swift*-XRT flux of the small-scale components (namely the core, the HST-1 knot and inner jet) is significantly lower in 2018 than in 2017. For both years these fractional variabilities are slightly higher than in the VHE $\gamma$-ray band, although well compatible within statistical uncertainties. Further information on our calculation of the fractional variability are reoperted in Appendix D. The resulting fractional variability values are given in Table D.1.

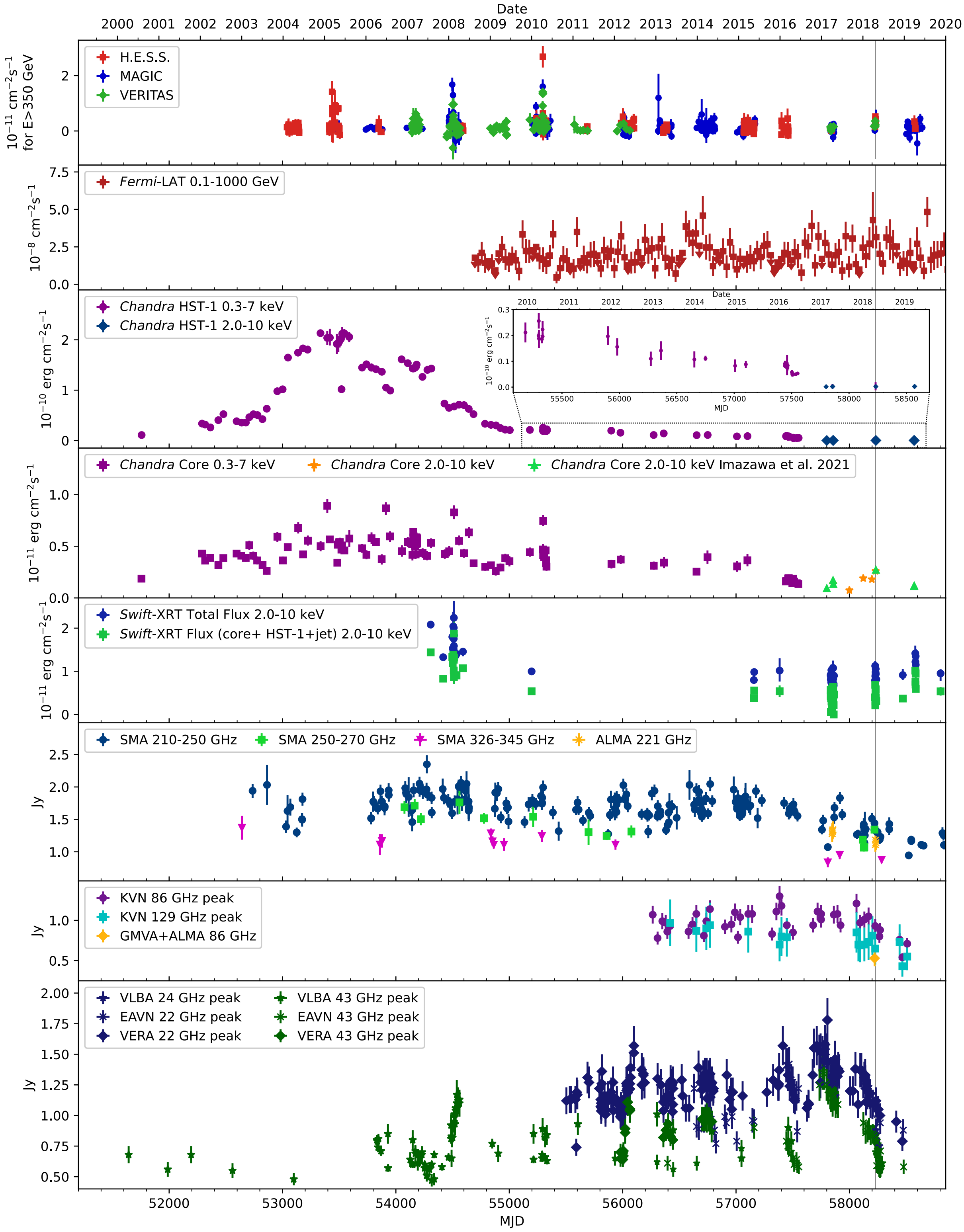

**Fig. 14.** Long-term MWL light curves of M87 from the last two decades taken by the instruments participating in the 2018 observational campaign. Here are listed the observations used in the plot and their reference (in case of already published results): *Chandra* (0.3–7 keV) (Sun et al. 2018), (2.0–10 keV) (Imazawa et al. 2021); MAGIC, H.E.S.S., VERITAS (2004–2011), *Fermi*-LAT (2008–2011), VLBA (43 GHz peak) (2006–2011) (Abramowski et al. 2012); H.E.S.S. (before April 2004, 2001–2016, after 2019) (H.E.S.S. Collaboration 2023); VERITAS (2011-2012) (Beilicke & VERITAS Collaboration 2012); VERA (2011–2012), MAGIC (2012–2015) (MAGIC Collaboration et al. 2020); KVN (2012–2016) recalculated from (Kim et al. 2018); VLBA, EAVN, VERA (2000–2018) (Cui et al. 2023); All instruments 2017 data (M87 MWL2017). We mark the VHE flare in 2018 with a grey-shaded line in the background. The inset in the third row provides a zoomed-in view of the *Chandra* measurement of HST-1.

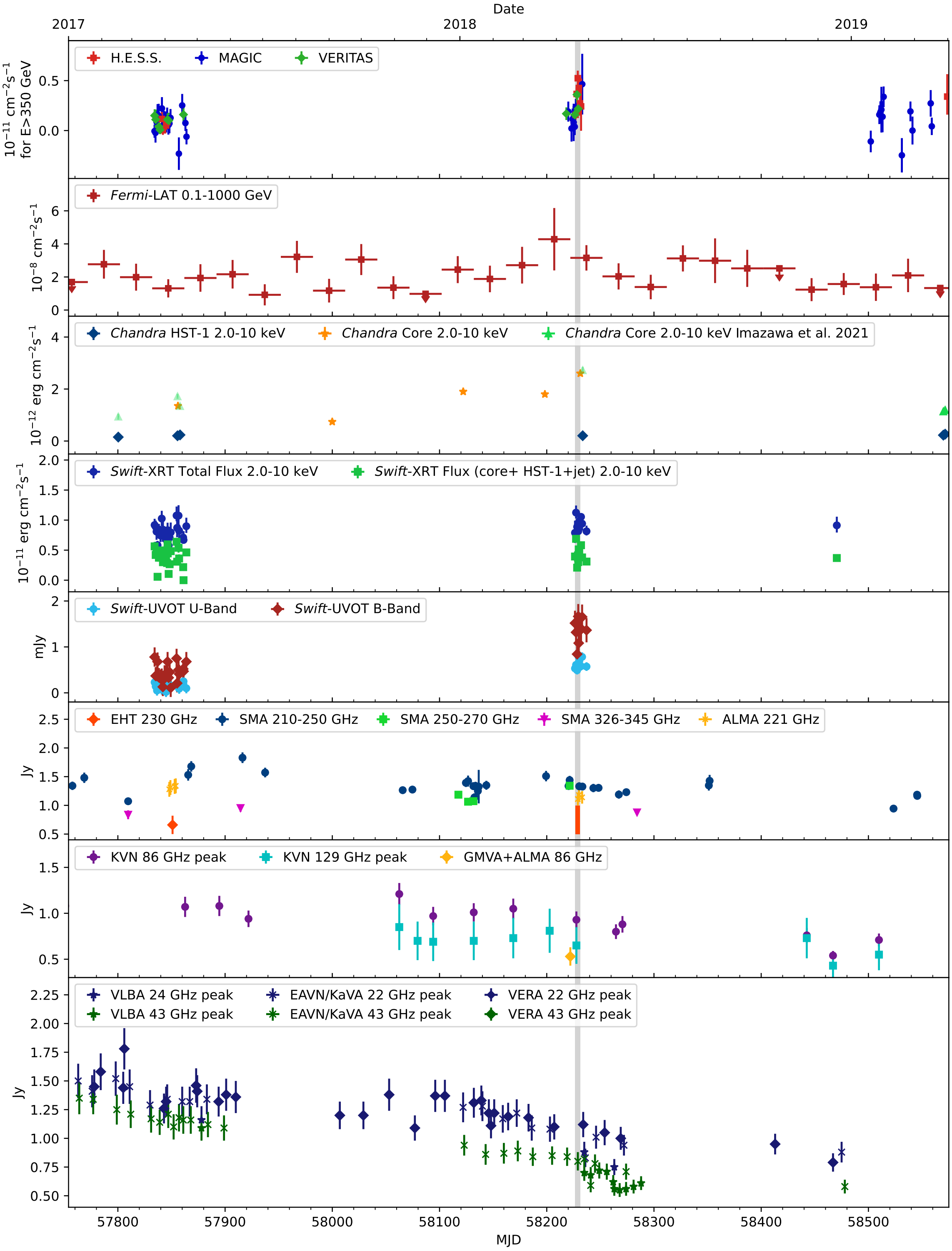

**Fig. 15.** Mid-term MWL light curves of M87 taken during the observational campaign covering January 2017 to April 2019. Here are listed the observations used in the plot and their reference (in case of already published results): *Chandra* (0.3–7 keV) (Sun et al. 2018), (2.0–10 keV) (Imazawa et al. 2021) H.E.S.S. (2019) (H.E.S.S. Collaboration 2023); VLBA, EAVN, VERA (2017–2018) (Cui et al. 2023); All instruments 2017 data (M87 MWL2017). We mark the VHE flare in 2018 with a grey-shaded region in the background.

We combined the spectral slopes and flux normalisations from the three VHE energy spectra obtained in this campaign with the historical measurements to test whether there is a general harder-when-brighter behaviour in the VHE band. The photon indices are plotted against the flux normalisations in Fig. 17, in which it is possible to see that there is no dependence of the spectral index on the flux. However, this comparison is limited by the fact that the VHE spectra shown cover different energy ranges (H.E.S.S. Collaboration 2006a; Acciari et al. 2008; Albert et al. 2008; Acciari et al. 2010; Aliu et al. 2012; Aleksić et al. 2012; Beilicke & VERITAS Collaboration 2012; MAGIC Collaboration et al. 2020; EHT MWL Science Working Group et al. 2021).

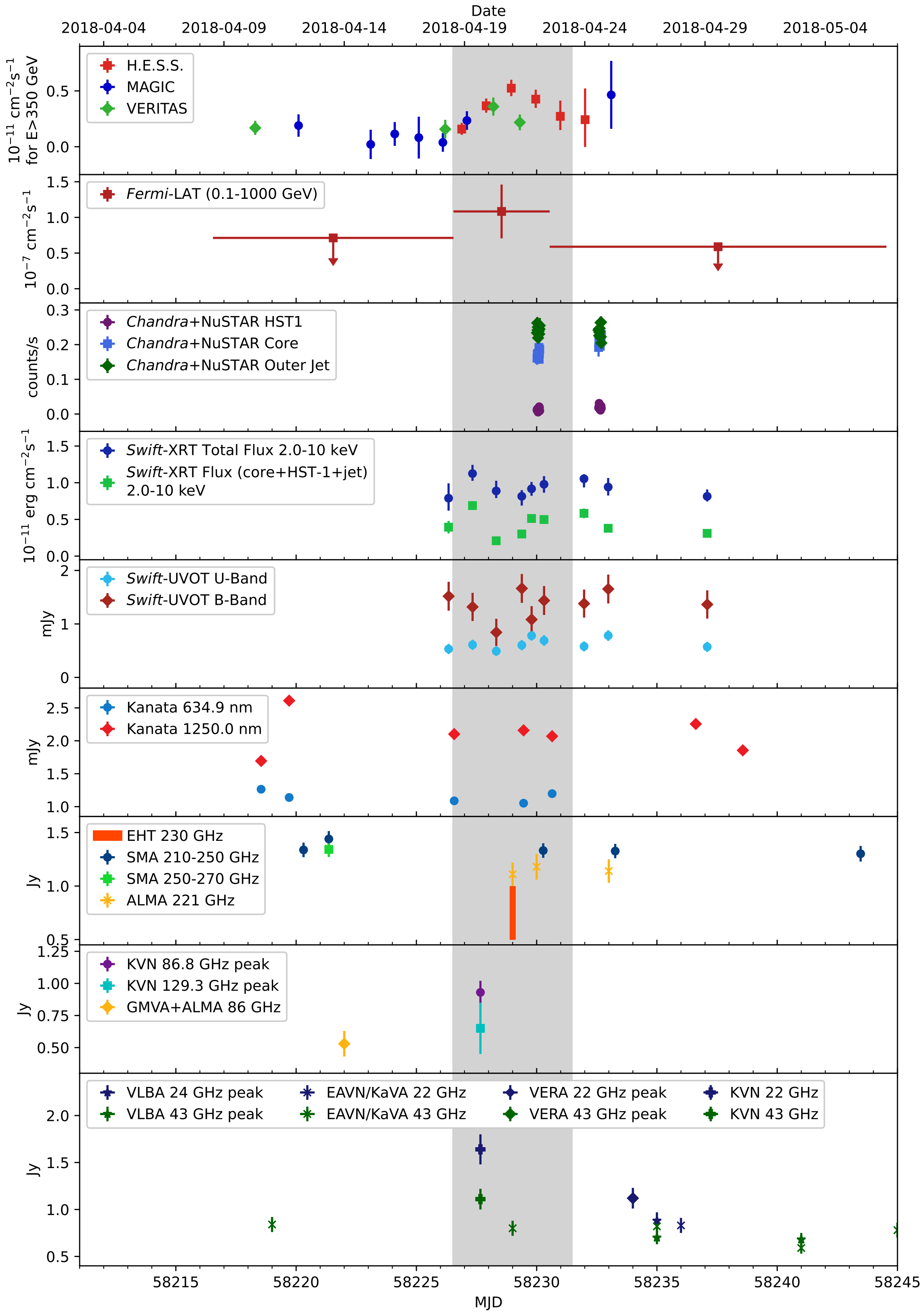

**Fig. 16.** Multi-wavelength light curves of M87 taken during the observational campaign covering MJD range 58211-–58245. The time period containing the 2008 VHE flare is shaded in grey.

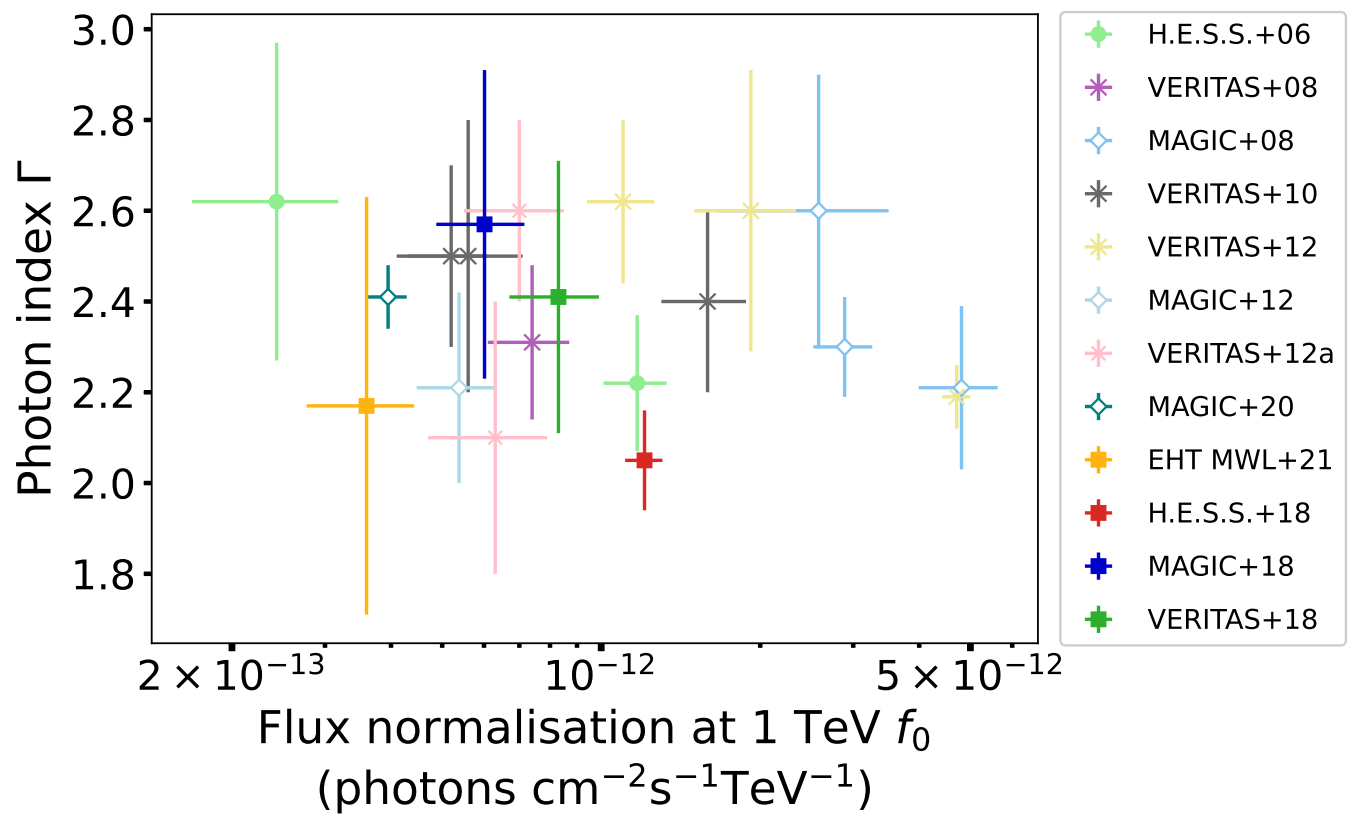

**Fig. 17.** Photon PL index Γ compared to the VHE flux normalisation at 1 TeV. Data from H.E.S.S. Collaboration (2006a); Acciari et al. (2008); Albert et al. (2008); Acciari et al. (2010); Aliu et al. (2012); Aleksić et al. (2012); Beilicke & VERITAS Collaboration (2012); MAGIC Collaboration et al. (2020); EHT MWL Science Working Group et al. (2021). Flux normalisations at different energies have been recalculated at 1 TeV using the respective power-law index. No dependence of the spectral index on the flux was observed.

### 3.3.1. VHE gamma ray flaring episode

The first strong indication of VHE $\gamma$-ray emission from M87 was found in 1998 with the High Energy Gamma Ray Astronomy (HEGRA) telescopes (Aharonian et al. 2003). Since 2004, frequent monitoring (Acciari et al. 2009; Abramowski et al. 2012) of M87 has been conducted by H.E.S.S. (H.E.S.S. Collaboration 2006a), MAGIC (Albert et al. 2008; Aleksić et al. 2012; MAGIC Collaboration et al. 2020), and VERITAS (Acciari et al. 2008, 2010; Aliu et al. 2012; Beilicke & VERITAS Collaboration 2012). These monitoring campaigns have yielded detections of VHE flaring activity from M87 in 2005 (H.E.S.S. Collaboration 2006a), 2008 (Albert et al. 2008), and 2010 (Abramowski et al. 2012). Furthermore, while VERITAS observations carried out in 2011 and 2012 did not show bright flares, they reported a hint of an underlying quiescent increase of the VHE emission evolving on longer time scales than day-scale rapid flares (Beilicke & VERITAS Collaboration 2012).

Despite deep investigations of past flaring episodes of the source as well as of the quiescent state during the 2017 EHT campaign (M87 MWL2017), the origin of the VHE $\gamma$-ray emission from M87 is still unclear. A unique opportunity to address this question is provided by the correlated temporal variations between the VHE and lower-energy MWL emission, particularly when the source can be spatially resolved (Rieger & Aharonian 2012; Rieger & Levinson 2018). Fortuitously, as shown in Sec. 2.4.2, the 2018 MWL campaign detected the first fast $\gamma$-ray variability observed since 2010, providing a chance to investigate the origin of the $\gamma$-ray emission in M87.

During the 2018 MWL campaign the VHE flux increased by a factor ~3.3 with respect to April 2018 low state (see Sect. 2.4.2, Fig. 9) reaching a peak flux that is still far below the flares observed in the past but a factor ~4 higher than the quiescent state flux measured during the 2017 campaign (M87 MWL2017). In 2005 the peak flux at energies > 350 GeV reached $(1.41 \pm 0.39) \times 10^{-11}$ ph $cm^{-2}$ $s^{-1}$, almost 20 times the quiescent state flux as measured during the 2017 EHT-MWL campaign (M87 MWL2017). Similarly, in 2008 and in 2010, the observed fluxes were 23 and 37 times the quiescent state flux (see Fig. 14), respectively.

The variability time scales observed during the past VHE flares are similar to the 2018 flare (FWHM approximately three days) presented here, albeit somewhat faster. The 2005 and 2008 flares showed a roughly two-day and one-day variability, respectively (H.E.S.S. Collaboration 2006a; Albert et al. 2008). In 2010 a two-sided exponential fit revealed a rise time of $1.69 \pm 0.30$ days and a decay time of $0.611 \pm 0.080$ days (Abramowski et al. 2012). The general light curve shape of the 2018 flare is similar to the flare seen in 2010, with only one isolated peak, while the flares observed in 2005 and 2008 comprise multiple separated high states during a similar time period.

### 3.3.2. Flux enhancement in the X-ray band

The X-ray flux from the unresolved core as observed with *Chandra* (within 0.4 arcsec radius), the combined core, HST-1, and jet fluxes from *Swift*-XRT, and the Swift-UVOT B- and U-band fluxes are all generally higher in April 2018 compared to both 2017 and 2019, as shown in Fig. 15. However, without additional X-ray/UV/optical observations immediately before and/or after the gamma-ray flare period, it is difficult to establish a firm link between the elevated X-ray/UV/optical fluxes and the flare itself. Figure 14 shows that M87's core X-ray flux can vary to a similar degree even in the absence of a flare, so the elevated X-ray/UV/optical flux near the 2018 flare cannot be definitively linked to the gamma-ray flare itself.

It is, anyway, tempting to relate the observed X-ray flux enhancement to the flaring observed in the VHE band despite the very sparse time coverage of *Chandra* observations. The difference in core flux between the two *Chandra* observations during the EHT 2018 campaign is clear but small (≃ 20 %; see Fig. 5) and rising away from the VHE flare that peaked approximately one day and three days before the first and the second *Chandra* observations, respectively. Two AstroSat-SXT observations, that do not directly resolve the core, suggest a modest increase in flux between February and April 2018.

Denser sampling in time is afforded by our *Swift*-XRT observations (Fig. 7), providing a measure of variability in the joint flux of the small-scale components (core, the HST-1 knot, and the resolved jet), that cannot be resolved individually. This flux, modelled as a fraction of the total flux observed by *Swift*-XRT does not show any coherent flux increases comparable to the VHE flare within the ≃ 11-day campaign window covered with roughly daily cadence. The variance over this period (variability amplitude of 5–10%) is consistent with that of long-term *Swift*-XRT monitoring. Comparing the *Swift*-XRT small-scale component fluxes to the *Chandra* core flux, we find that the *Swift*-XRT small-scale components fluxes produced by our analysis are typically about double the core fluxes reported in Table 1 of Cheng et al. (2023). Our *Swift*-XRT small-scale component fluxes and the *Chandra* core fluxes show a similar trend, presenting a small flux increase between 22 and 24 April 2018.

A model-dependent way to test for an association between the X-ray brightening and the VHE flare would be to search for changes in the X-ray spectrum. For example, a harder X-ray spectrum than observed in non-flaring states (e.g., the 2017 EHT campaign) might be expected in some broadband SED models (discussed in detail in Sect. 4.3). While Cheng et al. (2023) claim to find hardening in *Chandra* observations from April 2018, we find no evidence for this effect in our joint spectral analysis with *NuSTAR* (possibly because we account for pileup and use a broader energy range to measure the photon index). In our power law model, we find $\Gamma_{\rm core} = 2.03^{+0.12}_{-0.07}$ in 2018, statistically consistent with $\Gamma_{\rm core} = 2.06^{+0.10}_{-0.07}$ from the 2017 EHT observations. Thus the only statistically significant change in the X-ray core is the flux increase described above. Finally, we note that this flux enhancement is insensitive to the X-ray spectral model: because the PL break lies above 10 keV, it does not dramatically alter the inferred 2–10 keV core flux.

### 3.4. Multi-wavelength SED

As discussed in the previous subsections, the source underwent a $\gamma$-ray flaring episode during the 2018 EHT-MWL campaign, with a hint of spectral hardening towards the highest TeV energies. While a moderate flux enhancement is also seen at X-ray energies with respect to 2017 EHT-MWL observations (M87 MWL2017), in the radio band, the source presents a comparable, or even lower, emission. These features can be seen in Fig. 18 that shows the 2018 MWL broadband SED (reported also in Table C.1), where the 2017 EHT-MWL campaign SED is shown in the background in grey for comparison. The fact that the source was in a low state at all frequencies while the HST-1 knot emission was subdominant to the core in the X-ray band allowed us to confidently associate the *Chandra* and *NuSTAR* total fluxes with the core emission in 2017 data; however, the observed X-ray and VHE variability seen in 2018 leads to uncertainties regarding their localisation.

In Table C.1 we present the legacy data set with near-simultaneous broadband SED results collected over a broad range between a frequency of $\simeq$ 20 GHz and photon energy $\simeq$ 1 TeV. Similar to M87 MWL2017, we report for each given energy or frequency range the associated spatial scale of the total emission, the time ranges of the observations and fluxes (providing both $F_\nu$ and $\nu F_\nu$). As described in Sect. 2, all flux points are based on statistically significant detections ($> 3\,\sigma$ for all instruments except $> 2\,\sigma$ for IACTs) and provided with uncertainties equivalent to the $1\,\sigma$ confidence level. Upper limits on flux are given at the $2\,\sigma$ confidence level. The angular scale of the emission measured by the different instruments involved in this campaign, ranges between $\simeq 20\,\mu$as (EHT) and $\simeq 2^\circ$ (*Fermi*-LAT in low-energy $\gamma$-rays), namely a factor $\gtrsim 10^8$ in angular scale.

Besides the EHT observation (for which we list the estimated total flux within a $60 \times 60$ $\mu$as$^2$ region as described in Event Horizon Telescope Collaboration et al. 2019d), for other radio observations if the core is not resolved we use the flux peak obtained from the beam size as an upper limit on the core emission. For elliptical beams, we list the average of the axes as a representative angular scale. For the observations carried out in the optical and UV bands, due to the difference between Kanata and Swift-UVOT angular resolution (~10″and ~2.5″, respectively), we consider the first as an UL and the second as flux points, similarly to what was done for the Swift-UVOT observation carried out in M87 MWL2017. This difference is related to the larger uncertainties in the subtraction of the host galaxy contribution in the Kanata optical observations. However, we want to highlight that, considering its angular resolution, Swift-UVOT measurements include both the core and the HST-1 knot contribution. At higher energies, the scale typically corresponds to the point spread function (PSF) within a given band or the diameter of the region used for data extraction. For *Chandra* and *NuSTAR*, as in EHT MWL Science Working Group et al. (2021), we show the inferred core spectrum from the power-law and broken power-law models described in Sect. 2.3.1. We discuss the implications of the inferred X-ray spectrum on the SED modelling in Sect. 4.Despite the fact that *NuSTAR* can only separate out the core emission spectroscopically (see Sect. 2.3.1), the reported scale for *Chandra* and *NuSTAR* data analysed jointly is related to *Chandra*'s greater resolving power. Given that *Swift*-XRT cannot distinguish the core from other components detected by *Chandra*, we treat its unresolved flux as an upper limit.

Considering the timing relative to the observed $\gamma$-ray flare it is important to note that the majority of the observations arranged during our 2018 campaign have been carried out either during or within a few days of the flaring episode, well within the minimum dynamical timescale (5 - 61 days; Satapathy et al. 2022) for the innermost stable circular orbit of M87. The only exception is represented by global 43 GHz VLBI observations, collected in February 2018, that have been included taking into account the non-variability seen in the radio band on short and long time scales (see Sect. 3.3). Although these observations were not all strictly simultaneous with the $\gamma$-ray flare and EHT observations, the non-variability observed for the radio, optical and X-ray bands, suggests that they effectively provide a snapshot of the broadband SED of the M87 core. These results can be used as constraints for any model seeking to explain the source behaviour around the time of the observed $\gamma$-ray flare and EHT 2018 image acquisition.

In addition to highlighting the various observatories and instruments with different colours and labels, we also report the upper limits on the size of the emission region for the VLBI measurements. These data are provided to the community in machine-readable form via the EHTC Data Webpage[4] (or directly via DOI «TBD») along with associated documentation and files needed for spectral modelling of the X-ray data. A complete list of all supplementary material published along with this paper is provided in Appendix C.

## 4. Heuristic modelling of the 2018 SED

A variety of theoretical models have been proposed to explain MWL emission from M87 (e.g., Rieger & Aharonian 2012; Prieto et al. 2016, for review). Given the M87 polarisation imaging constraints (Event Horizon Telescope Collaboration et al. 2021c,d), the innermost, ionised regions must be moving along strong field lines. While the accretion flow could contribute in some parts of the SED, the jet radiation is more likely to be dominant in the SED as a whole (e.g., EHT MWL Science Working Group et al. 2021, and references therein). For this work, we choose to model the SED based on single zones, a common approach to many AGN blazar studies; we defer more sophisticated modelling such as structured jet models (e.g., Potter & Cotter 2013; Lucchini et al. 2019) to future works. Models based on single zones enable the baseline properties of the emitting regions to be constrained. In addition, this approach allows for comparisons to other M87 epochs, including the 2017 campaign, as well as to the other AGN observed by the EHT.

The origin of the VHE $\gamma$-ray emission from M87 is still unclear. The MWL observation campaign described above (see Sect. 2) offers a unique opportunity to tackle this issue, given the following key points: (i) the duration of the VHE $\gamma$-ray flaring episode is well constrained, while we cannot detect similar ac-tivity in the other energy bands, and (ii) EHT spatially resolved the photon ring and measured its flux during the flaring episode. With these data, the origin of the VHE $\gamma$-ray emission from M87 can in principle be better constrained. We briefly describe a heuristic model to explore this. First, as in M87 MWL2017, we introduce the EHT-oriented model. This model uses the flux density and the size of the emitting region obtained from actual EHT observations and allowed us to constrain the average physical state of the EHT emitting region near the black hole, assuming it is uniform within the region. Within the context of this approach, we find that the high energy $\gamma$-ray emission cannot be explained only by the EHT-oriented model. Therefore, the portion of the radiation that cannot be explained by the EHT-oriented model will be explained using a high-energy(HE)-oriented model, whose size is set by the VHE $\gamma$-ray flare duration.

[4] https://eventhorizontelescope.org/for-astronomers/data

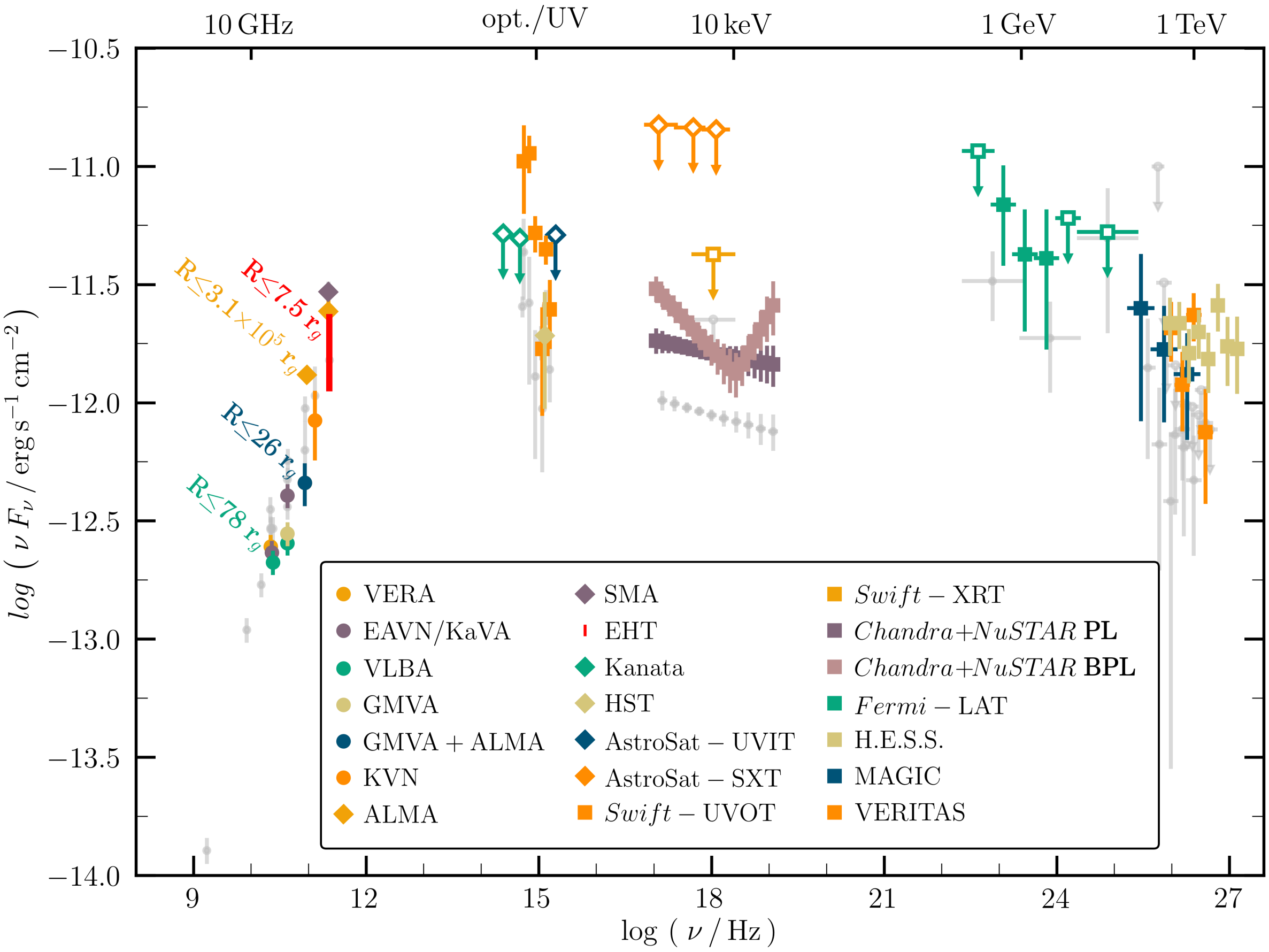

**Fig. 18.** Observed broadband SED of M87 contemporaneous with the EHT campaign in April 2018, with the exception of GMVA (carried on February 2018) and HST (carried on July 2018) observations (see Table C.1). The fluxes were measured by various instruments, and they are highlighted with different colours and markers. Filled markers represent flux point estimates, while empty markers indicates ULs. In the X-ray band, the two sets of points represent the core flux from the models with a power law (dark purple) and broken power law (light purple; see Table C.12). Grey points represent the observed broadband SED of the 2017 EHT campaign (EHT MWL Science Working Group et al. 2021). MAGIC observations were performed before the $\gamma$-ray flare, VERITAS observations only partially overlap the flare, while all H.E.S.S. observations have been taken during the VHE $\gamma$-ray flaring episode (see Table C.1). For the radio/millimetre VLBI and interferometric data, the upper limits on emission size (radius) for several representative frequencies, that are obtained by following the procedures described in M87 MWL2017, are labelled to clarify the constraints used in Sect. 4.

Even with the unique 2018 observational data, simple HE-oriented models are degenerate in interpretation. We therefore present two model scenarios (Model A and B), both of which assume a TeV $\gamma$-ray flux increase during the flare. Model A is a more straightforward two-zone model that is a combination of the EHT-oriented model and the HE-oriented model. It is characterised by having mildly relativistic bulk flow and an internal emitting particle distribution (assumed leptonic) to explain the observed TeV $\gamma$-ray data. In contrast, Model B is a three-zone model with an additional fast-moving component (corresponding to a part of the accelerating jet) that explicitly accounts for the $\Gamma_{\rm VHE}\approx2.0$ component of the TeV $\gamma$-ray data. In this scenario, the $\gamma$-ray emission originates from the fast-moving component, as often proposed in the literature (e.g., Ghisellini et al. 2005; Giannios et al. 2009, and references therein).

Here the shape of the X-ray spectrum also becomes relevant. As discussed in 2.3.1 (see also Sheridan et al., in preparation), a stacked analysis of *NuSTAR* observations of M87 including the 2018 data suggests a break in the X-ray PL spectrum that (i) appears to be stable from year to year and (ii) bends upward towards the $\gamma$-ray band. Extrapolating this broken power-law (BPL) to higher energies results in a substantially larger MeV energy flux compared to the decaying PL model. Because the 2018 data are statistically consistent with both the PL model and the BPL model, in subsequent sections we consider both scenarios when modelling the full SED with Model A (see Table 1, Sect. 4.4, and Fig. 21). We are left with three separate models explored in the sections below: model-A with PL, model-A with BPL, and model-B with PL. Unfortunately, the most notable difference between the PL-based models and BPL-based models is a large MeV bump that is not sampled by our observing campaign (see, e.g., Fig. 21), and so ultimately we are not able to rule out either X-ray model based on SED considerations.

The SED models include parameters relating to the source bulk physical properties as well as the internal emitting particle properties. The microphysics regarding the acceleration of leptons is phenomenologically included in non-thermal electron distributions (hereafter we refer for ease to these as electron distribution functions; eDFs). For this paper, GR and $\gamma$-$\gamma$ absorption effects are not considered in order to enable comparisons with the models in M87 MWL2017 and other single-zone models. In fact, H.E.S.S. Collaboration (2024) demonstrate that the absorption caused by the Extra-galactic Background Light (EBL) only becomes notable ($\tau_{\mathrm{EBL}}$>0.2) in M87 for $\gamma$-ray energies exceeding 10 TeV. Additionally, the absorption attributable to the infrared light from the galaxy is also found to be negligible ($\tau_{\mathrm{IR}}$<0.001) for TeV $\gamma$-rays. Lastly, the absorption by photons in the vicinity of the SMBH at 10-50 $r_g$, where $r_g = GM/c^2$ is the gravitational radius of the black hole of M87*, might have a more significant impact (Brodatzki et al. 2011; Cui et al. 2012). We have assessed the influence of synchrotron self-absorption (SSA) on the energy flux of TeV photons, utilising the best-fit parameters from our models (see Sect. 4.3). Our analysis indicates that SSA, at most, affects the energy flux by less than an order of magnitude. Consequently, this phenomenon does not significantly affect the interpretation of the findings presented in this study. However, we acknowledge that the exact locations of the individual zones proposed in our models relative to each other are unclear and could potentially impact the overall SED. Due to the complexity involved, this aspect remains unexplored in the current work for the sake of simplicity.

### 4.1. Observed differences between 2017 and 2018 source emission influencing the SED modelling

Before we begin to explain the SED modelling in detail, it is worth summarising the differences between the 2017 and 2018 source emission that influence the SED modelling. The most notable differences are the flux increase and a hint of a spectral hardening ($\Gamma_{\mathrm{VHE}}\approx2.0$) observed in the VHE $\gamma$-ray energy band during the April 2018 campaign. As reported in Sect. 2.4.2, the FWHM time scale of the VHE flux variation was ~ 3 days, providing an important constraint on the size of the $\gamma$-ray emitting region in April 2018 that was not available in 2017, albeit degenerate with relativistic beaming effects.

Our interpretation of the X-ray data for the modelling also differs somewhat compared to M87 MWL2017. While some variability is implied by the flux enhancement in the X-ray band in 2018 in comparison with 2017 (see Fig. 15 as well as Sect. 3.3.2), there is no definitive evidence of its association with the VHE $\gamma$-ray flare. Therefore, we will examine both of the cases where the X-ray flux is associated with the EHT-oriented model and high-energy ($E$ > 100 MeV) oriented model.

Finally, we treat the optical/IR flux observations as upper limits because we cannot separate out the galaxy contamination due to the inadequate resolution of the Kanata optical telescope. While we choose a value of the SED parameters to keep the model within the range consistent with the *Swift*-UVOT emission.

### 4.2. EHT oriented model

The following free parameters characterise the spherical, single-zone emission region: the radius ($R$), the Doppler factor of the bulk motion ($\delta$), and the averaged magnetic field strength ($B$). The Doppler factor is defined as $\delta = 1/\Gamma_L(1 - \beta\cos\theta_{\mathrm{view}})$, where $\Gamma_L$ is the bulk Lorentz factor of the emission region, and we set $\theta_{\mathrm{view}} = 17°$ (Event Horizon Telescope Collaboration et al. 2019a). The nonthermal eDF within the emission region is characterised by the electron number density ($n_e$), the spectral indices of the broken power law ($p_1$ and $p_2$), and the minimum/break/maximum Lorentz factors ($\gamma_{\min}$, $\gamma_{\mathrm{brk}}$, and $\gamma_{\max}$), respectively.

Overall, the characteristics of the radio-to-millimetre emission in the observed broadband SED in April 2018 are similar to the SED in April 2017 (M87 MWL2017). Therefore, we followed a similar modelling methodology. First, we set $R$ to be equivalent to the photon-ring size (Event Horizon Telescope Collaboration et al. 2019a), that is, the single-zone effectively encompasses a sphere around the SMBH out to $R \approx 5r_g$. Since the bulk speed of the jet has not yet become relativistic, we fix $\delta \sim 1$. We set $\gamma_{\min} = 1$ as it has little influence on the SED, that is, the value of $\gamma_{\min}$ affects the energy density of nonthermal electrons $U_e$ only if the photon index is softer than 2 (e.g., Kino et al. 2002; Finke et al. 2008). Therefore, we fixed the conservative $\gamma_{\min}$ for simplicity. Based on observed radio core shift measurements (Hada et al. 2011), the frequencies below ~ 230 GHz are optically thick to synchrotron self-absorption (SSA), that constrains the magnetic field strength $B$ > a few G (Event Horizon Telescope Collaboration et al. 2021d). Keeping this limit in mind, one can fit the SED by only adjusting $B$ and the eDF parameters. The best fit model parameters are summarised in Table 1. There are no essential differences between 2017 and 2018 for the EHT-oriented model.

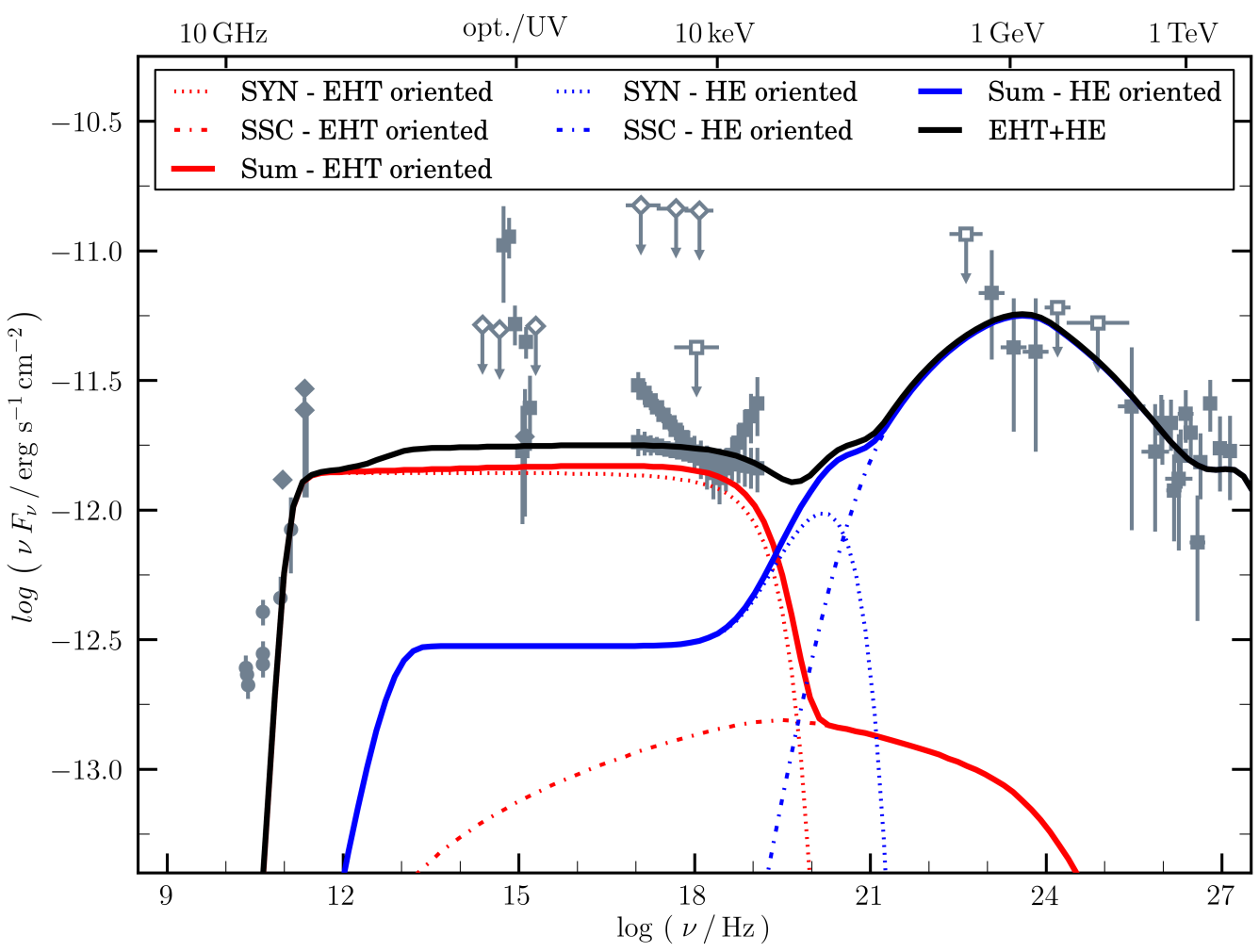

**Fig. 19.** Model A fitted to the observed contemporaneous broadband SED of M87 from the EHT campaign in April 2018. For each considered region (EHT and HE), the synchrotron (labeled as SYN) and synchrotron self-Compton (labeled as SSC) models as well as the summation of the two models (labeled as Sum) are reported.

Similar to the results obtained in M87 MWL2017, we find that the entire MWL radiation cannot be explained by the EHT-oriented model alone. In this model $U_e$ and $U_B$ are approximately comparable in the EHT emission region when $\gamma_{\min} = 1$. Since we adopt $\gamma_{min} = 1$, $U_e$ provides an upper limit. Although a detailed parameter survey will be left to future studies, $U_B > U_e$ is anticipated for $\gamma_{\rm min} > 1$ in the EHT emission region. This would align more closely with the scenario of a magnetically accelerated jet.

### 4.3. High-energy oriented model

Here we introduce the HE-oriented model to fit both the observed HE and VHE $\gamma$-ray spectrum. The variability time scale indicates that the characteristic size of the VHE $\gamma$-ray emitting region is limited to $R_{\rm HE} \lesssim 8 r_g \delta \left(\frac{\Delta t}{3\ \rm days}\right)$. From this, it is clear that there is a degeneracy between the observed variability timescale and $\delta$. Hence, we will also consider cases of $\delta > 1$ below. Similar to the EHT-oriented model, we first focus on the case of $\delta \sim 1$, because both a jet kinematic analysis using VLBI monitoring observations and an analysis of the brightness ratio of the approaching-jet and counter-jet suggest that the M87 jet is moving at subluminal speeds within the deprojected distance about $10^3\ r_g$ from the central SMBH (Mertens et al. 2016; Park et al. 2019, and references therein). We took as reference for the Doppler factor a value $\delta = 1.82$, that is consistent with the range allowed by the assumed viewing angle ($\theta_{\rm view} = 17°$) and able to reproduce the SED.

The 2018 SED VHE active phase further motivates us to explore the case of $\delta > 1$. It is not straightforward to fit both the VHE $\gamma$-ray spectrum with $\Gamma_{VHE} \approx 2.0$ and the Fermi-LAT spectrum with a single emitting zone. We therefore considered two approaches: In Model A, we modified the eDF parameters such that the modelled emission accounts for the tentative hardening of the VHE spectrum in April 2018 (i.e. $p_1 > p_2$), and in Model B, we took into account a fast moving component. [5]

The best-fit for Model A is shown in Fig. 19. We note that the physical origin is not obvious and further theoretical work is needed to assess its feasibility in the case of M87.

In Model B, we retain the loss-dominated approach in the single-zone for the higher energy index (i.e. $p_1 < p_2$). We examine the phenomenological eDF with $p_1 < p_2$, that is complementary to the model A, since the radiative cooling rate may dominate the acceleration rate. In this model, we consider a fast moving blob as an additional component of the SED to explain the hard H.E.S.S. $\gamma$-ray spectrum in a flaring epoch. It should be mentioned that we need this additional component not due to the softening of the high-energy eDF, but due to the appearance of the decrease of VHE $\gamma$-ray emission because of the Klein-Nishina effect. To reproduce the H.E.S.S. $\gamma$-ray flux and avoid overshooting the observed X-ray fluxes during the flare, we set a very weak magnetic field strength of $\boldsymbol{B} = 4$ mG and $\boldsymbol{\delta} = 2.55$ in the high-energy components. This model also looks consistent with the observed SED including the VHE $\gamma$-ray flare in Fig. 20. The resultant $U_{\rm e}/U_{\rm B} \simeq 5.8 \times 10^4$ shown in Table 1 is very high, requiring an extremely weakly magnetised plasma. The extent to which this could be attributed to magnetic reconnection is not yet clear as current reconnection models tend to suggest that the emission region is likely to be near equipartition (e.g., Petropoulou et al. 2016). Another possibility is that the interaction between the highly magnetised jets and the weakly magnetised wind (e.g., within the jet sheath) can result in the dissipation of the magnetic energy (Kusafuka et al. 2023, see also Komissarov 2012).

The estimated Doppler factor of Model B's VHE-flare component is $\delta = 2.55$ (the corresponding bulk Lorentz factor is $\lesssim 2$ if the flaring component is moving along the global jet). The assumed Lorentz factor is consistent with the mildly high bulk Lorentz factor proposed by observations of M87: the gradual acceleration of the jet is suggested by high-cadence VLBI observations (Park et al. 2019), where the estimated Lorentz factor is $\sim 2$ at the distance $\gtrsim 10^3 r_{\rm g}$, and even at the HST-1 knot, the X-ray observations also suggest that the Lorentz factor of the jet is high $\sim 6$ (Snios et al. 2019). The required jet power of the VHE flare component of the model B is $\sim 3.7 \times 10^{42}$ ergs$^{-1}$. This is compatible with the jet power exhibited in GRMHD simulations.

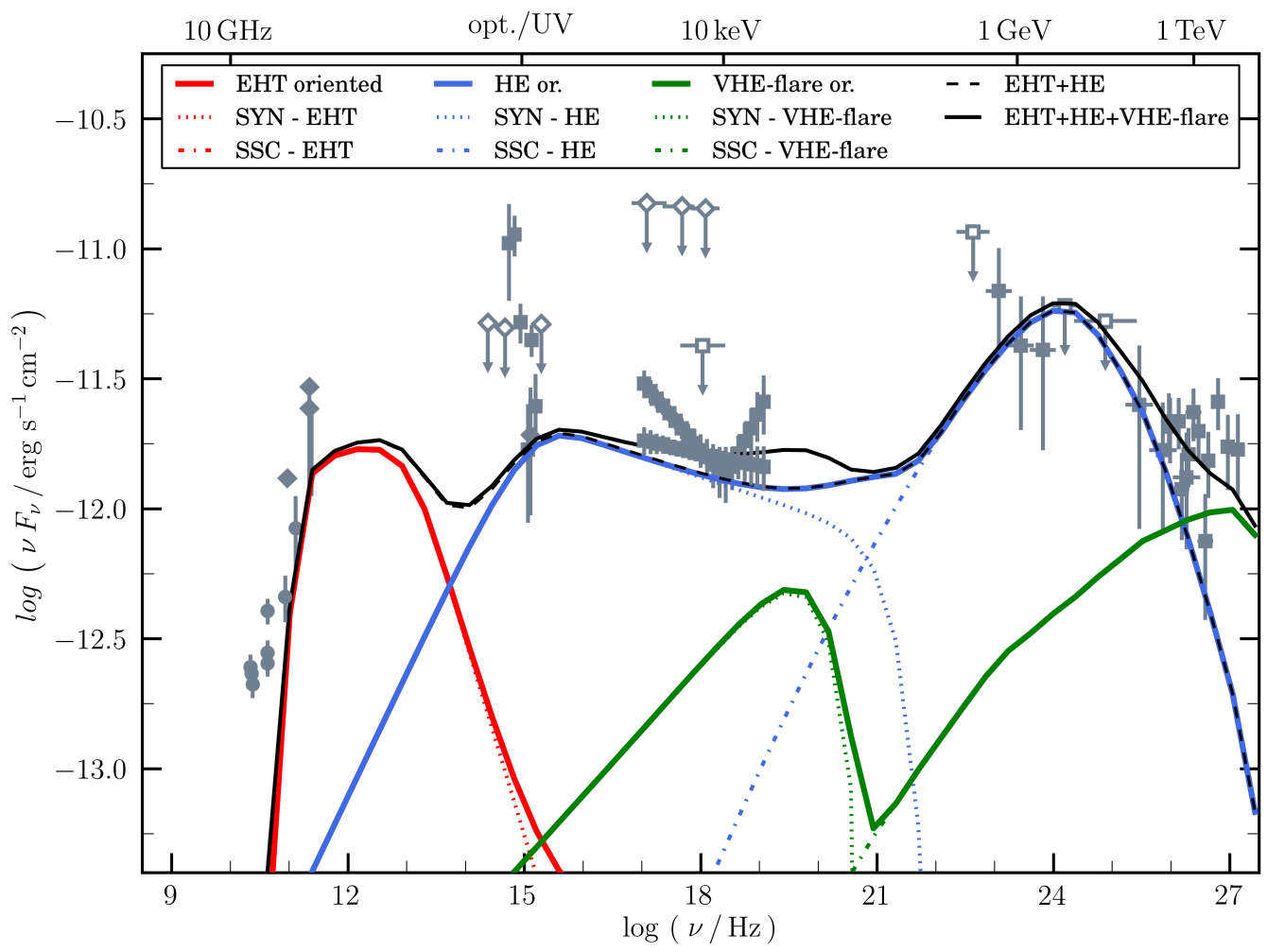

**Fig. 20.** Model B fitted to the observed contemporaneous broadband SED of M87 for the EHT campaign in April 2018. For each considered region (EHT, high-energy, and very-high-energy flare), the synchrotron (labeled as SYN) and synchrotron self-Compton (labeled as SSC) models as well as the summation of the SYN and SSC models (solid line) are reported.

### 4.4. X-ray broken power law oriented model

Fig. 21 shows the Model A fit for the case of BPL X-ray radiation. The inverted component on the HE side of the X-ray spectrum can be fitted by increasing the synchrotron flux in the HE-oriented model region. This leads to slightly different best fit values, since the eDF must be fine-tuned to reduce the increase in the synchrotron self-Compton (SSC) emission component corresponding to the increased synchrotron flux. It is clear that model A explains both PL and BPL cases naturally without altering the essential parts of the model. As shown by the figures in Table 1, the difference between the two cases can be attributed to the difference in the physical state of the HE region. When $n'_e$ and $B'$ in the HE region increase by several times, the low-energy tail part of the emission from the HE region contributes to the X-ray energy band, resulting in the realisation of BPL. Furthermore, it should be emphasised that when BPL is realised in the X-ray energy region, the advantage is that tighter constraints can be placed on the model parameters of the HE emission region through SED fitting. As can be seen from the figure, the high-energy side of the BPL spectrum is a combination of the high-energy bump of the SYN component from

[5] For Model A, we used `agnpy`, an open-source `Python` package that includes the radiative processes of jetted AGN (Nigro et al. 2022), while Model B is based on a GR MWL radiative transfer code RAIKOU (Kawashima et al. 2023), where the GR effects are turned off in this paper.

**Table 1.** SED model fitting parameters for three different models used (A-PL, A-BPL and B model - for models and parameters details see Sect. 4.2 and Sect. 4.3).

| Model | A (EHT) [5] | A (HE) | A (EHT/BPL) | A (HE/BPL) | B (EHT) | B (HE) | B (VHE-flare) |
|---|---|---|---|---|---|---|---|
| $\delta$ | 1 | 1.82 | 1 | 1.82 | 1 | 1.82 | 2.55 |
| $R$ $[r_g]$ [1 2] | 5.0 | 10.0 | 5.0 | 10.0 | 5.0 | 10.0 | 20.0 |
| $n'_e$ [cm$^{-3}$] [3] | $2.0\times10^{6}$ | $1.7\times10^{1}$ | $1.8\times10^{6}$ | $6.6\times10^{1}$ | $4.0\times10^{5}$ | $1.6\times10^{3}$ | $1.5\times10^{1}$ |
| $B'$ [G] | 5.3 | $2.3\times10^{-2}$ | 4.6 | $8.0\times10^{-2}$ | 10 | $2.5\times10^{-2}$ | $4.0\times10^{-3}$ |
| $\gamma_{\rm min}$ | 1 | $5\times10^{3}$ | 1 | $1.2\times10^{3}$ | 1 | 30 | $10^{3}$ |
| $\gamma_{\rm br}$ | – | $7\times10^{6}$ | – | $7\times10^{6}$ | $4\times10^{2}$ | $3\times10^{5}$ | – |
| $\gamma_{\rm max}/10^{6}$ | 1.0 | 50 | 0.19 | 45 | 10 | 100 | 60 |
| $p_1$ | 3.0 | 3.0 | 2.9 | 3.0 | 2.8 | 2.1 | 2.5 |
| $p_2$ | – | 2.0 | – | 2.0 | 4.5 | 3.15 | – |
| $U_e/U_B$ [4] | 2.9 | $6.7\times10^{3}$ | 3.6 | $5.2\times10^{2}$ | 0.18 | $7.6\times10^{3}$ | $5.8\times10^{4}$ |
| $L_e$ [erg s$^{-1}$] | $2.0\times10^{42}$ | $2.1\times10^{42}$ | $1.9\times10^{42}$ | $2.0\times10^{42}$ | $1.6\times10^{42}$ | $2.5\times10^{42}$ | $3.7\times10^{42}$ |
| $L_{poy}$ [erg s$^{-1}$] | $7.0\times10^{41}$ | $3.2\times10^{38}$ | $5.2\times10^{41}$ | $3.8\times10^{39}$ | $9.0\times10^{42}$ | $3.4\times10^{38}$ | $6.4\times10^{37}$ |

**Notes.**
(1) For M87*, $r_g = 9.8\times10^{14}$ cm.
(2) Parameter pegged to the observed ring size for the EHT-oriented models and the upper limit of the allowed interval for the HE-oriented models.
(3) The number density of total non-thermal electrons or electron/positron pairs.
(4) The ratio of the nonthermal electron (or electron/positron pair) energy to the magnetic energy, that is derived for comparison between models.
(5) For the EHT-MWL 2017 modelling, the inferred parameters were $\delta = 1$, $R = 5.2r_g$, $n'_e = 3.6\times10^{5}$ cm$^{-3}$, $B' = 4.7$ G, $\gamma_{\rm min} = 1$, $\gamma_{\rm max} = 3.5\times10^{6}$, $p_1 = 2.2$, and resulting $U_e/U_B = 2.3$, that are similar to those obtained for the EHT-oriented model A here.

the HE region and the low-energy side of the SSC component. This SED fitting can only be accurately reproduced with a delicate balance of model parameters. In other words, it can be said that this imposes stricter constraints on the model parameters. As mentioned in Sect. 4.1, the presence of the BPL spectrum requires the bump to be located in the MeV energy band. However, we face the lack of data at MeV energies that could help provide crucial information to distinguish between the proposed models.

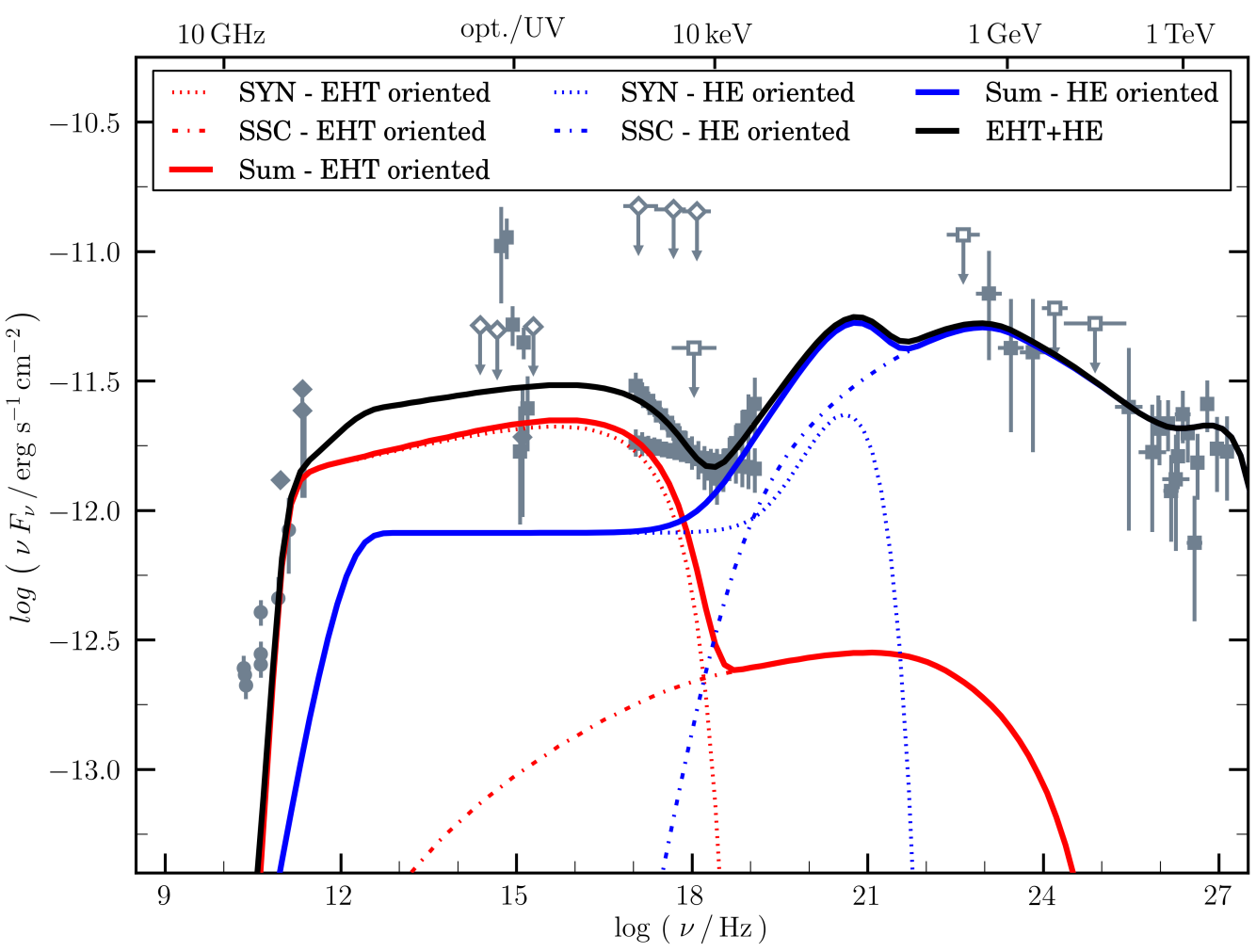

**Fig. 21.** Modified Model A fitted describing the BPL emission seen in the *Chandra-NuSTAR* X-ray spectrum.

### 4.5. The possible origin of $\gamma$-ray emission

One motivation to release this data set is to aid in the development of more complex models that can address the origin of the $\gamma$-ray emission, at the same time satisfying the new constraints from VLBI imaging. We discuss here potential scenarios for the generation of HE non-thermal electrons (and positrons) in a compact region with the size of the order of $\sim 10r_g$ or a larger region $\sim 100r_g$.

The potential production and acceleration of high-energy electron-positron pairs at the vacuum gap formed at the stagnation surface within an accreting BH magnetosphere, could offer an explanation for the observed gamma-ray flare emission (e.g., Levinson & Rieger 2011; Broderick & Tchekhovskoy 2015; Hirotani 2018; Chen & Yuan 2020; Crinquand et al. 2020; Katsoulakos & Rieger 2020; Kisaka et al. 2022, and references therein). Semi-analytical calculations based on this class of models indicate that the potential VHE emission power from this gap region is sufficient to explain the day-scale VHE variability from M87. While this gap-driven black hole magnetosphere model can explain TeV $\gamma$ rays, it seems difficult to reproduce GeV $\gamma$-ray flux due to a hard photon spectrum and a very high maximum energy of electrons (e.g., Levinson & Rieger 2011; Hirotani 2018). Therefore, other radiation mechanisms seem necessary to reproduce the observed GeV $\gamma$-ray emission.

Magnetic reconnection, a commonly occurring process in magnetised plasma, is capable of accelerating particles to high energies (e.g. Zenitani & Hoshino 2001; Lyubarsky & Liverts 2008; Sironi & Spitkovsky 2014; Guo et al. 2014; Werner et al. 2016). Comparison of the linearly polarised, horizon-scale emission of M87 detected by EHT with GRMHD simulations strongly suggests that M87*'s accretion flow is currently in a magnetically arrested disk (MAD) state (Event Horizon Telescope Collaboration et al. 2021a,b). The MAD state (Igumenshchev & Narayan 2002; Narayan et al. 2003, e.g.) occurs when magnetic fields carried in by the accretion flow build up on the horizon, to the point where the magnetic pressure can push back, and truncate, the accretion flow.

The preference for a MAD-like state is also supported by other recent investigations on the properties of magnetic fields and velocity fields downstream of the M87 jet (Chael et al. 2019; Cruz-Osorio et al. 2022; Yuan et al. 2022; Kino et al. 2022; Fromm et al. 2022). Recent ultra-high resolution GRMHD simulations of MAD-state flows have shown that magnetic reconnec-tion can occur in an equatorial current sheet just outside the event horizon as a result of magnetic flux eruptions; this process has been proposed as a possible origin of VHE $\gamma$-rays (e.g. Ripperda et al. 2020, 2022; Chashkina et al. 2021; Hakobyan et al. 2023), presenting models that are qualitatively consistent with the M87 MWL campaign observation in 2018. Furthermore, the size of the transient current layer can become of the order of 10 $r_g$ (Ripperda et al. 2022), that is qualitatively consistent with the HE-oriented model of M87 2018 data. Recent studies of Gelles et al. (2022) and Jia et al. (2023) showed that during a reconnection-powered VHE flare, the sub-millimetre/radio core flux can decline due to the temporarily ejected accretion disk, that may be similar to what we observe (see light curves in Figs. 15 and 16).

The flaring VHE $\gamma$-rays could also originate via an additional fast moving component (i.e. a structured, multi-zone scenario). There are mainly two types of models for representing the fast moving component: (i) the structured jet with fast spine and slow sheath components[6] (e.g. Ghisellini et al. 2005) and (ii) a fast moving blob (or mini jet) component (e.g. Lenain, J.-P. et al. 2008; Giannios et al. 2009). The latter is considered in Model B (Sect. 4.3) to account for the VHE $\gamma$ rays. There are two possible origins of this fast moving component: magnetic reconnection or shock.

Whether magnetic reconnection can account for the implied high $U_{\rm e}/U_{\rm B}$ (Table 1) remains to be explored. Current (2d PIC) simulations of relativistic reconnection (e.g., Sironi et al. 2015; Petropoulou et al. 2016) rather suggest $U_{\rm e} \sim U_{\rm B}$. The shock formed via the interaction between strongly magnetised jets and weakly magnetised winds is another possible origin for the blob component of the Model B. Kusafuka et al. (2023) performed one-dimensional spherically symmetric special-relativistic MHD simulations and found that $\sim$ 10% of the magnetic energy can be dissipated. This resulted in the acceleration of a weakly magnetised wind, that may be the emission site of the VHE $\gamma$-rays. Since the very high $U_{\rm e}/U_{\rm B} \sim 10^{4-6}$ is also reported in some blazars (e.g., 1ES 1101-232 and 3C 66A), with simple analysis using single-zone SSC models (Domínguez et al. 2013), it will be important to further explore to what extent a high $U_{\rm e}/U_{\rm B}$ can be realised under different conditions.

Although we primarily focus on the leptonic model, hadronic scenarios might also play a non-negligible role in explaining the M87 SED (Reimer et al. 2004; Reynoso et al. 2011; Rieger & Aharonian 2012; Kimura & Toma 2020; Boughelilba et al. 2022; H.E.S.S. Collaboration 2023).

### 4.6. Observational aid towards unveiling gamma ray origin in M87

While Model A agrees with magnetic reconnection on the point of the variation timescale if the reconnection takes place in the vicinity of the black hole in the MAD (e.g., see Sect. 4.1 in Ripperda et al. 2022), Model B may require shock acceleration because of a large $U_e/U_B$. X-ray polarisation information can help discriminate these two model scenarios by constraining the non-thermal electron production mechanism. If the flare is caused by magnetic reconnection, the position angle may drastically flip with time, whereas a shock-induced flare may not produce this strong effect. Observations with IXPE or other future X-ray polarimetry missions (such as eXTP; Zhang et al. 2019) could provide important information for unveiling the emission mechanisms in M87.

In addition, observations carried with more sensitive VHE instruments like the upcoming Cherenkov Telescope Array (Cherenkov Telescope Array Consortium et al. 2019; Zanin et al. 2022) are of fundamental interest. Thanks to its better short-timescale sensitivity and spectral resolution, CTA may provide new insights on future observed $\gamma$-ray flaring episodes.

Moreover, the sparse sampling and limited data qualities in the currently available EHT observations make it difficult to draw a conclusion on the origin of the long-term PA variation at $\mu$as scale. Further EHT images in the subsequent years will be needed to constrain the actual duration of PA variations at EHT scales and their physical origin.

More extended MWL monitoring with higher observation frequency is essential for capturing future MWL flares beyond the VHE one detected from M87 in 2018. In future campaigns, we aim to extend the monitoring period, increasing the chances of identifying potential correlations between the MWL light curves and improving sensitivity to detect up to weeks-long variability. Finally, an enhanced EHT array, as well as potentially future next-generation experiments (e.g., Lico et al. 2023), will allow us to increase the observed field-of-view. This will improve our ability to probe regions near the core (e.g., the inner jet) as potential sites for $\gamma$-ray flares.

## 5. Summary

We have presented the results of the second MWL observational campaign on M87 together with M87* observations performed by the EHT in April 2018. In addition to obtaining the second most extensive quasi-simultaneous broadband spectra of M87 taken yet, we have provided details of the individual observations and light curves, with the aim of investigating the emission variability of M87 and constraining the origin of the $\gamma$-ray emission in M87. Similar to M87 MWL2017, the primary outcome of this campaign is represented by the incredible legacy data set, for which all data and some analysis scripts are available to the community via a Cyverse repository (see Appendix 5). While we defer detailed modelling to future works by the broader community, we draw preliminary conclusions using a heuristic approach to the SED modelling. As seen in 2017, the HST-1 knot component was subdominant, and the core was dominating the total flux in the radio through the X-ray bands, providing ideal conditions for combining MWL data over a large range of spatial resolutions.

The main results of the 2018 EHT-MWL observational campaign on M87 can be summarised as follows:

- We detected a short (approximately three days) VHE $\gamma$-ray flare, the first since 2010. The combined data from all IACTs involved in the campaign, H.E.S.S., MAGIC, and VERITAS, helped fully describe the flare duration and the shape of the light curve and indicated spectral hardening during the flare.
- An HE $\gamma$-ray flux increase was also detected by *Fermi*-LAT during the same days.
- A likely longer-term core flux enhancement was observed in the X-ray band by *Chandra*, presenting an X-ray spectrum harder than in non-flaring states, as was observed during the EHT 2017 campaign.
- While radio and millimetre core fluxes are compatible with (or even lower than) the emission seen in April 2017, VLBA

[6] The spine-sheath structure is proposed by the radio observations of M87 (Kovalev et al. 2007; Asada et al. 2016; Hada et al. 2016; Lu et al. 2023) and may be related to the VHE $\gamma$-ray emission. On the other hand, there is a caveat that weakly de-beamed dense infrared emission could result in considerable $\gamma\gamma$ absorption and a softer SED, that may be difficult to explain the hard VHE components (e.g., Rieger & Aharonian 2012, and references therein).

(centimetre to millimetre) radio observations present a clear change, on an annual basis, of the jet position angle within a few milli-arcsecond from the core. The EHT (millimetre) radio observations also revealed a significant variation in the position angle of the asymmetry of the ring, indicating that this might be related to changes seen at larger scales in the jet's position angle.
– Although the presence of the flaring episodes allowed us to constrain the size of the VHE $\gamma$-ray emitting region in the SED modelling, the location is still uncertain. M87*'s complex, broadband SED cannot be modelled by a one-zone model. Instead, an additional second component (or even a third one as obtained for Model B) is needed to capture its emission.

The results of this broadband MWL campaign of M87 offer a unique opportunity to investigate the origin of the $\gamma$-ray emission during a flaring episode. In particular, they highlight the importance of having coordinated MWL observations with well-sampled coverage to fully characterise the spectral variability in M87*, that likely spans several different timescales. The data sets made public as a part of this effort also offer a rich archive for future investigations. As outlined in Sect. 1, these data add to the growing suite of observational constraints that inform GR models as well as phenomenological models, such as those presented in this work. The ~17% uncertainty on testing GR is currently dominated by theoretical uncertainties, but combining the M87 SED with fits to the EHT image can reduce these errors and improve our understanding of the underlying astrophysics.

The EHT-MWL observational campaigns performed with a more sensitive array in 2021 and 2022 as well as those planned for upcoming years will enable further insights for constraining the physics around M87*, such as the disk-jet connection, and the origin and mechanisms responsible for the emission of $\gamma$-ray photons. In addition, the results from the EHT observations of other sources such as Sgr A* (Event Horizon Telescope Collaboration et al. 2022a,b), 3C 279 (Kim et al. 2020), Centaurus A (Janssen et al. 2021), OJ 287, and other targets planned for the upcoming campaigns, together with their contemporaneous MWL coverage, will greatly expand the sample of objects and the impact of the MWL studies undertaken within the EHT-MWL working group. The high resolution EHT millimetre-VLBI images and the broadband SED information will enable deep investigation of the emission mechanisms driving AGN.

## Data availability

Data products presented in this paper are available for download through the EHT Collaboration Data Webpage (https://eventhorizontelescope.org/for-astronomers/data), or directly from the CyVerse repository. The repository contains the following data products:

– Broadband spectrum table with frequency, flux density, its uncertainty, and instrument index (format: DAT)
– Sampled posterior distributions of the SED broadband spectral model (A-PL) described in Sect. 4.2 (format: DAT).
– Sampled posterior distributions of the SED broadband spectral model (A-BPL) described in Sect. 4.3 (format: DAT).
– Sampled posterior distributions of the SED broadband spectral model (B) described in Sect. 4.3 (format: DAT).

*Acknowledgements.* The full acknowledgements are available in Appendix A.

[1] Department of Physics, Faculty of Science, University of Malaya, 50603 Kuala Lumpur, Malaysia
[2] Yale Center for Astronomy & Astrophysics, 52 Hillhouse Avenue, New Haven, CT 06511, USA
[3] Department of Physics, Yale University, P.O. Box 2018120, New Haven, CT 06520, USA
[4] Centre for Space Research, North-West University, Potchefstroom 2520, South Africa
[5] Physical Research Laboratory, Ahmedabad, Gujarat, 380009, India
[6] Korea Astronomy and Space Science Institute, Daedeok-daero 776, Yuseong-gu, Daejeon 34055, Republic of Korea
[7] University of Science and Technology, Gajeong-ro 217, Yuseong-gu, Daejeon 34113, Republic of Korea
[8] Research Center for Astronomical Computing, Zhejiang Lab, Hangzhou 311100, People's Republic of China
[9] INAF Istituto di Radioastronomia, Via P. Gobetti, 101, I-40129 Bologna, Italy
[10] 525 Davey Laboratory, Department of Astronomy and Astrophysics, Pennsylvania State University, University Park, PA 16802, USA
[11] Department of Physics, McGill University, 3600 University Street, Montréal, QC H3A 2T8, Canada
[12] Trottier Space Institute at McGill, 3550 University Street, Montréal, QC H3A 2A7, Canada
[13] Department of Astrophysics, Institute for Mathematics, Astrophysics and Particle Physics (IMAPP), Radboud University, P.O. Box 9010, 6500 GL Nijmegen, The Netherlands
[14] Leiden Observatory—Allegro, Leiden University, P.O. Box 9513, 2300 RA Leiden, The Netherlands
[15] Center for Astrophysics | Harvard & Smithsonian, Cambridge, MA 02138, USA
[16] Mizusawa VLBI Observatory, National Astronomical Observatory of Japan, 2-12 Hoshigaoka, Mizusawa, Oshu, Iwate 023-0861, Japan
[17] Astronomical Science Program, The Graduate University for Advanced Studies (SOKENDAI), 2-21-1 Osawa, Mitaka, Tokyo 181-8588, Japan
[18] Graduate School of Science, Nagoya City University, Yamanohata 1, Mizuho-cho, Mizuho-ku, Nagoya, 467-8501, Aichi, Japan
[19] McGill Space Institute, McGill University, 3550 University Street, Montréal, QC H3A 2A7, Canada
[20] Institute for Astrophysical Research, Boston University, 725 Commonwealth Ave., Boston 02215, MA, USA
[21] Department of Astronomy and Astrophysics, Pennsylvania State University, University Park, PA 16802, USA
[22] Institute for Cosmic Ray Research, The University of Tokyo, 5-1-5 Kashiwanoha, Kashiwa, Chiba 277-8582, Japan
[23] Max-Planck-Institut für Radioastronomie, Auf dem Hügel 69, D-53121 Bonn, Germany
[24] National Astronomical Observatory of Japan, 2-21-1 Osawa, Mitaka, Tokyo 181-8588, Japan
[25] Kogakuin University of Technology & Engineering, Academic Support Center, 2665-1 Nakano, Hachioji, Tokyo 192-0015, Japan
[26] Shanghai Astronomical Observatory, Chinese Academy of Sciences, 80 Nandan Road, Shanghai 200030, People's Republic of China
[27] Key Laboratory of Radio Astronomy, Chinese Academy of Sciences, Nanjing 210008, People's Republic of China
[28] API–Anton Pannekoek Institute for Astronomy, University of Amsterdam, Science Park 904, 1098 XH Amsterdam, The Netherlands
[29] GRAPPA–Gravitation and AstroParticle Physics Amsterdam, University of Amsterdam, Science Park 904, 1098 XH Amsterdam, The Netherlands
[30] Department of Physics and Astronomy, Northwestern University, 2145 Sheridan Rd, Evanston, IL 60208, USA
[31] Center for Interdisciplinary Exploration and Research in Astrophysics (CIERA), Northwestern University, 1800 Sherman Ave, Evanston, IL 60201, USA

[32] Department of Physics, Villanova University, 800 E. Lancaster Avenue, Villanova, PA 19085, USA
[33] Physics Department, Washington University CB 1105, St Louis, MO 63130, USA
[34] Dipartimento di Fisica, Università di Trieste, I-34127 Trieste, Italy
[35] Istituto Nazionale di Fisica Nucleare, Sezione di Trieste, I-34127 Trieste, Italy
[36] Finnish Centre for Astronomy with ESO, FI-20014 University of Turku, Finland
[37] Aalto University Metsähovi Radio Observatory, Metsähovintie 114, FI-02540 Kylmälä, Finland
[38] Center for Computational Astrophysics, Flatiron Institute, 162 Fifth Avenue, New York, NY 10010, USA
[39] Department of Astrophysical Sciences, Peyton Hall, Princeton University, Princeton, NJ 08544, USA
[40] Hiroshima Astrophysical Science Center, Hiroshima University, 1-3-1 Kagamiyama, Higashi-Hiroshima, Hiroshima 739-8526, Japan
[41] Massachusetts Institute of Technology Haystack Observatory, 99 Millstone Road, Westford, MA 01886, USA
[42] Black Hole Initiative at Harvard University, 20 Garden Street, Cambridge, MA 02138, USA
[43] Instituto de Astrofísica de Andalucía-CSIC, Glorieta de la Astronomía s/n, E-18008 Granada, Spain
[44] Max-Planck-Institut für Radioastronomie, Auf dem Hügel 69, D-53121 Bonn, Germany
[45] Department of Physics & Astronomy, The University of Texas at San Antonio, One UTSA Circle, San Antonio, TX 78249, USA
[46] Institute of Astronomy and Astrophysics, Academia Sinica, 11F of Astronomy-Mathematics Building, AS/NTU No. 1, Sec. 4, Roosevelt Rd., Taipei 10617, Taiwan, R.O.C.
[47] Departament d'Astronomia i Astrofísica, Universitat de València, C. Dr. Moliner 50, E-46100 Burjassot, València, Spain
[48] Observatori Astronòmic, Universitat de València, C. Catedrático José Beltrán 2, E-46980 Paterna, València, Spain
[49] Department of Space, Earth and Environment, Chalmers University of Technology, Onsala Space Observatory, SE-43992 Onsala, Sweden
[50] Steward Observatory and Department of Astronomy, University of Arizona, 933 N. Cherry Ave., Tucson, AZ 85721, USA
[51] Astronomy Department, Universidad de Concepción, Casilla 160-C, Concepción, Chile
[52] Department of Physics, University of Illinois, 1110 West Green Street, Urbana, IL 61801, USA
[53] Fermi National Accelerator Laboratory, MS209, P.O. Box 500, Batavia, IL 60510, USA
[54] Department of Astronomy and Astrophysics, University of Chicago, 5640 South Ellis Avenue, Chicago, IL 60637, USA
[55] East Asian Observatory, 660 N. A'ohoku Place, Hilo, HI 96720, USA
[56] James Clerk Maxwell Telescope (JCMT), 660 N. A'ohoku Place, Hilo, HI 96720, USA
[57] California Institute of Technology, 1200 East California Boulevard, Pasadena, CA 91125, USA
[58] Institute of Astronomy and Astrophysics, Academia Sinica, 645 N. A'ohoku Place, Hilo, HI 96720, USA
[59] Department of Physics and Astronomy, University of Hawaii at Manoa, 2505 Correa Road, Honolulu, HI 96822, USA
[60] Department of Physics, McGill University, 3600 rue University, Montréal, QC H3A 2T8, Canada
[61] Trottier Space Institute at McGill, 3550 rue University, Montréal, QC H3A 2A7, Canada
[62] Institut de Radioastronomie Millimétrique (IRAM), 300 rue de la Piscine, F-38406 Saint Martin d'Hères, France
[63] Perimeter Institute for Theoretical Physics, 31 Caroline Street North, Waterloo, ON N2L 2Y5, Canada
[64] Department of Physics and Astronomy, University of Waterloo, 200 University Avenue West, Waterloo, ON N2L 3G1, Canada
[65] Waterloo Centre for Astrophysics, University of Waterloo, Waterloo, ON N2L 3G1, Canada
[66] Department of Astronomy, University of Massachusetts, Amherst, MA 01003, USA
[67] Kavli Institute for Cosmological Physics, University of Chicago, 5640 South Ellis Avenue, Chicago, IL 60637, USA
[68] Department of Physics, University of Chicago, 5720 South Ellis Avenue, Chicago, IL 60637, USA
[69] Enrico Fermi Institute, University of Chicago, 5640 South Ellis Avenue, Chicago, IL 60637, USA
[70] Princeton Gravity Initiative, Jadwin Hall, Princeton University, Princeton, NJ 08544, USA
[71] Data Science Institute, University of Arizona, 1230 N. Cherry Ave., Tucson, AZ 85721, USA
[72] Program in Applied Mathematics, University of Arizona, 617 N. Santa Rita, Tucson, AZ 85721, USA
[73] Cornell Center for Astrophysics and Planetary Science, Cornell University, Ithaca, NY 14853, USA
[74] Department of Astronomy, Yonsei University, Yonsei-ro 50, Seodaemun-gu, 03722 Seoul, Republic of Korea
[75] Physics Department, Fairfield University, 1073 North Benson Road, Fairfield, CT 06824, USA
[76] Department of Astronomy, University of Illinois at Urbana-Champaign, 1002 West Green Street, Urbana, IL 61801, USA
[77] Instituto de Astronomía, Universidad Nacional Autónoma de México (UNAM), Apdo Postal 70-264, Ciudad de México, México
[78] Institut für Theoretische Physik, Goethe-Universität Frankfurt, Max-von-Laue-Straße 1, D-60438 Frankfurt am Main, Germany
[79] Research Center for Intelligent Computing Platforms, Zhejiang Laboratory, Hangzhou 311100, China
[80] Tsung-Dao Lee Institute, Shanghai Jiao Tong University, Shengrong Road 520, Shanghai, 201210, People's Republic of China
[81] Department of Astronomy and Columbia Astrophysics Laboratory, Columbia University, 500 W. 120th Street, New York, NY 10027, USA
[82] Center for Computational Astrophysics, Flatiron Institute, 162 Fifth Avenue, New York, NY 10010, USA
[83] Dipartimento di Fisica "E. Pancini", Università di Napoli "Federico II", Compl. Univ. di Monte S. Angelo, Edificio G, Via Cinthia, I-80126, Napoli, Italy
[84] INFN Sez. di Napoli, Compl. Univ. di Monte S. Angelo, Edificio G, Via Cinthia, I-80126, Napoli, Italy
[85] Wits Centre for Astrophysics, University of the Witwatersrand, 1 Jan Smuts Avenue, Braamfontein, Johannesburg 2050, South Africa
[86] Department of Physics, University of Pretoria, Hatfield, Pretoria 0028, South Africa
[87] Centre for Radio Astronomy Techniques and Technologies, Department of Physics and Electronics, Rhodes University, Makhanda 6140, South Africa
[88] ASTRON, Oude Hoogeveensedijk 4, 7991 PD Dwingeloo, The Netherlands
[89] LESIA, Observatoire de Paris, Université PSL, CNRS, Sorbonne Université, Université de Paris, 5 place Jules Janssen, F-92195 Meudon, France
[90] JILA and Department of Astrophysical and Planetary Sciences, University of Colorado, Boulder, CO 80309, USA
[91] National Astronomical Observatories, Chinese Academy of Sciences, 20A Datun Road, Chaoyang District, Beijing 100101, PR China
[92] Las Cumbres Observatory, 6740 Cortona Drive, Suite 102, Goleta, CA 93117-5575, USA
[93] Department of Physics, University of California, Santa Barbara, CA 93106-9530, USA
[94] National Radio Astronomy Observatory, 520 Edgemont Road, Charlottesville, VA 22903, USA
[95] Department of Electrical Engineering and Computer Science, Massachusetts Institute of Technology, 32-D476, 77 Massachusetts Ave., Cambridge, MA 02142, USA
[96] Google Research, 355 Main St., Cambridge, MA 02142, USA
[97] Institut für Theoretische Physik und Astrophysik, Universität Würzburg, Emil-Fischer-Str. 31, D-97074 Würzburg, Germany

[98] Department of History of Science, Harvard University, Cambridge, MA 02138, USA
[99] Department of Physics, Harvard University, Cambridge, MA 02138, USA
[100] NCSA, University of Illinois, 1205 W. Clark St., Urbana, IL 61801, USA
[101] Instituto de Astronomia, Geofísica e Ciências Atmosféricas, Universidade de São Paulo, R. do Matão, 1226, São Paulo, SP 05508-090, Brazil
[102] CP3-Origins, University of Southern Denmark, Campusvej 55, DK-5230 Odense M, Denmark
[103] Instituto Nacional de Astrofísica, Óptica y Electrónica. Apartado Postal 51 y 216, 72000. Puebla Pue., México
[104] Consejo Nacional de Humanidades, Ciencia y Tecnología, Av. Insurgentes Sur 1582, 03940, Ciudad de México, México
[105] Key Laboratory for Research in Galaxies and Cosmology, Chinese Academy of Sciences, Shanghai 200030, People's Republic of China
[106] NOVA Sub-mm Instrumentation Group, Kapteyn Astronomical Institute, University of Groningen, Landleven 12, 9747 AD Groningen, The Netherlands
[107] Department of Astronomy, School of Physics, Peking University, Beijing 100871, People's Republic of China
[108] Kavli Institute for Astronomy and Astrophysics, Peking University, Beijing 100871, People's Republic of China
[109] Department of Astronomy, Graduate School of Science, The University of Tokyo, 7-3-1 Hongo, Bunkyo-ku, Tokyo 113-0033, Japan
[110] The Institute of Statistical Mathematics, 10-3 Midori-cho, Tachikawa, Tokyo, 190-8562, Japan
[111] Department of Statistical Science, The Graduate University for Advanced Studies (SOKENDAI), 10-3 Midori-cho, Tachikawa, Tokyo 190-8562, Japan
[112] Kavli Institute for the Physics and Mathematics of the Universe, The University of Tokyo, 5-1-5 Kashiwanoha, Kashiwa, 277-8583, Japan
[113] Leiden Observatory, Leiden University, Postbus 2300, 9513 RA Leiden, The Netherlands
[114] ASTRAVEO LLC, PO Box 1668, Gloucester, MA 01931
[115] Applied Materials Inc., 35 Dory Road, Gloucester, MA 01930
[116] Joint Institute for VLBI ERIC (JIVE), Oude Hoogeveensedijk 4, 7991 PD Dwingeloo, The Netherlands
[117] Department of Physics, Korea Advanced Institute of Science and Technology (KAIST), 291 Daehak-ro, Yuseong-gu, Daejeon 34141, Republic of Korea
[118] Graduate School of Science and Technology, Niigata University, 8050 Ikarashi 2-no-cho, Nishi-ku, Niigata 950-2181, Japan
[119] Physics Department, National Sun Yat-Sen University, No. 70, Lien-Hai Road, Kaosiung City 80424, Taiwan, R.O.C.
[120] School of Astronomy and Space Science, Nanjing University, Nanjing 210023, People's Republic of China
[121] Key Laboratory of Modern Astronomy and Astrophysics, Nanjing University, Nanjing 210023, People's Republic of China
[122] INAF-Istituto di Radioastronomia, Via P. Gobetti 101, I-40129 Bologna, Italy
[123] INAF-Istituto di Radioastronomia & Italian ALMA Regional Centre, Via P. Gobetti 101, I-40129 Bologna, Italy
[124] Department of Physics, National Taiwan University, No. 1, Sec. 4, Roosevelt Rd., Taipei 10617, Taiwan, R.O.C
[125] Instituto de Radioastronomía y Astrofísica, Universidad Nacional Autónoma de México, Morelia 58089, México
[126] Yunnan Observatories, Chinese Academy of Sciences, 650011 Kunming, Yunnan Province, People's Republic of China
[127] Center for Astronomical Mega-Science, Chinese Academy of Sciences, 20A Datun Road, Chaoyang District, Beijing, 100012, People's Republic of China
[128] Key Laboratory for the Structure and Evolution of Celestial Objects, Chinese Academy of Sciences, 650011 Kunming, People's Republic of China
[129] Department of Astrophysical Sciences, Peyton Hall, Princeton University, Princeton, NJ 08544, USA
[130] School of Physics and Astronomy, Shanghai Jiao Tong University, 800 Dongchuan Road, Shanghai, 200240, People's Republic of China
[131] Institut de Radioastronomie Millimétrique (IRAM), Avenida Divina Pastora 7, Local 20, E-18012, Granada, Spain
[132] National Institute of Technology, Hachinohe College, 16-1 Uwanotai, Tamonoki, Hachinohe City, Aomori 039-1192, Japan
[133] Astronomy Department, Universidad de Concepción, Casilla 160-C, Concepción, Chile Research Center for Astronomy, Academy of Athens, Soranou Efessiou 4, 115 27 Athens, Greece
[134] School of Physics, Georgia Institute of Technology, 837 State St NW, Atlanta, GA 30332, USA
[135] Department of Astronomy and Space Science, Kyung Hee University, 1732, Deogyeong-daero, Giheung-gu, Yongin-si, Gyeonggi-do 17104, Republic of Korea
[136] Canadian Institute for Theoretical Astrophysics, University of Toronto, 60 St. George Street, Toronto, ON M5S 3H8, Canada
[137] Dunlap Institute for Astronomy and Astrophysics, University of Toronto, 50 St. George Street, Toronto, ON M5S 3H4, Canada
[138] Canadian Institute for Advanced Research, 180 Dundas St West, Toronto, ON M5G 1Z8, Canada
[139] Department of Physics, National Taiwan Normal University, No. 88, Sec. 4, Tingzhou Rd., Taipei 116, Taiwan, R.O.C.
[140] Center of Astronomy and Gravitation, National Taiwan Normal University, No. 88, Sec. 4, Tingzhou Road, Taipei 116, Taiwan, R.O.C.
[141] Aalto University Metsähovi Radio Observatory, Metsähovintie 114, FI-02540 Kylmälä, Finland
[142] Gemini Observatory/NSF's NOIRLab, 670 N. Aòhōkū Place, Hilo, HI 96720, USA
[143] Frankfurt Institute for Advanced Studies, Ruth-Moufang-Strasse 1, D-60438 Frankfurt, Germany
[144] School of Mathematics, Trinity College, Dublin 2, Ireland
[145] Department of Physics, University of Toronto, 60 St. George Street, Toronto, ON M5S 1A7, Canada
[146] Aalto University Department of Electronics and Nanoengineering, PL 15500, FI-00076 Aalto, Finland
[147] Jeremiah Horrocks Institute, University of Central Lancashire, Preston PR1 2HE, UK
[148] National Biomedical Imaging Center, Peking University, Beijing 100871, People's Republic of China
[149] College of Future Technology, Peking University, Beijing 100871, People's Republic of China
[150] Tokyo Electron Technology Solutions Limited, 52 Matsunagane, Iwayado, Esashi, Oshu, Iwate 023-1101, Japan
[151] Department of Physics and Astronomy, University of Lethbridge, Lethbridge, Alberta T1K 3M4, Canada
[152] Netherlands Organisation for Scientific Research (NWO), Postbus 93138, 2509 AC Den Haag, The Netherlands
[153] Frontier Research Institute for Interdisciplinary Sciences, Tohoku University, Sendai 980-8578, Japan
[154] Astronomical Institute, Tohoku University, Sendai 980-8578, Japan
[155] Department of Physics and Astronomy, Seoul National University, Gwanak-gu, Seoul 08826, Republic of Korea
[156] University of New Mexico, Department of Physics and Astronomy, Albuquerque, NM 87131, USA
[157] Physics Department, Brandeis University, 415 South Street, Waltham, MA 02453, USA
[158] Tuorla Observatory, Department of Physics and Astronomy, University of Turku, Finland
[159] Radboud Excellence Fellow of Radboud University, Nijmegen, The Netherlands
[160] School of Natural Sciences, Institute for Advanced Study, 1 Einstein Drive, Princeton, NJ 08540, USA
[161] School of Physics, Huazhong University of Science and Technology, Wuhan, Hubei, 430074, People's Republic of China
[162] Mullard Space Science Laboratory, University College London, Holmbury St. Mary, Dorking, Surrey, RH5 6NT, UK

[163] Center for Astronomy and Astrophysics and Department of Physics, Fudan University, Shanghai 200438, People's Republic of China
[164] Astronomy Department, University of Science and Technology of China, Hefei 230026, People's Republic of China
[165] Department of Physics and Astronomy, Michigan State University, 567 Wilson Rd, East Lansing, MI 48824, USA
[166] Istituto Nazionale di Fisica Nucleare, Sezione di Pisa, I-56127 Pisa, Italy
[167] California State University, Los Angeles, Department of Physics and Astronomy, Los Angeles, CA 90032, USA
[168] Dipartimento di Fisica "M. Merlin" dell'Università e del Politecnico di Bari, via Amendola 173, I-70126 Bari, Italy
[169] Istituto Nazionale di Fisica Nucleare, Sezione di Bari, I-70126 Bari, Italy
[170] W. W. Hansen Experimental Physics Laboratory, Kavli Institute for Particle Astrophysics and Cosmology, Department of Physics and SLAC National Accelerator Laboratory, Stanford University, Stanford, CA 94305, USA
[171] Istituto Nazionale di Fisica Nucleare, Sezione di Torino, I-10125 Torino, Italy
[172] Dipartimento di Fisica, Università degli Studi di Torino, I-10125 Torino, Italy
[173] Laboratoire Leprince-Ringuet, CNRS/IN2P3, 'Ecole polytechnique, Institut Polytechnique de Paris, 91120 Palaiseau, France
[174] INAF-Istituto di Astrofisica Spaziale e Fisica Cosmica Milano, via E. Bassini 15, I-20133 Milano, Italy
[175] Italian Space Agency, Via del Politecnico snc, 00133 Roma, Italy
[176] Space Science Division, Naval Research Laboratory, Washington, DC 20375-5352, USA
[177] Istituto Nazionale di Fisica Nucleare, Sezione di Roma "Tor Vergata", I-00133 Roma, Italy
[178] Space Science Data Center - Agenzia Spaziale Italiana, Via del Politecnico, snc, I-00133, Roma, Italy
[179] Dipartimento di Fisica, Università degli Studi di Perugia, I-06123 Perugia, Italy
[180] Istituto Nazionale di Fisica Nucleare, Sezione di Perugia, I-06123 Perugia, Italy
[181] Grupo de Altas Energ'ias, Universidad Complutense de Madrid, E-28040 Madrid, Spain
[182] Ruhr University Bochum, Faculty of Physics and Astronomy, Astronomical Institute (AIRUB), 44780 Bochum, Germany
[183] Department of Physical Sciences, Hiroshima University, Higashi-Hiroshima, Hiroshima 739-8526, Japan
[184] Dipartimento di Fisica, Università di Roma "Tor Vergata", I-00133 Roma, Italy
[185] Universit'e Paris Cité, Université Paris-Saclay, CEA, CNRS, AIM, F-91191 Gif-sur-Yvette, France
[186] NASA Goddard Space Flight Center, Greenbelt, MD 20771, USA
[187] Department of Physics, KTH Royal Institute of Technology, AlbaNova, SE-106 91 Stockholm, Sweden
[188] The Oskar Klein Centre for Cosmoparticle Physics, AlbaNova, SE-106 91 Stockholm, Sweden
[189] NASA Marshall Space Flight Center, Huntsville, AL 35812, USA
[190] The Aerospace Corporation, 14745 Lee Rd, Chantilly, VA 20151, USA
[191] Department of Physics and Center for Space Sciences and Technology, University of Maryland Baltimore County, Baltimore, MD 21250, USA
[192] Hiroshima Astrophysical Science Center, Hiroshima University, Higashi-Hiroshima, Hiroshima 739-8526, Japan
[193] Vatican Observatory, V-00120 Castel Gandolfo, Vatican City State
[194] Department of physics and Astronomy, Louisiana State University, Baton Rouge, LA 70803, USA
[195] INAF-Astronomical Observatory of Padova, Vicolo dell'Osservatorio 5, I-35122 Padova, Italy
[196] Institut für Astro- und Teilchenphysik, Leopold-Franzens-Universität Innsbruck, A-6020 Innsbruck, Austria
[197] Instituto de Física Teórica UAM/CSIC, Universidad Autónoma de Madrid, E-28049 Madrid, Spain
[198] Departamento de Física Teórica, Universidad Autónoma de Madrid, 28049 Madrid, Spain
[199] Santa Cruz Institute for Particle Physics Department of Physics and Department of Astronomy and Astrophysics, University of California at Santa Cruz, Santa Cruz, CA 95064, USA
[200] NYCB Real-Time Computing Inc., Lattingtown, NY 11560-1025, USA
[201] Purdue University Northwest, Hammond, IN 46323, USA
[202] Nagoya University, Institute for Space-Earth Environmental Research, Furo-cho, Chikusa-ku, Nagoya 464-8601, Japan
[203] Kobayashi-Maskawa Institute for the Origin of Particles and the Universe, Nagoya University, Furo-cho, Chikusa-ku, Nagoya, Japan
[204] Institute of Space Sciences (ICE, CSIC), Campus UAB, Carrer de Magrans s/n, E-08193 Barcelona, Spain; and Institut d'Estudis Espacials de Catalunya (IEEC), E-08034 Barcelona, Spain
[205] Instituci'o Catalana de Recerca i Estudis Avanccats (ICREA), E-08010 Barcelona, Spain
[206] Center for Astrophysics and Cosmology, University of Nova Gorica, Nova Gorica, Slovenia
[207] Dublin Institute for Advanced Studies, 31 Fitzwilliam Place, Dublin 2, Ireland
[208] Max-Planck-Institut für Kernphysik, Saupfercheckweg 1, 69117 Heidelberg, Germany
[209] Yerevan State University, 1 Alek Manukyan St, Yerevan 0025, Armenia
[210] Yerevan State University, 1 Alek Manukyan St., Yerevan 0025, Armenia
[211] Landessternwarte, Universität Heidelberg, Königstuhl, D 69117 Heidelberg, Germany
[212] Kapteyn Astronomical Institute, University of Groningen, Landleven 12, 9747 AD Groningen, The Netherlands
[213] Laboratoire Leprince-Ringuet, École Polytechnique, CNRS, Institut Polytechnique de Paris, F-91128 Palaiseau, France
[214] Deutsches Elektronen-Synchrotron DESY, Platanenallee 6, 15738 Zeuthen, Germany
[215] Institut für Physik und Astronomie, Universität Potsdam, Karl-Liebknecht-Strasse 24/25, D 14476 Potsdam, Germany
[216] Université de Paris, CNRS, Astroparticule et Cosmologie, F-75013 Paris, France
[217] Department of Physics and Electrical Engineering, Linnaeus University, 351 95 Växjö, Sweden
[218] Institut für Physik, Humboldt-Universität zu Berlin, Newtonstr. 15, D 12489 Berlin, Germany
[219] Laboratoire Univers et Théories, Observatoire de Paris, Université PSL, CNRS, Université Paris Cité, 5 Pl. Jules Janssen, 92190 Meudon, France
[220] Sorbonne Université, CNRS/IN2P3, Laboratoire de Physique Nucléaire et de Hautes Energies, LPNHE, 4 place Jussieu, 75005 Paris, France
[221] IRFU, CEA, Université Paris-Saclay, F-91191 Gif-sur-Yvette, France
[222] University of Oxford, Department of Physics, Denys Wilkinson Building, Keble Road, Oxford OX1 3RH, UK
[223] Friedrich-Alexander-Universität Erlangen-Nürnberg, Erlangen Centre for Astroparticle Physics, Nikolaus-Fiebiger-Str. 2, 91058 Erlangen, Germany
[224] Astronomical Observatory, The University of Warsaw, Al. Ujazdowskie 4, 00-478 Warsaw, Poland
[225] Instytut Fizyki Jądrowej PAN, ul. Radzikowskiego 152, 31-342 Kraków, Poland
[226] Universität Hamburg, Institut für Experimentalphysik, Luruper Chaussee 149, D 22761 Hamburg, Germany
[227] School of Physics, University of the Witwatersrand, 1 Jan Smuts Avenue, Braamfontein, Johannesburg, 2050 South Africa
[228] Laboratoire Univers et Particules de Montpellier, Université Montpellier, CNRS/IN2P3, CC 72, Place Eugène Bataillon, F-34095 Montpellier Cedex 5, France
[229] School of Physical Sciences, University of Adelaide, Adelaide 5005, Australia

[230] Aix Marseille Université, CNRS/IN2P3, CPPM, Marseille, France
[231] Institut für Astronomie und Astrophysik, Universität Tübingen, Sand 1, D 72076 Tübingen, Germany
[232] Universität Innsbruck, Institut für Astro- und Teilchenphysik, Technikerstraße 25, 6020 Innsbruck, Austria
[233] Obserwatorium Astronomiczne, Uniwersytet Jagielloński, ul. Orla 171, 30-244 Kraków, Poland
[234] University of Namibia, Department of Physics, Private Bag 13301, Windhoek 10005, Namibia
[235] Institute of Astronomy, Faculty of Physics, Astronomy and Informatics, Nicolaus Copernicus University, Grudziadzka 5, 87-100 Torun, Poland
[236] Nicolaus Copernicus Astronomical Center, Polish Academy of Sciences, ul. Bartycka 18, 00-716 Warsaw, Poland
[237] Université Bordeaux, CNRS, LP2I Bordeaux, UMR 5797, F-33170 Gradignan, France
[238] University of Southern Denmark, DK-5230 Odense, Denmark
[239] Department of Physics and Astronomy, The University of Leicester, University Road, Leicester, LE1 7RH, United Kingdom
[240] Department of Physics, Konan University, 8-9-1 Okamoto, Higashinada, Kobe, Hyogo 658-8501, Japan
[241] Kavli Institute for the Physics and Mathematics of the Universe (WPI), The University of Tokyo Institutes for Advanced Study (UTIAS), The University of Tokyo, 5-1-5 Kashiwa-no-Ha, Kashiwa, Chiba, 277-8583, Japan
[242] Japanese MAGIC Group: Institute for Cosmic Ray Research (ICRR), The University of Tokyo, Kashiwa, 277-8582 Chiba, Japan
[243] ETH Zürich, CH-8093 Zürich, Switzerland
[244] Università di Siena and INFN Pisa, I-53100 Siena, Italy
[245] Institut de Física d'Altes Energies (IFAE), The Barcelona Institute of Science and Technology (BIST), E-08193 Bellaterra (Barcelona), Spain
[246] Universitat de Barcelona, ICCUB, IEEC-UB, E-08028 Barcelona, Spain
[247] Instituto de Astrofísica de Andalucía-CSIC, Glorieta de la Astronomía s/n, 18008, Granada, Spain
[248] National Institute for Astrophysics (INAF), I-00136 Rome, Italy
[249] Università di Udine and INFN Trieste, I-33100 Udine, Italy
[250] Max-Planck-Institut für Physik, D-85748 Garching, Germany
[251] Università di Padova and INFN, I-35131 Padova, Italy
[252] Croatian MAGIC Group: University of Zagreb, Faculty of Electrical Engineering and Computing (FER), 10000 Zagreb, Croatia
[253] Centro Brasileiro de Pesquisas Físicas (CBPF), 22290-180 URCA, Rio de Janeiro (RJ), Brazil
[254] IPARCOS Institute and EMFTEL Department, Universidad Complutense de Madrid, E-28040 Madrid, Spain
[255] Instituto de Astrofísica de Canarias and Dpto. de Astrofísica, Universidad de La Laguna, E-38200, La Laguna, Tenerife, Spain
[256] University of Lodz, Faculty of Physics and Applied Informatics, Department of Astrophysics, 90-236 Lodz, Poland
[257] Centro de Investigaciones Energéticas, Medioambientales y Tecnológicas, E-28040 Madrid, Spain
[258] Departament de Física, and CERES-IEEC, Universitat Autònoma de Barcelona, E-08193 Bellaterra, Spain
[259] Università di Pisa and INFN Pisa, I-56126 Pisa, Italy
[260] INFN MAGIC Group: INFN Sezione di Bari and Dipartimento Interateneo di Fisica dell'Università e del Politecnico di Bari, I-70125 Bari, Italy
[261] Department for Physics and Technology, University of Bergen, Norway
[262] INFN MAGIC Group: INFN Sezione di Torino and Università degli Studi di Torino, I-10125 Torino, Italy
[263] Croatian MAGIC Group: University of Rijeka, Faculty of Physics, 51000 Rijeka, Croatia
[264] Universität Würzburg, D-97074 Würzburg, Germany
[265] Technische Universität Dortmund, D-44221 Dortmund, Germany
[266] Japanese MAGIC Group: Physics Program, Graduate School of Advanced Science and Engineering, Hiroshima University, 739-8526 Hiroshima, Japan
[267] Armenian MAGIC Group: ICRANet-Armenia, 0019 Yerevan, Armenia
[268] Croatian MAGIC Group: University of Split, Faculty of Electrical Engineering, Mechanical Engineering and Naval Architecture (FESB), 21000 Split, Croatia
[269] Croatian MAGIC Group: Josip Juraj Strossmayer University of Osijek, Department of Physics, 31000 Osijek, Croatia
[270] Finnish MAGIC Group: Finnish Centre for Astronomy with ESO, Department of Physics and Astronomy, University of Turku, FI-20014 Turku, Finland
[271] Japanese MAGIC Group: Department of Physics, Tokai University, Hiratsuka, 259-1292 Kanagawa, Japan
[272] University of Geneva, Chemin d'Ecogia 16, CH-1290 Versoix, Switzerland
[273] Saha Institute of Nuclear Physics, A CI of Homi Bhabha National Institute, Kolkata 700064, West Bengal, India
[274] Inst. for Nucl. Research and Nucl. Energy, Bulgarian Academy of Sciences, BG-1784 Sofia, Bulgaria
[275] Japanese MAGIC Group: Department of Physics, Yamagata University, Yamagata 990-8560, Japan
[276] Finnish MAGIC Group: Space Physics and Astronomy Research Unit, University of Oulu, FI-90014 Oulu, Finland
[277] Japanese MAGIC Group: Chiba University, ICEHAP, 263-8522 Chiba, Japan
[278] Japanese MAGIC Group: Institute for Space-Earth Environmental Research and Kobayashi-Maskawa Institute for the Origin of Particles and the Universe, Nagoya University, 464-6801 Nagoya, Japan
[279] Japanese MAGIC Group: Department of Physics, Kyoto University, 606-8502 Kyoto, Japan
[280] INFN MAGIC Group: INFN Roma Tor Vergata, I-00133 Roma, Italy
[281] Japanese MAGIC Group: Department of Physics, Konan University, Kobe, Hyogo 658-8501, Japan
[282] also at International Center for Relativistic Astrophysics (ICRA), Rome, Italy
[283] also at Port d'Informació Científica (PIC), E-08193 Bellaterra (Barcelona), Spain
[284] also at Department of Physics, University of Oslo, Norway
[285] Max-Planck-Institut für Physik, D-85748 Garching, Germany
[286] also at INAF Padova, I-35122 Padova, Italy
[287] Japanese MAGIC Group: Institute for Cosmic Ray Research (ICRR), The University of Tokyo, Kashiwa, 277-8582 Chiba, Japan
[288] CP3-Origins, University of Southern Denmark, Campusvej 55, 5230 Odense M, Denmark
[289] Physics Department, Columbia University, New York, NY 10027, USA
[290] Department of Physics and Astronomy and the Bartol Research Institute, University of Delaware, Newark, DE 19716, USA
[291] Department of Physics and Astronomy, University of Utah, Salt Lake City, UT 84112, USA
[292] Physics Department, California Polytechnic State University, San Luis Obispo, CA 94307, USA
[293] Department of Physics, Washington University, St. Louis, MO 63130, USA
[294] Department of Physics, California State University - East Bay, Hayward, CA 94542, USA
[295] School of Physics and Astronomy, University of Minnesota, Minneapolis, MN 55455, USA
[296] Santa Cruz Institute for Particle Physics and Department of Physics, University of California, Santa Cruz, CA 95064, USA
[297] Department of Physics and Astronomy, University of Iowa, Van Allen Hall, Iowa City, IA 52242, USA
[298] Department of Physics and Astronomy, Iowa State University, Ames, IA 50011, USA
[299] Department of Physics and Astronomy, DePauw University, Greencastle, IN 46135-0037, USA
[300] School of Physics, University College Dublin, Belfield, Dublin 4, Ireland
[301] Department of Physics and Astronomy, Dartmouth College, 6127 Wilder Laboratory, Hanover, NH 03755, USA

[302] Department of Physics, University of Maryland, College Park, MD, USA
[303] School of Natural Sciences, University of Galway, University Road, Galway, H91 TK33, Ireland
[304] Physics Department, McGill University, Montreal, QC H3A 2T8, Canada
[305] Department of Physics and Astronomy, Purdue University, West Lafayette, IN 47907, USA
[306] Department of Physics and Astronomy, University of California, Los Angeles, CA 90095, USA
[307] Arthur B. McDonald Canadian Astroparticle Physics Research Institute, 64 Bader Lane, Queen's University, Kingston, ON Canada, K7L 3N6
[308] Institute of Physics and Astronomy, University of Potsdam, 14476 Potsdam-Golm, Germany
[309] Fakultät für Physik & Astronomie, Ruhr-Universität Bochum, D-44780 Bochum, Germany
[310] School of Physics, University College Dublin, Belfield, Dublin 4, Ireland
[311] Department of Physical Sciences, Munster Technological University, Bishopstown, Cork, T12 P928, Ireland
[312] Department of Physics, Indiana University-Purdue University Indianapolis, Indianapolis, IN 46202, USA
[313] Department of Physics, University of Maryland, Baltimore County, Baltimore MD 21250, USA
[314] Department of Physics and Astronomy, Barnard College, Columbia University, NY 10027, USA
[315] Department of Physics and Astronomy, University of Alabama, Tuscaloosa, AL 35487, USA
[316] NASA GSFC, Greenbelt, MD 20771, USA
[317] Key Laboratory of Radio Astronomy and Technology, Chinese Academy of Sciences, A20 Datun Road, Chaoyang District, Beijing 100101, PR China
[318] Xinjiang Astronomical Observatory, Chinese Academy of Sciences, 150 Science 1-Street, Urumqi 830011, People's Republic of China
[319] Graduate School of Sciences and Technology for Innovation, Yamaguchi University, J-753-0841 Yamaguchi, Japan
[320] The Research Institute for Time Studies, J-753-0841 Yamaguchi, Japan University
[321] Korea Astronomy and Space Science Institute, Daedeok-daero 776, Yuseong-gu, Daejeon 34055, Republic of Korea
[322] National Astronomical Research Institute of Thailand (NARIT), 260 Moo 4, T. Donkaew, A. Maerim, 50180 Chiangmai, Thailand
[323] Graduate School of Science, Osaka Metropolitan University, 1-1 Gakuen-cho, Naka-ku, Sakai, Osaka 599-8531, Japan
[324] Center for Astronomy, Ibaraki University, 2-1-1 Bunkyo, Mito, Ibaraki 310-8512, Japan

## Appendix A: Acknowledgements

Individual author contribution: G. Principe: project leadership, Fermi-LAT data analysis, MWL variability and SED study, results interpretation, paper writing; J. C. Algaba: MWL images study; C. Arcaro: H.E.S.S. analysis; M. Balokovic: X-ray data analysis and interpretation ; V. Barbosa Martins: H.E.S.S. analysis, results interpretation, and paper writing; S. Chandra: AstroSAT data analysis; Y.-Z. Cui: VERA, EAVN/KaVA data analysis, and radio properties study; F. D'Ammando: Swift-UVOT; A. D. Falcone: Swift-XRT analysis; N. M. Ford: X-ray results interpretation; D. Glawion: H.E.S.S. analysis and paper writing; M. Giroletti: Fermi-LAT data crosscheck and paper writing; C. Goddi: ALMA data analysis; M. A. Gurwell: SMA data analysis; K. Hada: VERA, EAVN/KaVA data analysis, radio results interpretation, and paper writing; D. Haggard: X-ray analysis, MWL results interpretation and paper writing; A. Hahn: MAGIC data analysis, MWL variability study, paper writing, and IACT cross-normalisation analysis; W. Jin: VERITAS data analysis, IACT cross-normalisation analysis, and paper writing; S. Jorstad: Swift-UVOT data analysis; A. Kaur: Swift-XRT data analysis; T. Kawashima: SED Modelling and interpretation; S. Kerby: Swift-XRT data analysis; J.-Y. Kim: KVN data analysis; M. Kino: SED modelling and interpretation; E. V. Kravchenko: VLBA data analysis; S.-S. Lee: KVN data analysis; R.-S. Lu: GMVA data analysis; S. Markoff: MWL results interpretation and paper writing ; D. Mazin: MAGIC data analysis; J. Michail: paper writing; J. Neilsen: Chandra and NuSTAR data analysis and X-ray results interpretation; M. A. Nowak: Swift-XRT analysis; G. Pühlhofer: results interpretation and IACT cross-normalisation analysis; V. Ramakrishnan: paper writing; B. Ripperda: MWL results interpretation and paper writing; M. Santander: VERITAS data analysis; M. Sasada: SED modelling; S. S. Savchenko: Swift-UVOT analysis crosscheck; C. Sheridan: paper writing; all remaining authors: contributed with technical tasks, development, maintenance, and operations of the instruments as well as with reviewing the draft.

The Event Horizon Telescope Collaboration thanks the following organisations and programs: the Academia Sinica; the Academy of Finland (projects 274477, 284495, 312496, 315721); the Agencia Nacional de Investigación y Desarrollo (ANID), Chile via NCN19_058 (TITANs), Fondecyt 1221421 and BASAL FB210003; the Alexander von Humboldt Stiftung; an Alfred P. Sloan Research Fellowship; Allegro, the European ALMA Regional Centre node in the Netherlands, the NL astronomy research network NOVA and the astronomy institutes of the University of Amsterdam, Leiden University, and Radboud University; the ALMA North America Development Fund; the Astrophysics and High Energy Physics programme by MCIN (with funding from European Union NextGenerationEU, PRTR-C17I1); the Black Hole Initiative, that is funded by grants from the John Templeton Foundation (60477, 61497, 62286) and the Gordon and Betty Moore Foundation (Grant GBMF-8273) - although the opinions expressed in this work are those of the author and do not necessarily reflect the views of these Foundations; the Brinson Foundation; "la Caixa" Foundation (ID 100010434) through fellowship codes LCF/BQ/DI22/11940027 and LCF/BQ/DI22/11940030; Chandra DD7-18089X and TM6-17006X; the China Scholarship Council; the China Postdoctoral Science Foundation fellowships (2020M671266, 2022M712084); Consejo Nacional de Humanidades, Ciencia y Tecnología (CONAHCYT, Mexico, projects U0004-246083, U0004-259839, F0003-272050, M0037-279006, F0003-281692, 104497, 275201, 263356); the Colfuturo Scholarship; the Consejería de Economía, Conocimiento, Empresas y Universidad of the Junta de Andalucía (grant P18-FR-1769), the Consejo Superior de Investigaciones Científicas (grant 2019AEP112); the Delaney Family via the Delaney Family John A. Wheeler Chair at Perimeter Institute; Dirección General de Asuntos del Personal Académico-Universidad Nacional Autónoma de México (DGAPA-UNAM, projects IN112820 and IN108324); the Dutch Organization for Scientific Research (NWO) for the VICI award (grant 639.043.513), the grant OCENW.KLEIN.113, and the Dutch Black Hole Consortium (with project No. NWA 1292.19.202) of the research programme the National Science Agenda; the Dutch National Supercomputers, Cartesius and Snellius (NWO grant 2021.013); the EACOA Fellowship awarded by the East Asia Core Observatories Association, which consists of the Academia Sinica Institute of Astronomy and Astrophysics, the National Astronomical Observatory of Japan, Center for Astronomical Mega-Science, Chinese Academy of Sciences, and the Korea Astronomy and Space Science Institute; the European Research Council (ERC) Synergy Grant "BlackHoleCam: Imaging the Event Horizon of Black Holes" (grant 610058); the European Union Horizon 2020 research and innovation programme under grant agreements RadioNet (No. 730562) and M2FINDERS (No. 101018682); the European Research Council for advanced grant 'JETSET: Launching, propagation and emission of relativistic jets from binary mergers and across mass scales' (grant No. 884631); the Horizon ERC Grants 2021 programme under grant agreement No. 101040021; the FAPESP (Fundação de Amparo á Pesquisa do Estado de São Paulo) under grant 2021/01183-8; the Fondo CAS-ANID folio CAS220010; the Generalitat Valenciana (grants APOSTD/2018/177 and ASFAE/2022/018) and GenT Program (project CIDEGENT/2018/021); the Gordon and Betty Moore Foundation (GBMF-3561, GBMF-5278, GBMF-10423); the Institute for Advanced Study; the Istituto Nazionale di Fisica Nucleare (INFN) sezione di Napoli, iniziative specifiche TEONGRAV; the International Max Planck Research School for Astronomy and Astrophysics at the Universities of Bonn and Cologne; DFG research grant "Jet physics on horizon scales and beyond" (grant No. 443220636); Joint Columbia/Flatiron Postdoctoral Fellowship (research at the Flatiron Institute is supported by the Simons Foundation); the Japan Ministry of Education, Culture, Sports, Science and Technology (MEXT; grant JPMXP1020200109); the Japan Society for the Promotion of Science (JSPS) Grant-in-Aid for JSPS Research Fellowship (JP17J08829); the Joint Institute for Computational Fundamental Science, Japan; the Key Research Program of Frontier Sciences, Chinese Academy of Sciences (CAS, grants QYZDJ-SSW-SLH057, QYZDJSSW-SYS008, ZDBS-LY-SLH011); the Leverhulme Trust Early Career Research Fellowship; the Max-Planck-Gesellschaft (MPG); the Max Planck Partner Group of the MPG and the CAS; the MEXT/JSPS KAKENHI (grants 18KK0090, JP21H01137, JP18H03721, JP18K13594, 18K03709, JP19K14761, 18H01245, 25120007, 19H01943, 21H01137, 21H04488, 22H00157, 23K03453); the MICINN Research Project PID2019-108995GB-C22; the MIT International Science and Technology Initiatives (MISTI) Funds; the Ministry of Science and Technology (MOST) of Taiwan (103-2119-M-001-010-MY2, 105-2112-M-001-025-MY3, 105-2119-M-001-042, 106-2112-M-001-011, 106-2119-M-001-013, 106-2119-M-001-027, 106-2923-M-001-005, 107-2119-M-001-017, 107-2119-M-001-020, 107-2119-M-001-041, 107-2119-M-110-005, 107-2923-M-001-009, 108-2112-M-001-048, 108-2112-M-001-051, 108-2923-M-001-002, 109-

2112-M-001-025, 109-2124-M-001-005, 109-2923-M-001-001, 110-2112-M-001-033, 110-2124-M-001-007, 110-2923-M-001-001, and 112-2112-M-003-010-MY3); the National Science and Technology Council (NSTC) of Taiwan (111-2124-M-001-005, 112-2124-M-001-014); the Ministry of Education (MoE) of Taiwan Yushan Young Scholar Program; the Physics Division, National Center for Theoretical Sciences of Taiwan; the National Aeronautics and Space Administration (NASA, Fermi Guest Investigator grant 80NSSC23K1508, NASA Astrophysics Theory Program grant 80NSSC20K0527, NASA NuSTAR award 80NSSC20K0645); NASA Hubble Fellowship grants HST-HF2-51431.001-A, HST-HF2-51482.001-A awarded by the Space Telescope Science Institute, which is operated by the Association of Universities for Research in Astronomy, Inc., for NASA, under contract NAS5-26555; the National Institute of Natural Sciences (NINS) of Japan; the National Key Research and Development Program of China (grant 2016YFA0400704, 2017YFA0402703, 2016YFA0400702); the National Science and Technology Council (NSTC, grants NSTC 111-2112-M-001 -041, NSTC 111-2124-M-001-005, NSTC 112-2124-M-001-014); the US National Science Foundation (NSF, grants AST-0096454, AST-0352953, AST-0521233, AST-0705062, AST-0905844, AST-0922984, AST-1126433, OIA-1126433, AST-1140030, DGE-1144085, AST-1207704, AST-1207730, AST-1207752, MRI-1228509, OPP-1248097, AST-1310896, AST-1440254, AST-1555365, AST-1614868, AST-1615796, AST-1715061, AST-1716327, AST-1726637, OISE-1743747, AST-1743747, AST-1816420, AST-1935980, AST-1952099, AST-2034306, AST-2205908, AST-2307887); NSF Astronomy and Astrophysics Postdoctoral Fellowship (AST-1903847); the Natural Science Foundation of China (grants 11650110427, 10625314, 11721303, 11725312, 11873028, 11933007, 11991052, 11991053, 12192220, 12192223, 12273022, 12325302); the Natural Sciences and Engineering Research Council of Canada (NSERC, including a Discovery Grant and the NSERC Alexander Graham Bell Canada Graduate Scholarships-Doctoral Program); the National Research Foundation of Korea (the Global PhD Fellowship Grant: grants NRF-2015H1A2A1033752, the Korea Research Fellowship Program: NRF-2015H1D3A1066561, Brain Pool Program: 2019H1D3A1A01102564, Basic Research Support Grant 2019R1F1A1059721, 2021R1A6A3A01086420, 2022R1C1C1005255, 2022R1F1A1075115); Netherlands Research School for Astronomy (NOVA) Virtual Institute of Accretion (VIA) postdoctoral fellowships; NOIRLab, which is managed by the Association of Universities for Research in Astronomy (AURA) under a cooperative agreement with the National Science Foundation; Onsala Space Observatory (OSO) national infrastructure, for the provisioning of its facilities/observational support (OSO receives funding through the Swedish Research Council under grant 2017-00648); the Perimeter Institute for Theoretical Physics (research at Perimeter Institute is supported by the Government of Canada through the Department of Innovation, Science and Economic Development and by the Province of Ontario through the Ministry of Research, Innovation and Science); the Princeton Gravity Initiative; the Spanish Ministerio de Ciencia e Innovación (grants PGC2018-098915-B-C21, AYA2016-80889-P, PID2019-108995GB-C21, PID2020-117404GB-C21); the University of Pretoria for financial aid in the provision of the new Cluster Server nodes and SuperMicro (USA) for a SEEDING GRANT approved toward these nodes in 2020; the Shanghai Municipality orientation program of basic research for international scientists (grant no. 22JC1410600); the Shanghai Pilot Program for Basic Research, Chinese Academy of Science, Shanghai Branch (JCYJ-SHFY-2021-013); the State Agency for Research of the Spanish MCIU through the "Center of Excellence Severo Ochoa" award for the Instituto de Astrofísica de Andalucía (SEV-2017- 0709); the Spanish Ministry for Science and Innovation grant CEX2021-001131-S funded by MCIN/AEI/10.13039/501100011033; the Spinoza Prize SPI 78-409; the South African Research Chairs Initiative, through the South African Radio Astronomy Observatory (SARAO, grant ID 77948), which is a facility of the National Research Foundation (NRF), an agency of the Department of Science and Innovation (DSI) of South Africa; the Swedish Research Council (VR); the Taplin Fellowship; the Toray Science Foundation; the UK Science and Technology Facilities Council (grant no. ST/X508329/1); the US Department of Energy (USDOE) through the Los Alamos National Laboratory (operated by Triad National Security, LLC, for the National Nuclear Security Administration of the USDOE, contract 89233218CNA000001); the YCAA Prize Postdoctoral Fellowship; and Conicyt through Fondecyt Postdoctorado (project 3220195).

We thank the staff at the participating observatories, correlation centres, and institutions for their enthusiastic support. This paper makes use of the following ALMA data: ADS/JAO.ALMA#2016.1.01154.V and ADS/JAO.ALMA#2017.1.00841.V. ALMA is a partnership of the European Southern Observatory (ESO; Europe, representing its member states), NSF, and National Institutes of Natural Sciences of Japan, together with National Research Council (Canada), Ministry of Science and Technology (MOST; Taiwan), Academia Sinica Institute of Astronomy and Astrophysics (ASIAA; Taiwan), and Korea Astronomy and Space Science Institute (KASI; Republic of Korea), in cooperation with the Republic of Chile. The Joint ALMA Observatory is operated by ESO, Associated Universities, Inc. (AUI)/NRAO, and the National Astronomical Observatory of Japan (NAOJ). The NRAO is a facility of the NSF operated under cooperative agreement by AUI. This research used resources of the Oak Ridge Leadership Computing Facility at the Oak Ridge National Laboratory, which is supported by the Office of Science of the U.S. Department of Energy under contract No. DE-AC05-00OR22725; the ASTROVIVES FEDER infrastructure, with project code IDIFEDER-2021-086; the computing cluster of Shanghai VLBI correlator supported by the Special Fund for Astronomy from the Ministry of Finance in China; We also thank the Center for Computational Astrophysics, National Astronomical Observatory of Japan. This work was supported by FAPESP (Fundacao de Amparo a Pesquisa do Estado de Sao Paulo) under grant 2021/01183-8.

APEX is a collaboration between the Max-Planck-Institut für Radioastronomie (Germany), ESO, and the Onsala Space Observatory (Sweden). The SMA is a joint project between the SAO and ASIAA and is funded by the Smithsonian Institution and the Academia Sinica. The JCMT is operated by the East Asian Observatory on behalf of the NAOJ, ASIAA, and KASI, as well as the Ministry of Finance of China, Chinese Academy of Sciences, and the National Key Research and Development Program (No. 2017YFA0402700) of China and Natural Science Foundation of China grant 11873028. Additional funding support for the JCMT is provided by the Science and Technologies Facility Council (UK) and participating universities in the UK and Canada. The LMT is a project operated by the Instituto Nacional de Astrófisica, Óptica, y Electrónica (Mexico) and the University of Massachusetts at Amherst (USA). The IRAM 30-m telescope on Pico Veleta, Spain is operated by IRAM and sup-ported by CNRS (Centre National de la Recherche Scientifique, France), MPG (Max-Planck-Gesellschaft, Germany), and IGN (Instituto Geográfico Nacional, Spain). The SMT is operated by the Arizona Radio Observatory, a part of the Steward Observatory of the University of Arizona, with financial support of operations from the State of Arizona and financial support for instrumentation development from the NSF. Support for SPT participation in the EHT is provided by the National Science Foundation through award OPP-1852617 to the University of Chicago. Partial support is also provided by the Kavli Institute of Cosmological Physics at the University of Chicago. The SPT hydrogen maser was provided on loan from the GLT, courtesy of ASIAA.

This work used the Extreme Science and Engineering Discovery Environment (XSEDE), supported by NSF grant ACI-1548562, and CyVerse, supported by NSF grants DBI-0735191, DBI-1265383, and DBI-1743442. XSEDE Stampede2 resource at TACC was allocated through TG-AST170024 and TG-AST080026N. XSEDE JetStream resource at PTI and TACC was allocated through AST170028. This research is part of the Frontera computing project at the Texas Advanced Computing Center through the Frontera Large-Scale Community Partnerships allocation AST20023. Frontera is made possible by National Science Foundation award OAC-1818253. This research was done using services provided by the OSG Consortium (Pordes et al. 2007; Sfiligoi et al. 2009), which is supported by the National Science Foundation award Nos. 2030508 and 1836650. Additional work used ABACUS2.0, which is part of the eScience center at Southern Denmark University, and the Kultrun Astronomy Hybrid Cluster (projects Conicyt Programa de Astronomia Fondo Quimal QUIMAL170001, Conicyt PIA ACT172033, Fondecyt Iniciacion 11170268, Quimal 220002). Simulations were also performed on the SuperMUC cluster at the LRZ in Garching, on the LOEWE cluster in CSC in Frankfurt, on the HazelHen cluster at the HLRS in Stuttgart, and on the Pi2.0 and Siyuan Mark-I at Shanghai Jiao Tong University. The computer resources of the Finnish IT Center for Science (CSC) and the Finnish Computing Competence Infrastructure (FCCI) project are acknowledged. This research was enabled in part by support provided by Compute Ontario (http://computeontario.ca), Calcul Quebec (http://www.calculquebec.ca), and Compute Canada (http://www.computecanada.ca).

The EHTC has received generous donations of FPGA chips from Xilinx Inc., under the Xilinx University Program. The EHTC has benefited from technology shared under open-source license by the Collaboration for Astronomy Signal Processing and Electronics Research (CASPER). The EHT project is grateful to T4Science and Microsemi for their assistance with hydrogen masers. This research has made use of NASA's Astrophysics Data System. We gratefully acknowledge the support provided by the extended staff of the ALMA, from the inception of the ALMA Phasing Project through the observational campaigns of 2017 and 2018. We would like to thank A. Deller and W. Brisken for EHT-specific support with the use of DiFX. We thank Martin Shepherd for the addition of extra features in the Difmap software that were used for the CLEAN imaging results presented in this paper. We acknowledge the significance that Maunakea, where the SMA and JCMT EHT stations are located, has for the indigenous Hawaiian people.

This research has made use of data obtained with 12 radio telescopes from the East Asian VLBI Network (EAVN): 2 stations from the Chinese VLBI Network (the Tianma radio telescope operated by Shanghai Astronomical Observatory of Chinese Academy of Sciences (CAS) and the Nanshan radio telescope operated by Xinjiang Astronomical Observatory of CAS), 2 stations from the Japanese VLBI Network (the Hitachi station operated by Ibaraki University and the Kashima station operated by the National Institute of Information and Communications Technology), the Korean VLBI Network operated by the Korea Astronomy and Space Science Institute (KASI), the VLBI Exploration of Radio Astrometry (VERA) operated by the National Astronomical Observatory of Japan (NAOJ), and the Nobeyama 45-meter radio telescope operated by NAOJ. We thank VERA staff members who helped the operation. We are grateful to KVN staff who helped operate the array and to correlate the data for quasi-simultaneous multi-wavelength observations of M87. The KVN is a facility operated by KASI. The KVN operations are supported by KREONET (Korea Research Environment Open NETwork), which is managed and operated by KISTI (Korea Institute of Science and Technology Information)

The VLBA is an instrument of the National Radio Astronomy Observatory. The National Radio Astronomy Observatory is a facility of the National Science Foundation operated by Associated Universities, Inc.

The Fermi LAT Collaboration acknowledges generous ongoing support from a number of agencies and institutes that have supported both the development and the operation of the LAT as well as scientific data analysis. These include the National Aeronautics and Space Administration and the Department of Energy in the United States, the Commissariat à l'Energie Atomique and the Centre National de la Recherche Scientifique / Institut National de Physique Nucléaire et de Physique des Particules in France, the Agenzia Spaziale Italiana and the Istituto Nazionale di Fisica Nucleare in Italy, the Ministry of Education, Culture, Sports, Science and Technology (MEXT), High Energy Accelerator Research Organization (KEK) and Japan Aerospace Exploration Agency (JAXA) in Japan, and the K. A. Wallenberg Foundation, the Swedish Research Council and the Swedish National Space Board in Sweden. Additional support for science analysis during the operations phase is gratefully acknowledged from the Istituto Nazionale di Astrofisica in Italy and the Centre National d'Etudes Spatiales in France. This work is performed in part under DOE Contract DE-AC02-76SF00515.

The support of the Namibian authorities and of the University of Namibia in facilitating the construction and operation of H.E.S.S. is gratefully acknowledged, as is the support by the German Ministry for Education and Research (BMBF), the Max Planck Society, the German Research Foundation (DFG), the Helmholtz Association, the Alexander von Humboldt Foundation, the French Ministry of Higher Education, Research and Innovation, the Centre National de la Recherche Scientifique (CNRS/IN2P3 and CNRS/INSU), the Commissariat à l'énergie atomique et aux énergies alternatives (CEA), the U.K. Science and Technology Facilities Council (STFC), the Knut and Alice Wallenberg Foundation, the National Science Centre, Poland grant no. 2016/22/M/ST9/00382, the South African Department of Science and Technology and National Research Foundation, the University of Namibia, the National Commission on Research, Science & Technology of Namibia (NCRST), the Austrian Federal Ministry of Education, Science and Research and the Austrian Science Fund (FWF), the Australian Research Council (ARC), the Japan Society for the Promotion of Science and by the University of Amsterdam. We appreciate the excellent work of the technical support staff in Berlin, Zeuthen, Heidelberg, Palaiseau, Paris, Saclay, Tübingen and in Namibia in the construction and operation of the equipment. This work benefited from services provided by the H.E.S.S. Virtual Organisa-tion, supported by the national resource providers of the EGI Federation.

We would like to thank the Instituto de Astrofísica de Canarias for the excellent working conditions at the Observatorio del Roque de los Muchachos in La Palma. The financial support of the German BMBF, MPG and HGF; the Italian INFN and INAF; the Swiss National Fund SNF; the grants PID2019-104114RB-C31, PID2019-104114RB-C32, PID2019-104114RB-C33, PID2019-105510GB-C31, PID2019-107847RB-C41, PID2019-107847RB-C42, PID2019-107847RB-C44, PID2019-107988GB-C22, PID2022-136828NB-C41, PID2022-137810NB-C22, PID2022-138172NB-C41, PID2022-138172NB-C42, PID2022-138172NB-C43, PID2022-139117NB-C41, PID2022-139117NB-C42, PID2022-139117NB-C43, PID2022-139117NB-C44 funded by the Spanish MCIN/AEI/ 10.13039/501100011033 and "ERDF A way of making Europe"; the Indian Department of Atomic Energy; the Japanese ICRR, the University of Tokyo, JSPS, and MEXT; the Bulgarian Ministry of Education and Science, National RI Roadmap Project DO1-400/18.12.2020 and the Academy of Finland grant nr. 320045 is gratefully acknowledged. This work was also been supported by Centros de Excelencia "Severo Ochoa" y Unidades "María de Maeztu" program of the Spanish MCIN/AEI/ 10.13039/501100011033 (CEX2019-000920-S, CEX2019-000918-M, CEX2021-001131-S) and by the CERCA institution and grants 2021SGR00426 and 2021SGR00773 of the Generalitat de Catalunya; by the Croatian Science Foundation (HrZZ) Project IP-2022-10-4595 and the University of Rijeka Project uniri-prirod-18-48; by the Deutsche Forschungsgemeinschaft (SFB1491) and by the Lamarr-Institute for Machine Learning and Artificial Intelligence; by the Polish Ministry Of Education and Science grant No. 2021/WK/08; and by the Brazilian MCTIC, CNPq and FAPERJ.

The Medicina and Noto radio telescopes are funded by the Ministry of University and Research (MUR) and are operated as National Facility by the National Institute for Astrophysics (INAF).

VERITAS is supported by grants from the U.S. Department of Energy Office of Science, the U.S. National Science Foundation and the Smithsonian Institution, by NSERC in Canada, and by the Helmholtz Association in Germany. This research used resources provided by the Open Science Grid, which is supported by the National Science Foundation and the U.S. Department of Energy's Office of Science, and resources of the National Energy Research Scientific Computing Center (NERSC), a U.S. Department of Energy Office of Science User Facility operated under Contract No. DE-AC02-05CH11231. We acknowledge the excellent work of the technical support staff at the Fred Lawrence Whipple Observatory and at the collaborating institutions in the construction and operation of the instrument.

S.M. is thankful for support from an NWO (Netherlands Organisation for Scientific Research) VICI award, grant Nr. 639.043.513.

D.H. acknowledges support from the Natural Sciences and Engineering Research Council of Canada (NSERC) Discovery Grant and the Canada Research Chairs program.

G.P. acknowledges support by ICSC – Centro Nazionale di Ricerca in High Performance Computing, Big Data and Quantum Computing, funded by European Union – NextGenerationEU.

J.-C.A. acknowledges support from the Malaysian Fundamental Research Grant Scheme (FRGS) FRGS/1/2019/STG02/UM/02/6.

M.B. acknowledges support from the YCAA Prize Postdoctoral Fellowship and the Black Hole Initiative at Harvard University, which is funded in part by the Gordon and Betty Moore Foundation (grant GBMF8273) and in part by the John Templeton Foundation.

K.H. acknowledges support from JSPS KAKENHI grant Nos. JJP18H03721, JP19H01943, and JP18KK0090.

T.K. acknowledges support from JSPS KAKENHI grant Nos. JP18K13594, JP19H01908, JP19H01906, JP23K03448, JP23H00117, MEXT as "Program for Promoting Researches on the Supercomputer Fugaku" (Structure and Evolution of680 the Universe Unraveled by Fusion of Simulation and AI;681 Grant Number JPMXP1020230406; Project ID hp230204 , hp240219), by the HPCI System Research Project (Project ID: hp230116, hp240054), and JICFuS. A part of the calculations were carried out on the XC50 at the Center for Computational Astrophysics, National Astronomical Observatory of Japan.

R.-S.L. is supported by the National Science Fund for Distinguished Young Scholars of China (Grant No. 12325302), the Key Program of the National Natural Science Foundation of China (NSFC, grant No. 11933007), the Key Research Program of Frontier Sciences, CAS (grant No. ZDBS-LY-SLH011), the Shanghai Pilot Program for Basic Research, CAS, Shanghai Branch (JCYJ-SHFY-2021-013) and the Max Planck Partner Group of the MPG and the CAS.

J.N. acknowledges support from SAO award DD7-18089X and NASA award 80NSSC20K0645.

J.P. acknowledges financial support from the Korean National Research Foundation (NRF) via Global PhD Fellowship grant 2014H1A2A1018695 and support through the EACOA Fellowship awarded by the East Asia Core Observatories Association, which consists of the Academia Sinica Institute of Astronomy and Astrophysics, the National Astronomical Observatory of Japan, Center for Astronomical Mega-Science, Chinese Academy of Sciences, and the Korea Astronomy and Space Science Institute.

Y.-Z.C. acknowledges support from the Natural Science Foundation of China (grant 12303021) and the China Postdoctoral Science Foundation (No. 2024T170845).

J.Y.K. is supported for this research by the National Research Foundation of Korea (NRF) grant funded by the Korean government (Ministry of Science and ICT; grant no. 2022R1C1C1005255).

J.M. is supported by an NSF Astronomy and Astrophysics Postdoctoral Fellowship under award AST-2401752.

## Appendix B: Detailed analysis descriptions

### *Appendix B.1: Radio observations*

#### Appendix B.1.1: VERA 22 GHz

All the data were analysed in with standard VERA data reduction procedures (see Nagai et al. 2013; Hada et al. 2014, for more details). We note that VERA recovers only part of the extended jet emission due to the lack of short baselines, so the total VERA fluxes listed in Table C.1 significantly underestimate the actual total jet fluxes.

#### Appendix B.1.2: EAVN/KaVA 22 and 43 GHz

Especially between March and May of 2017 and 2018 when EHT M87 observations were performed, the observing cadence of EAVN monitoring was intensified. The default array configurations were KVN+VERA+Tianma+Nanshan+Ibaraki at 22 GHz and KVN+VERA+Tianma at 43 GHz respectively, while occasionally a few more stations in East Asia (Sejong, Kashima, Nobeyama) and Italy (Medicina, Sardinia) joined if available. During the period outside March–May of each year, the observations were mostly conducted with the KVN and VERA Array (KaVA; Niinuma et al. 2014; Hada et al. 2017; Park et al. 2019), a core array of EAVN.

Each of the KaVA/EAVN sessions was made in a 5–7-hour continuous run at a data recording rate of 1 Gbps (a total bandwidth of 256 MHz in a single polarisation mode). All the data were correlated at the Daejeon hardware correlator installed at KASI. All the EAVN data were calibrated in the standard manner of VLBI data reduction procedures. We used the `AIPS` (Greisen 2003) software package for the initial calibration of visibility amplitude, bandpass and phase calibration. Imaging with the CLEAN algorithm (Högbom 1974) and self-calibration were performed with the `Difmap` (Shepherd 1997) software.

### Appendix B.1.3: VLBA 24 and 43 GHz

The total on-source time during the full-track segments is about 1.7 hours at 24 GHz and 6 hours at 43 GHz, while for the short segments it amounts to 30 minutes at 43 GHz. The sources OJ 287 and 3C 279 were observed to use as fringe finders and bandpass calibrators. In each band eight 32 MHz-wide frequency channels were recorded in both right and left circular polarisations at a rate of 2048 Mbps, and correlated with the VLBA software correlator in Socorro (Deller et al. 2011). The initial data reduction was conducted using `AIPS`, following the standard calibration procedures for VLBI data (Crossley et al. 2012; Walker 2014). Deconvolution and self-calibration algorithms, implemented in `Difmap`, were used for phase and amplitude calibration and for constructing the final images. Amplitude calibration accuracy of 10% is adopted for both frequencies. Images and polarimetric results from part of these VLBA observations are published in Kravchenko et al. (2020) and Park et al. (2021).

### Appendix B.1.4: GMVA+ALMA 86 GHz

The participating GMVA stations include the eight VLBA antennas equipped with 3.5 mm receivers, the Effelsberg 100 m telescope, the 100 m Green Bank Telescope, the Metsähovi 14 m telescope, the Onsala 20 m telescope, the IRAM 30 m telescope, and the Yebes 40m telescope (YS). We note that the participation of ALMA and the GLT has improved the north-south resolution by a factor of ∼4 for the observations of M87.

The observations (GMVA project code ML005) were performed for a total of ∼13 h (∼6 h with five telescopes in Europe (Effelsberg, Metstähovi, Onsala, IRAM30m and Yebes), ∼7 h with ALMA, VLBA and GBT, and ∼12 h with the GLT). Two bright calibration sources (3C273 and 3C279) were observed every 20-40 minutes in interleaved scans. The data were correlated with the VLBI correlator at the Max Planck Institute for Radio Astronomy using DiFX (Deller et al. 2011).

During these observations, most stations recorded the signals on a circular polarisation basis. However, the phased ALMA recorded the signals on a linear polarisation basis. Consequently, the visibilities between ALMA and the other stations were correlated on a mixed polarisation basis. After correlation, these visibilities were converted into a pure circular basis using the internal calibration of the ALMA interferometric data with the `PolConvert` algorithm (Martí-Vidal et al. 2016). In addition, GLT participated in these observations on a best-effort basis during its commissioning phase. The waveguide phase-shifter of the GLT that was used to convert the on-sky circular-polarisation signals into linear polarisation for detection at the backend was erroneously configured to apply a rotation of 45° (instead of 90°) between the polarisation channels. The instrumental polarisation effects caused by this rotation were removed with a dedicated algorithm after correlation. After this instrumental polarisation calibration, the data were then further calibrated following typical VLBI data reduction procedures in `AIPS`. Hybrid imaging was performed in the `Difmap` software with CLEAN and self-calibration. We refer the reader to Lu et al. (2023) for further details of data calibration. Based on the calibration and imaging of the calibrator 3C273, we consider an error budget of 25 % for the flux density estimate of M87. While a dedicated study based on these data is presented in Lu et al. (2023), here we provide a peak flux density of the core and VLBI-scale total flux density (Tables C.1 and C.2).

### Appendix B.1.5: KVN 22, 43, 86, and 129 GHz

The data were calibrated using a modified version of the KVN pipeline (Hodgson et al. 2016) to include ionospheric delay corrections with total electron content (TEC) maps, parallactic angle corrections, and Earth orientation parameter corrections. Prior to global fringe fitting, the 43, 86, and 129 GHz data were calibrated with the 22 GHz data by means of frequency phase transfer. After the pipeline calibration with `AIPS`, the inner 75% of each subband (due to flux loss at the bandpass edges) was used for further analysis. Following Kim et al. (2018), the data were averaged by 30 seconds for the 22 GHz and 43 GHz data, and by 10 seconds for the 86 GHz and 129 GHz data. As in M87 MWL2017, we assume a flux density measurement uncertainty of 10% at 43 GHz and 86 GHz, and 30% at 129 GHz. During the observations, a phase instability issue occurred with the 22 GHz receiver at the KVN Yonsei site (KYS) between January and March 2018, and between November 2018 and February 2019. This resulted in a ∼ 50% flux loss for all KYS baselines. From a sample of compact sources within the iMOGABA program, we derived and applied empirical correction factors to the impacted data. An uncertainty of 15% is assumed for the 22 GHz flux density measurements made during observations where the empirical corrections were applied. For observations that were not impacted by the flux loss, an uncertainty of 10% is adopted. A common 1 mas FWHM circular Gaussian beam is used to convolve the 22, 43, 86, 129 GHz CLEAN maps when determining the peak flux density of the core. This is closest to the natural beams of the KVN at 86 and 129 GHz. In Table C.1, we summarise the data obtained during 2018. In Fig.14, we also add the earlier (2012–2016) data that were originally published in the literature (Kim et al. 2018) but the peak flux densities are recalculated using the procedures applied for 2017–2018 data.

### Appendix B.1.6: Global 43 GHz VLBI

The participating stations include Effelsberg, Yebes 40 m, Metsahovi, KVN (three stations), 100-m Green Bank Telescope, VLBA (10 stations), and the phased Jansky Very Large Array (JVLA). The data were recorded with a total recording rate of 2 Gbps, corresponding to 256 MHz bandwidth per polarisation. The scheduling, observation, and data reduction followed the standard procedures using `AIPS`. Similar to other VLBI data reduction and imaging, the CLEAN imaging and self-calibration were made using procedures implemented in `Difmap`. The resulting FWHM angular resolutions are $0.2 \times 0.7$ ($0.12 \times 0.47$) $mas^2$ with the beam position angle $\sim$ 0 deg for the natural (uniform) weighting.

#### Appendix B.1.7: ALMA 93 GHz and 221 GHz

The VLBI observations were carried out while the array was in its most compact configuration and only antennas within a radius of 180-m (from the array centre) were used for phasing. The observations were performed in full-polarisation mode to supply the inputs to the polarisation conversion process at the VLBI correlators (Martí-Vidal et al. 2016). The setup included four spectral windows (SPWs) of 1875 MHz in each band, two in the lower and two in the upper sideband, centred at 86, 88, 98, and 100 GHz in Band 3, and 213, 215, 227, and 229 GHz in Band 6, respectively. The SPWs were correlated with 240 channels per SPW (corresponding to a spectral resolution of 7.8125 MHz). Details about ALMA operations in VLBI mode and a full description of the data processing and calibration can be found in Goddi et al. (2019).

Imaging was performed with the Common Astronomy Software Applications package (`CASA`, version 6.4.1 used; McMullin et al. 2007; CASA Team et al. 2022) task `tclean`. Only phased antennas were used to produce the final images (with baselines <360 m), yielding synthesised beam sizes ~2.5″ in Band 3 and ~ 1″ in Band 6, respectively. To isolate the core emission from the jet, we compute the sum of the central nine pixels of the model (cleaning component) map, an area of $3 \times 3$ pixels; the contribution from the jet is accounted for by also summing the clean components along the jet. While the extended emission accounts for less than 20% of the total emission at 1.3 mm, at 3mm it also includes emission from the radio lobes, accounting for more than 60% of the total emission. The absolute amplitude scale calibration has a 5%/10% uncertainty in Band 3/6, respectively. Details about the imaging and flux extraction methods can be found in Goddi et al. (2021).

#### Appendix B.1.8: SMA 200 & 345 GHz

Data from this program are updated regularly and are available at the SMA Observer Center website (SMAOC[7]). Data were primarily obtained in a compact configuration (baselines from 10 to 75 m) though some were obtained at longer baselines up to 210 m. The effective spatial resolution was therefore generally around 3″. We have limited analysis to projected baselines longer than 30 k$\lambda$ (effective spatial scales smaller than 6.8″). The flux density was obtained by vector averaging of the calibrated visibility data from each observation.

Interferometric observations of M87 were additionally conducted as part of the 2017 and 2018 EHT campaigns, concurrent with the SMA running in phased-array mode operating at 230 GHz. In 2017 the SMA was in compact configuration, while in 2018 the SMA was in extended configuration (baselines 25 m to 210 m). In both years, the interferometer operated in dual-polarisation mode. The SMA correlator produces four separate but contiguous 2 GHz spectral windows per sideband, resulting in frequency coverage of 208 – 216 and 224 – 232 GHz. Data were passband and amplitude calibrated using 3C 279, with flux calibration performed using either Callisto, Ganymede, or Titan. Phase calibration was done through self-calibration of the M87 data itself, assuming a point-source model.

[7] http://sma1.sma.hawaii.edu/callist/callist.html

### Appendix B.2: Optical and UV observations

#### Appendix B.2.1: Optical observations by the Kanata telescope

We adopted standard reduction procedures for the images: bias- or dark-frame subtraction, correction of bad pixels, flattening correction, removal of contamination of CR signals, sky-background subtraction, solving the corresponding coordinates in the World Coordinate System (WCS) using field stars, and stacking of the images aligned with the coordinates of WCS.

We performed an aperture photometry of field stars in the stacked images using `Source Extractor` software (Bertin & Arnouts 1996). The value of `FLUX_AUTO` was adopted as the photometric measurements. We estimated zero points of the images in the AB magnitude system using nearby stars whose actual magnitudes have been provided in the references: the Pan-STARRS1 catalogue (Flewelling et al. 2020) of the optical band after conversion of the filter system (Tonry et al. 2012) and the 2MASS catalogue (Skrutskie et al. 2006) of the near-infrared band.

#### Appendix B.2.2: Optical and UV observations from *Swift*-UVOT

UVOT data in all filters were analysed with the `uvotimsum` and `uvotsource` tasks included in the HEASoft package (v6.32) and the 20231208 CALDB-UVOTA release. Source counts were extracted from a circular region of 5 ″radius centred on the source, while background counts were derived from a circular region with 30 ″radius in a source-free region located outside the host galaxy radius (~ 120 ″). The UVOT magnitudes are corrected for the host galaxy contamination (see Table C.6 ) and the Galactic extinction using an E(B–V) value of 0.020 from Schlafly & Finkbeiner (2011) and the extinction laws from Cardelli et al. (1989) and converted to flux densities using the conversion factors from Breeveld et al. (2010). We have used the same host galaxy model as presented in Table A2 of EHT MWL Science Working Group et al. (2021). However, we have to note that flux densities of the host galaxy listed in Table A2 of EHT MWL Science Working Group et al. (2021) did not have aspect corrections although the correct values of the host galaxy as given in Table C.6 were used in the data reduction of *Swift*-UVOT data in EHT MWL Science Working Group et al. (2021).

#### Appendix B.2.3: Far-UV observations from AstroSat-UVIT

For the AstroSat-UVIT analysis, the VIS data are used for drift correction of the images in NUV and FUV channels. The FUV and NUV detectors work in photon-counting mode and the VIS detector works in integration mode. Each of these channels is equipped with a number of narrow and wide band filters (for details refer to Tandon et al. 2017; Tandon et al. 2020). Out of the two AstroSat pointing observations (i.e. A04_115T02_9000001900 and A04_115T02_9000002042), only the UVIT data of the second one are available. These observations are made in the BaF2 band (F154W: $\lambda_{eff}$ ~1541 Å, with $\Delta\lambda_{eff}$ ~380 Å). The level-1 data, downloaded from the AstroSat data archive 'astrobrowse', is used to create the image using the `CCDLAB` software package (Postma & Leahy 2017) incorporating the corrections for the spacecraft drift, geometrical distortions and flat-fielding.

A two-dimensional (2D) image decomposition method was used with IMFIT[8] (Erwin 2015) tool. The following combination of 2D models is used to create the model image of M87 using UVIT: PointSource + PointSource + Core-Sersic + SersicJet + Flat-Sky. The point sources used here are convolved with the PSF of the UVIT instrument.

The UVIT-BaF2 image, centred at M87 and zooming into a small region around it is shown in Fig. B.1. The two point source models are added to account for the contributions from M87*'s core and the HST-1 knot. The Core-Sersic model (Graham et al. 2004) is used to represent the emission from the host galaxy. In order to take the asymmetric jet into account we created a modified version of the Sersic model by adding an asymmetry term to the originally symmetric Sersic function. But even with the added asymmetry, the shape of the bright spots towards the end of the jet is too complex to develop a good analytical model to describe it. Motivated by the above we masked this irregular complex part of the jet with a rectangular mask. The Flat-Sky model (Mondal et al. 2018) was used to include a component for the sky background. The best fit composite model corresponds to a reduced $\chi^2$ value of 1.

This model is then used to represent the composite radial profile for the host and jet. We then performed aperture photometry of the image using the Python package photutils[9] with a source region of 5″ centred at the position of M87*. The contributions of other prominent components to this region were estimated using the modelled radial profile for composite emission from the jet, the host galaxy and the sky. The corrected magnitude is then converted to flux using an appropriate flux conversion factor as provided by the instrument team. The corresponding numbers are shown in the Table C.8.

### *Appendix B.3: X-ray observations*

#### Appendix B.3.1: *Chandra* and *NuSTAR* observations and joint spectral analysis

We keep the data processing and analysis as consistent as possible with our previous work on the *Chandra* and *NuSTAR* data taken in parallel with the 2017 EHT campaign described in M87 MWL2017. We only briefly summarise them here. For the *Chandra* data, we used standard data reduction procedures in CIAO (Fruscione et al. 2006) version 4.9 to extract 0.4–8 keV spectra from the spatially resolved components, the core, the HST-1 knot, and the outer jet, along with the corresponding instrumental response files. The *NuSTAR* observation was processed following standard procedures outlined in Perri et al. (2017), using NuSTARDAS version 1.8.0 to produce 3–79 keV spectra and instrumental responses with CALDB version 20180312. Source and background regions for Focal Plane Module A and B (FPMA, FPMB) were the same as in M87 MWL2017: 45″ circles that are large enough to include all of the *Chandra* spatial components mentioned above as well as much of the surrounding diffuse emission from the intracluster medium (ICM).

We proceed to joint spectral modelling of the *Chandra* and *NuSTAR* data in order to take advantage of the high spatial resolution of *Chandra* and the high-energy sensitivity of *NuSTAR*. Our modelling strategy, described in detail in M87 MWL2017, properly accounts for the effects of pileup in *Chandra* data and includes consideration of the cross-normalisation between the instruments with respect to the modelled diffuse ICM emission that is differently sampled by *Chandra* and *NuSTAR*. For the diffuse component, we adopt a spectral model with two APEC components (Smith et al. 2001) with different temperatures and variable elemental abundances, three additional emission lines, and a cutoff component for the low-mass X-ray binary contribution from within the host galaxy. For the core, the HST-1 knot, and the remainder of the jet, we modelled their continuum emission using PLs. All spectral models include interstellar absorption (Wilms et al. 2000). The absorption column towards the core was previously found to be higher (e.g. Perlman & Wilson 2005), that we allow for by including an extra layer of absorption in the model for this component.

We performed spectral fitting on spectra binned using a custom algorithm that ensures a minimum signal-to-noise ratio within each energy bin (as described in M87 MWL2017). To find the best-fit model parameters we minimised the Cash statistic (Cash 1979) as appropriate for the low-count regime and our binning strategy. Our best fit has a Cash statistic of 1219.023 for 1034 data bins and 37 free parameters. Our multi-component best-fit model is shown along with the data in Fig. 6; red lines represent the model for each dataset, and the *NuSTAR* model is the sum of all the other spatial components. We derive uncertainties using a Markov Chain Monte Carlo (MCMC) sampler emcee (Foreman-Mackey et al. 2013) set to run for 60,000 steps with 10 walkers for each of 37 parameters (a total of 22.2 million model evaluations). We discarded the first 185,000 evaluations for calculating chain statistics. Unless otherwise noted, quoted uncertainties represent minimum width intervals containing 90 % of the probability function for each spectral parameter.

#### Appendix B.3.2: *Swift*-XRT observations and analysis

The standard pipeline xrtpipeline version 0.13.5 was used to create level 3 cleaned event files for each *Swift*-XRT observation. We use a circular source region of radius 35″ for spectral extraction, encompassing all three of the X-ray sources examined individually in the *Chandra* analysis. For background extraction, we create an annular background region with inner and outer radii of 130″ and 150″, respectively. Using xselect (version 2.4k), we extract spectra for the source and background regions after correcting for pile-up in the source region. If the source region count rate exceeded 0.5 ct $s^{-1}$ for any observations, we applied a pile-up correction by using an annular region to exclude central pixels affected by pile-up.

Creating ancillary response files for each spectrum with the xrtmkarf task, we grouped the spectral files for each epoch to a minimum of 20 counts per bin using the grppha command to facilitate $\chi^2$ fitting using xspec. Since the spectral resolution of *Chandra* is better than *Swift*-XRT, we adopted the source and background model fitted in the *Chandra* analysis. All spectral parameters of that model were frozen at the *Chandra* best-fit values (see Sect. 2.3.1) except the normalisation of the APEC and line emission backgrounds and the joint normalisation of the three power law components for the core, the HST-1 knot, and jet emission. The relative normalisation between the two APEC background components and between the three PL source components were kept frozen at the *Chandra* values. We use $\chi^2$ fitting via xspec to conduct this X-ray analysis. In this way, we evaluate the changing total and small-scale components (core, the HST-1 knot and inner jet) fluxes from the source.

8 https://www.mpe.mpg.de/ erwin/code/imfit
9 https://photutils.readthedocs.io/en/stable/aperture.html

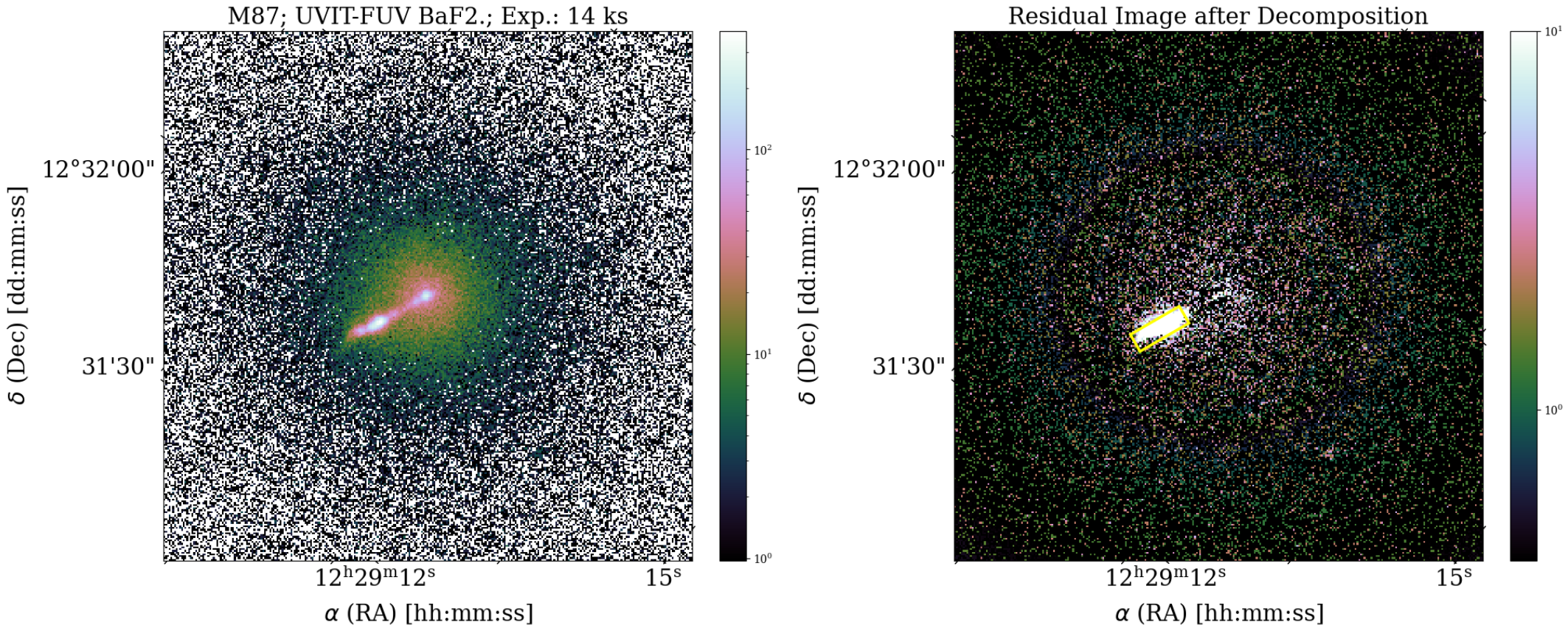

**Fig. B.1.** AstroSat UVIT observations of M87. *Left:* M87 in BaF2 band. *Right:* Difference image of M87. The patched area is basically the masked region of the jet.

#### Appendix B.3.3: AstroSat-SXT observations and analysis

We used `sxtpipeline` (version 1.4b) along with the standard filtering criteria to generate the cleaned events file. The standard procedure includes time-tagging of events, removing hot pixels, bias corrections, coordinate transformation, filtering (South Atlantic Anomaly passage, Sun-angle, Moon-angle), charge transfer inefficiency corrections, and generation of full-frame products. The processed event lists obtained in this way were then input to `xselect` to extract spectra, light curves, and images.

We used a circular source region of radius of 15′ centred on the brightest pixel in the SXT image ($\alpha$=12:30:50.41, $\delta$=+12:22:32.4), that is consistent with the peak of the flux distribution in *Chandra* and *NuSTAR* images. From the PSF size of the SXT we estimate that this region includes over 97 % of the total number of detected photons. The background spectrum extracted using deep blank sky observations and distributed by the instrument team[10] is used to model the spectrum. We used the tool `sxtARFModule` (version 0.3) along with the full-frame ARF (`sxt_pc_excl00_v04.arf`) distributed by the SXT Payload Operation Center[11] to create ARF for the specific source region selection. We utilised the RMF file `sxt_pc_mat_g0to12.rmf` corresponding to the default 0–12 grade selection.

For spectral fitting we used a model defined with the following expression in `xspec`: `TBabs` × `(vvapec + vvapec + cutoffpl + TBabs` × `powerlaw + powerlaw + powerlaw + Gaussian + vvapec)`. This is similar to the model adopted for *Chandra* and *NuSTAR* spectral analysis, except one Gaussian component, that is added in order to fit the broad hump-like residuals around 1.4 keV. An additional `vvapec` component is included to account for the diffuse emission from cluster gas in the outer regions because the SXT samples a larger portion of it (radius ≃15′) than other focusing X-ray instruments considered in this work. The parameters of the three `vvapec` components, except the temperature and normalisation, are fixed to the values obtained from joint *Chandra* and *NuSTAR* spectral modelling (Sect. 2.3.1). To reduce the number of free parameters, we tied together the temperature and normalisation parameters of the `vvapec` components with a multiplicative factor. We adopt the sum of these three components with different temperatures to represent the large-scale emission from hot gas in the galaxy and the surrounding cluster. Three PL components are representative of the core, the HST-1 knot, and jet contributions; their PL photon indices are kept fixed based on values determined from the analysis of *Chandra* and *NuSTAR* data. We keep the normalisation of the core contribution free in the fitting, while the normalisation of the HST-1 knot and jet components is tied with the core normalisation using the multiplicative factors also employed in *Swift*-XRT spectral modelling in Sect. 2.3.2, as well as in M87 MWL2017.

As in Sect. 2.3.1, we used the Cash statistic to find the best-fitting composite model. To account for the systematics introduced by the procedures related to ARF generation (such as vignetting model and PSF-based scaling of the effective area, etc.), we included an additional 1% uncertainty to the observed spectra. Constraints on the free parameters of the model are given in Table C.9. Figure B.2 shows the *AstroSat*-SXT individual PL components to be interpreted as upper limits due to the angular resolution of the instrument.

### *Appendix B.4: $\gamma$-ray observations*

#### Appendix B.4.1: *Fermi*-LAT observations

*Fermi*-LAT (Atwood et al. 2009) is a $\gamma$-ray telescope that detects photons by conversion into electron-positron pairs, and it has an operational energy range from 20 MeV to more than 300 GeV. Since its launch in June 2008, it mostly works in survey mode covering the whole sky in about three hours and with a remarkable overall uptime of 99.8% (Ajello et al. 2021).

Regarding the data selection, we used P8R3 (v3) `SOURCE` class events (Bruel et al. 2018), in the energy range between 100 MeV and 1 TeV, in a region of interest (ROI) of 15° × 15° centred on the position of M87. The low-energy threshold is motivated by the large uncertainties in the arrival directions of the photons below 100 MeV, leading to possible confusion between point-like sources and the Galactic diffuse component (see Principe et al. (2018) for a different analysis implementation to solve this and other issues at low energies with *Fermi*-LAT).

The analysis (that consists of the following steps: model optimisation, source localisation, spectrum and emission variability studies) was performed with `Fermipy`[12] (v1.0.1; Wood et al. 2017), that uses the Fermi Tools, version 2-0-18. We created counts maps using a pixel size of 0.1°. All $\gamma$-rays with zenith angle larger than 95° were excluded in order to limit the contamination from secondary $\gamma$-rays from the Earth's limb (Abdo et al. 2009). In order to optimise the analysis, we applied different cuts for the data selections at low energies and selected event types with the best PSF[13]. In particular, for energies below 300 MeV, we excluded events with zenith angle larger than 85°, as well as photons from PSF0 and PSF1 event types. Between 300 MeV and 1 GeV we excluded events with zenith angle larger than 95°, as well as photons from the PSF0 event type. Above 1 GeV we used all events with zenith angles less than 105°. The P8R3_Source_V3 instrument response functions (IRFs) were used. The model used to describe the sky includes all point-like and extended sources located at a distance < 15° from the source position and listed in the 4FGL-DR2 (Abdollahi et al. 2020) as well as the Galactic diffuse and isotropic emission. For the latter contributions, we used the same templates,[14] adopted to derive the 4FGL catalogue.

[10] AstroSat Science Support Cell: http://astrosat-ssc.iucaa.in
[11] TIFR, Mumbai: www.tifr.res.in/~astrosat_sxt
[12] http://fermipy.readthedocs.io/en/latest/

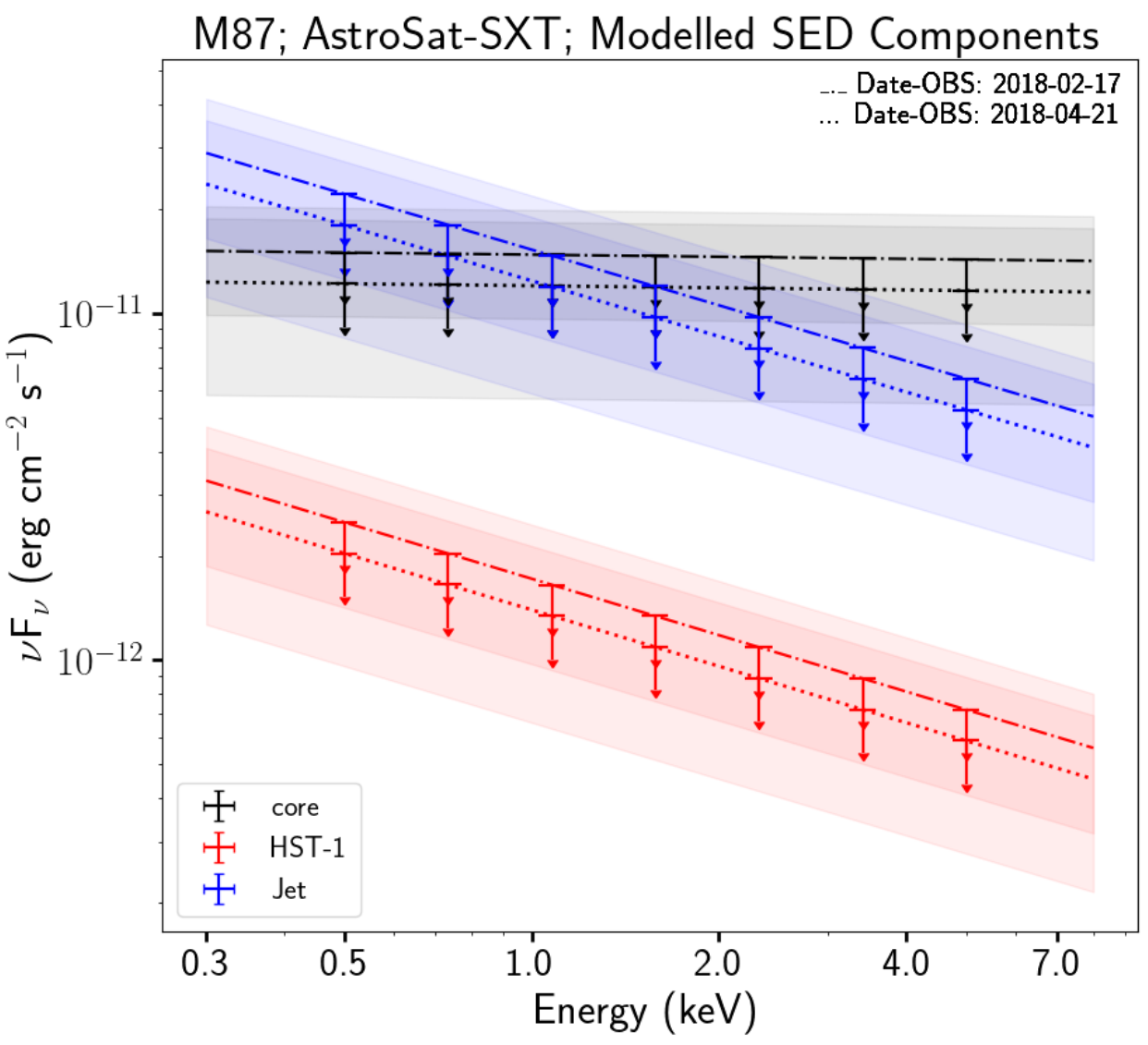

**Fig. B.2.** *AstroSat*-SXT observations of M87 with SED estimates from spectral modelling. Dashed-dotted and dotted lines are representative of observations made in February and March 2018, respectively. The grey, light-blue and light-red shaded regions around the lines are $1\sigma$ uncertainty regime derived using parameter estimations, each representing core, jet and the HST-1 knot sub-components, respectively. The downward arrows are indicative that given the limitations of the instrumental capabilities, these estimations should be treated as upper limits for further investigations.

We performed the spectral analysis, allowing the diffuse background template normalisation and the spectral parameters of the sources to vary within 3° from our targets. For the sources in a radius between 3°–5° and all variable sources only the normalisation was fit, while we fixed to their 4FGL values the parameters of all the remaining sources within the ROI at larger angular distances from our target. We modelled the spectrum of M87 with a PL function.

Finally, for the analysis on the 12 years of LAT data we extracted a light curve using time bins of one month. Similarly, for analysis on a one-month scale centred on EHT observations, the light curve analysis was repeated using time intervals of two days in order to investigate the emission variability on short time scales. We tried also different time intervals with both fixed (1 day and 1 week) or customised duration (discriminating between before, during and after the VHE $\gamma$-ray flare). The fluxes in each interval were obtained by leaving only the normalisation free to vary and freezing the other spectral parameters to the best-fit values obtained from the full range analysis.

[13] A measure of the quality of the direction reconstruction is used to assign events to four quartiles. $\gamma$-rays in Pass 8 data can be separated into 4 PSF event types: 0, 1, 2, 3, where PSF0 has the worst resolution and PSF3 has the best one.

[14] https://fermi.gsfc.nasa.gov/ssc/data/access/lat/BackgroundModels.html

### Appendix B.4.2: Very high energy observations: H.E.S.S., MAGIC, VERITAS

H.E.S.S. observations were performed with a wobble mode with an offset of 0.7° from the position of M87, that allows for a simultaneous background estimation. The analysis and reconstruction of the Cherenkov shower images were done with the ImPACT maximum likelihood-based technique (Parsons & Hinton 2014). The separation between $\gamma$-rays and hadrons was performed with a boosted decision tree classification method (Ohm et al. 2009). For the signal estimation, the background was estimated using the "ring-background model" whereas the calculation of the light curve and the spectrum was done using the "reflected-background model" (Berge et al. 2007). A source with 1% of the Crab Nebula's flux can be detected above 100 GeV in ~ 10 hours.

The stereoscopic MAGIC observations were conducted in wobble mode with an offset of 0.4°. Image reconstruction and data analysis were performed with the standard MAGIC analysis framework MARS (MAGIC Analysis and Reconstruction Software; Zanin 2013; Aleksić et al. 2016). To estimate the instrument's PSF in true energy we used the spatial likelihood analysis package `SkyPrism` from Vovk et al. (2018). The observations on MJD 58232 were taken with a strong moonlight background present and therefore analysed as described in Ahnen et al. (2017). MAGIC's integral sensitivity is 1% of the Crab Nebula's flux in ~26 hours above an analysis energy threshold of 290 GeV.

Data analysis of the observations in 2019 was performed identically to the aforementioned 2018 observations. The reconstructed flux levels in 2019 are compatible with the previously observed quiescent flux (see, e.g. MAGIC Collaboration et al. 2020; EHT MWL Science Working Group et al. 2021). The MAGIC fluxes published in MAGIC Collaboration et al. (2020) use an analysis threshold of 300 GeV for the light curve. Instead of scaling the VHE fluxes between 2012 and 2015 according to the PL spectrum, we repeated the flux calculations with a threshold of 350 GeV, using the original high-level analysis files of the authors of MAGIC Collaboration et al. (2020).

The stereoscopic VERITAS observations were conducted in wobble mode with an offset of 0.5°. The data were analysed using the VERITAS standard analysis procedure (Daniel 2008). The main software package used in this work is called *Eventdisplay*. An overview of this analysis package can be found in Maier & Holder (2018). In addition, a cross-check analysis was performed using an independent software package called *VEGAS* (Cogan 2007). VERITAS is designed to measure $\gamma$ rays with energies from ~85 GeV and it is able to detect a source with about 1% of the Crab Nebula within 25 hours (Park 2016).

To explore possible systematic flux scale offsets between VERITAS, MAGIC, and H.E.S.S. and investigate how the uncertain flux scales might affect the results in sec. 2.4.2, we applied two cross-normalisation factors ($f_1$ and $f_2$) to the VERITAS and MAGIC fluxes as extra fitting parameters. We note that statistical errors on individual flux points were not altered.

**Table B.1.** Best fit results for a constant and a Gaussian model (Eq. 2.4.2) describing the temporal progression of the flare as seen by VERITAS, MAGIC, and H.E.S.S.

| Model | $\chi^2$/d.o.f. | p value | $f_1$ | $f_2$ | $F_0$ ($10^{-12}$ ph cm$^{-2}$s$^{-1}$) | $F_1$ ($10^{-12}$ ph cm$^{-2}$s$^{-1}$) | $t_0$ (MJD) | $\sigma$ (days) |
|---|---|---|---|---|---|---|---|---|
| Constant | 33.66/14 | $2.3 \times 10^{-3}$ | 1.32 ± 0.41 | 1.35 ± 0.42 | - | - | - | - |
| Gaussian | 10.68/11 | 0.47 | 1.35 ± 0.25 | 1.30 ± 0.49 | (1.60 ± 0.62) | (3.18 ± 0.73) | 58228.99 ± 0.26 | 1.11 ± 0.36 |

We note that this choice is purely arbitrary and that the cross-normalisation factors could be applied to any two of the three IACTs. For an earlier study of systematic offsets between IACTs see Meyer et al. (2010). In our case, the maximum flux scale offsets between the instruments are set to 15%. These offsets are used to constrain the fitting range of $f_1$ and $f_2$. The results of the fit of a constant and a Gaussian function (Eq. 2.4.2) are shown in Table B.1.

## Appendix C: Tabulated data

This section contains the list of tabulated data from observations described in Sect. 2, as follows:

1. Summary of radio observations and flux densities at centimetre wavelengths in Table C.1 and millimetre wavelengths in Table C.2;
2. Kanata Telescope flux densities at the optical wavelength of 634.9 nm in Table C.3 and near-infrared wavelength of 1250 nm in Table C.4;
3. Optical and ultraviolet flux densities of Swift-UVOT in Table C.5, while Table C.6 contains of the host galaxy;
4. Photometry at the ultraviolet wavelength of 154 nm from AstroSat-UVIT in Table C.8;
5. Optical HST observations in Table C.7;
6. Parameters of the composite spectral model for AstroSat-SXT data in Table C.9;
7. Summary of *Swift*-XRT observations and fluxes in the 2–10 keV band in Table C.10;
8. Summary of *Chandra* core fluxes for the observations performed in 2017 and 2018. in Table C.11, while the *Chandra* +*NuSTAR* spectral values for stacked observation, normalised on 2018 emission are reported in Table C.12;
9. Summary of *Fermi*-LAT fluxes for the customised-bin light curve analysis on April 2018. in Table C.13;
10. Summary of VHE $\gamma$-ray observations and fluxes above energies of 350 GeV in Table C.14 and parameters of the spectral models fitted to data from H.E.S.S., MAGIC, and VERITAS in Table C.15;
11. Spectral energy distribution for M87 observed during the 2018 EHT campaign in Table C.1;
12. in addition the spectral models (Model-A-PL, Model-A, BPL, Model-B) used to describe the MWL SED (see Sect. ref-sec:sedmodel) are also reported as supplementary material.

Tables C.1 to C.15 are available in electronic form at the CDS via anonymous ftp to cdsarc.u-strasbg.fr (130.79.128.5) or via http://cdsweb.u-strasbg.fr/cgi-bin/qcat?J/A+A/. Furthermore, these data are provided to the community in machine-readable form via the EHTC Data Webpage[15] along with associated documentation and files needed for spectral modelling of the X-ray data.

[15] https://eventhorizontelescope.org/for-astronomers/data

**Table C.1.** Spectral Energy Distribution for M87 during the 2018 EHT-MWL campaign.

| Observatory | Band | $\nu$ [Hz] | Angular Scale[1] [″] | Obs. date [MJD] | Flux [Jy] | $\nu F_\nu$ [$\times 10^{-12}$ erg cm$^{-2}$ s$^{-1}$] | Sect. |
|---|---|---|---|---|---|---|---|
| VERA | 22 GHz | $2.2 \times 10^{10}$ | $1 \times 10^{-3}$ | 2018-04-26 / 58234 | $1.12 \pm 0.11$ | $0.246 \pm 0.024$ | 2.1.1. |
| EAVN/KaVA | 22 GHz | $2.2 \times 10^{10}$ | $1 \times 10^{-3}$ | 2018-05-08 / 58246 | $1.01 \pm 0.10$ | $0.232 \pm 0.023$ | 2.1.2. |
| EAVN/KaVA | 43 GHz | $4.3 \times 10^{10}$ | $1 \times 10^{-3}$ | 2018-04-21 / 58229 | $0.94 \pm 0.09$ | $0.404 \pm 0.039$ | 2.1.2. |
| VLBA | 24 GHz | $2.4 \times 10^{10}$ | $6 \times 10^{-4}$ | 2018-04-27 / 58235 | $0.88 \pm 0.09$ | $0.211 \pm 0.021$ | 2.1.3. |
| VLBA | 43 GHz | $4.3 \times 10^{10}$ | $4 \times 10^{-4}$ | 2018-04-27 / 58235 | $0.59 \pm 0.06$ | $0.254 \pm 0.024$ | 2.1.3. |
| Global VLBI | 43 GHz | $4.3 \times 10^{10}$ | $2.5 \times 10^{-4}$ | 2018-02-01 / 58150 | $0.65 \pm 0.06$ | $0.279 \pm 0.027$ | 2.1.6. |
| GMVA+ALMA | 86 GHz | $8.6 \times 10^{10}$ | $2.0 \times 10^{-4}$ | 2018-04-14 / 58222 | $0.53 \pm 0.10$ | $0.458 \pm 0.086$ | 2.1.5. |
| KVN | 86.8 GHz | $8.68 \times 10^{10}$ | $1.0 \times 10^{-3}$ | 2018-04-19 / 58227 | $0.93 \pm 0.09$ | $0.807 \pm 0.078$ | 2.1.5. |
| KVN | 129.3 GHz | $1.29 \times 10^{11}$ | $1.0 \times 10^{-3}$ | 2018-04-19 / 58227 | $0.65 \pm 0.20$ | $0.84 \pm 0.26$ | 2.1.5. |
| ALMA | 93 GHz | $9.3 \times 10^{10}$ | 2.4 | 2018-04-15 / 58223 | $1.41 \pm 0.07$ | $1.31 \pm 0.07$ | 2.1.7. |
| ALMA | 221 GHz | $2.21 \times 10^{11}$ | 0.85 | 2018-04-21 / 58229 | $1.10 \pm 0.11$ | $2.43 \pm 0.24$ | 2.1.7. |
| SMA | 221 GHz | $2.21 \times 10^{11}$ | $\simeq 3.0$ | 2018-04-22 / 58230 | $1.33 \pm 0.07$ | $2.94 \pm 0.15$ | 2.1.8. |
| EHT | 230 GHz | $2.3 \times 10^{11}$ | $2 \times 10^{-5}$ | 2018-04-21 / 58229 | 0.5–1.0 | $1.15 - 2.30$ | 1. |
| Kanata | 1250 nm | $2.4 \times 10^{14}$ | $\simeq 10$ | 2018-04-21 / 58229 | $< 2.16 \times 10^{-3}$ | $< 5.18$ | 2.2.1. |
| Kanata | 634.9 nm | $4.7 \times 10^{14}$ | $\simeq 10$ | 2018-04-21 / 58229 | $< 1.05 \times 10^{-3}$ | $< 4.96$ | 2.2.1. |
| HST | 236.0 nm | $1.27 \times 10^{15}$ | $\simeq 0.1$ | 2018-07-30/58329 | $(1.51 \pm 0.77) \times 10^{-4}$ | $1.92 \pm 0.96$ | 2.2.3. |
| AstroSat-UVIT | 173-135 nm | $1.95 \times 10^{15}$ | $\simeq 300$ | 2018-04-21 / 58229 | $< 1.05 \times 10^{-3}$ | $< 5.13$ | 2.2.4. |
| AstroSat-SXT | 0.2-1 keV | $1.21 \times 10^{17}$ | $\simeq 900$ | 2018-04-21 / 58229 | $< 7.75 \times 10^{-6}$ | $< 15.0$ | 2.2.4. |
| AstroSat-SXT | 1-3 keV | $4.84 \times 10^{17}$ | $\simeq 900$ | 2018-04-21 / 58229 | $< 3.01 \times 10^{-6}$ | $< 14.6$ | 2.2.4. |
| AstroSat-SXT | 3-8 keV | $7.26 \times 10^{17}$ | $\simeq 900$ | 2018-04-21 / 58229 | $< 1.18 \times 10^{-7}$ | $< 14.3$ | 2.2.4. |
| *Swift*-UVOT | 546.8 nm | $5.48 \times 10^{14}$ | $\simeq 3$ | 2018-04-21 / 58229 | $(1.92 \pm 0.79) \times 10^{-3}$ | $10.5 \pm 4.1$ | 2.2.2. |
| *Swift*-UVOT | 439.2 nm | $6.83 \times 10^{14}$ | $\simeq 3$ | 2018-04-21 / 58229 | $(1.66 \pm 0.27) \times 10^{-3}$ | $11.4 \pm 1.8$ | 2.2.2. |
| *Swift*-UVOT | 346.5 nm | $8.65 \times 10^{14}$ | $\simeq 3$ | 2018-04-21 / 58229 | $(6.01 \pm 0.09) \times 10^{-4}$ | $5.22 \pm 0.82$ | 2.2.2. |
| *Swift*-UVOT | 260.0 nm | $1.15 \times 10^{15}$ | $\simeq 3$ | 2018-04-21 / 58229 | $(1.47 \pm 0.69) \times 10^{-4}$ | $1.69 \pm 0.79$ | 2.2.2. |
| *Swift*-UVOT | 224.6 nm | $1.33 \times 10^{15}$ | $\simeq 3$ | 2018-04-21 / 58229 | $(3.35 \pm 0.41) \times 10^{-4}$ | $4.45 \pm 0.54$ | 2.2.2. |
| *Swift*-UVOT | 192.8 nm | $1.55 \times 10^{15}$ | $\simeq 3$ | 2018-04-21 / 58229 | $(1.60 \pm 0.51) \times 10^{-4}$ | $2.48 \pm 0.76$ | 2.2.2. |
| *Swift*-XRT | 2-10 keV | $1.08 \times 10^{18}$ | $\simeq 36$ | 2018-04-21 / 58229 | $(2.19 \pm 0.22) \times 10^{-7}$ | $4.24 \pm 0.43$ | 2.3.2. |
| *Chandra* + *NuSTAR* | 0.4 - 0.51 keV | $1.09 \times 10^{17}$ | 0.8 | 2018-04-22–24 / 58230–58232 | $(1.67 \pm 0.17) \times 10^{-6}$ | $1.83 \pm 0.19$ | 2.3.2. |
| *Chandra* + *NuSTAR* | 0.51 - 0.65 keV | $1.40 \times 10^{17}$ | 0.8 | 2018-04-22–24 / 58230–58232 | $(1.29 \pm 0.11) \times 10^{-6}$ | $1.81 \pm 0.16$ | 2.3.2. |
| *Chandra* + *NuSTAR* | 0.65 - 0.84 keV | $1.79 \times 10^{17}$ | 0.8 | 2018-04-22–24 / 58230–58232 | $(1.00 \pm 0.07) \times 10^{-6}$ | $1.78 \pm 0.13$ | 2.3.2. |
| *Chandra* + *NuSTAR* | 0.84 - 1.07 keV | $2.29 \times 10^{17}$ | 0.8 | 2018-04-22–24 / 58230–58232 | $(7.67 \pm 0.47) \times 10^{-7}$ | $1.77 \pm 0.11$ | 2.3.2. |
| *Chandra* + *NuSTAR* | 1.07 - 1.37 keV | $2.93 \times 10^{17}$ | 0.8 | 2018-04-22–24 / 58230–58232 | $(5.92 \pm 0.29) \times 10^{-7}$ | $1.73 \pm 0.08$ | 2.3.2. |
| *Chandra* + *NuSTAR* | 1.37 - 1.75 keV | $3.75 \times 10^{17}$ | 0.8 | 2018-04-22–24 / 58230–58232 | $(4.57 \pm 0.17) \times 10^{-7}$ | $1.71 \pm 0.06$ | 2.3.2. |
| *Chandra* + *NuSTAR* | 1.75 - 2.24 keV | $4.79 \times 10^{17}$ | 0.8 | 2018-04-22–24 / 58230–58232 | $(3.52 \pm 0.08) \times 10^{-7}$ | $1.69 \pm 0.05$ | 2.3.2. |
| *Chandra* + *NuSTAR* | 2.24 - 2.87 keV | $6.13 \times 10^{17}$ | 0.8 | 2018-04-22–24 / 58230–58232 | $(2.71 \pm 0.07) \times 10^{-7}$ | $1.67 \pm 0.04$ | 2.3.2. |
| *Chandra* + *NuSTAR* | 2.87 - 3.67 keV | $7.84 \times 10^{17}$ | 0.8 | 2018-04-22–24 / 58230–58232 | $(2.10 \pm 0.07) \times 10^{-7}$ | $1.64 \pm 0.06$ | 2.3.2. |
| *Chandra* + *NuSTAR* | 3.67 - 4.69 keV | $1.00 \times 10^{18}$ | 0.8 | 2018-04-22–24 / 58230–58232 | $(1.62 \pm 0.07) \times 10^{-7}$ | $1.62 \pm 0.07$ | 2.3.2. |
| *Chandra* + *NuSTAR* | 4.69 - 6.00 keV | $1.28 \times 10^{18}$ | 0.8 | 2018-04-22–24 / 58230–58232 | $(1.25 \pm 0.07) \times 10^{-7}$ | $1.60 \pm 0.09$ | 2.3.2. |
| *Chandra* + *NuSTAR* | 6.00 - 7.67 keV | $1.64 \times 10^{18}$ | 0.8 | 2018-04-22–24 / 58230–58232 | $(9.66 \pm 0.67) \times 10^{-8}$ | $1.59 \pm 0.11$ | 2.3.2. |
| *Chandra* + *NuSTAR* | 7.67 - 9.81 keV | $2.10 \times 10^{18}$ | 0.8 | 2018-04-22–24 / 58230–58232 | $(7.47 \pm 0.62) \times 10^{-8}$ | $1.57 \pm 0.13$ | 2.3.2. |
| *Chandra* + *NuSTAR* | 9.81 - 12.6 keV | $2.68 \times 10^{18}$ | 0.8 | 2018-04-22–24 / 58230–58232 | $(5.77 \pm 0.56) \times 10^{-8}$ | $1.55 \pm 0.15$ | 2.3.2. |
| *Chandra* + *NuSTAR* | 12.6 - 16.1 keV | $3.43 \times 10^{18}$ | 0.8 | 2018-04-22–24 / 58230–58232 | $(4.48 \pm 0.49) \times 10^{-8}$ | $1.53 \pm 0.17$ | 2.3.2. |
| *Chandra* + *NuSTAR* | 16.1 - 20.5 keV | $4.39 \times 10^{18}$ | 0.8 | 2018-04-22–24 / 58230–58232 | $(3.45 \pm 0.43) \times 10^{-8}$ | $1.51 \pm 0.19$ | 2.3.2. |
| *Chandra* + *NuSTAR* | 20.5 - 26.3 keV | $5.62 \times 10^{18}$ | 0.8 | 2018-04-22–24 / 58230–58232 | $(2.66 \pm 0.37) \times 10^{-8}$ | $1.50 \pm 0.21$ | 2.3.2. |
| *Chandra* + *NuSTAR* | 26.3 - 33.6 keV | $7.18 \times 10^{18}$ | 0.8 | 2018-04-22–24 / 58230–58232 | $(2.06 \pm 0.31) \times 10^{-8}$ | $1.48 \pm 0.23$ | 2.3.2. |
| *Chandra* + *NuSTAR* | 33.6 - 43.0 keV | $9.19 \times 10^{18}$ | 0.8 | 2018-04-22–24 / 58230–58232 | $(1.59 \pm 0.27) \times 10^{-8}$ | $1.47 \pm 0.24$ | 2.3.2. |
| *Chandra* + *NuSTAR* | 43.0 - 55.0 keV | $1.18 \times 10^{19}$ | 0.8 | 2018-04-22–24 / 58230–58232 | $(1.23 \pm 0.22) \times 10^{-8}$ | $1.45 \pm 0.26$ | 2.3.2. |
| *Fermi*-LAT | 0.1 - 0.32 GeV | $4.31 \times 10^{22}$ | ≃7000 | 2018-04-08–05-08 / 58216–58246 | $< 2.17 \times 10^{-11}$ | $< 11.6$ | 2.4.1. |
| *Fermi*-LAT | 0.32 - 0.75 GeV | $1.18 \times 10^{23}$ | ≃4500 | 2018-04-08–05-08 / 58216–58246 | $(6.61 \pm 2.89) \times 10^{-12}$ | $6.88 \pm 3.01$ | 2.4.1. |
| *Fermi*-LAT | 0.75 - 1.78 GeV | $2.79 \times 10^{23}$ | ≃2800 | 2018-04-08–05-08 / 58216–58246 | $(1.70 \pm 0.88) \times 10^{-12}$ | $4.24 \pm 2.20$ | 2.4.1. |
| *Fermi*-LAT | 1.78 - 4.22 GeV | $6.63 \times 10^{24}$ | ≃1100 | 2018-04-08–05-08 / 58216–58246 | $(6.91 \pm 4.01) \times 10^{-13}$ | $4.08 \pm 2.37$ | 2.4.1. |
| *Fermi*-LAT | 4.22 - 10 GeV | $1.57 \times 10^{24}$ | ≃700 | 2018-04-08–05-08 / 58216–58246 | $< 4.32 \times 10^{-11}$ | $< 6.04$ | 2.4.1. |
| *Fermi*-LAT | 10 - 100 GeV | $7.65 \times 10^{24}$ | ≃350 | 2018-04-08–05-08 / 58216–58246 | $< 2.42 \times 10^{-11}$ | $< 5.27$ | 2.4.1. |
| H.E.S.S. | 0.32 - 0.46 TeV | $9.35 \times 10^{25}$ | ≃ 360 | 2018-04-18–24 / 58226–58232 | $(6.40 \pm 1.65) \times 10^{-15}$ | $2.17 \pm 0.65$ | 2.4.2. |
| H.E.S.S. | 0.46 - 0.68 TeV | $1.37 \times 10^{26}$ | ≃ 360 | 2018-04-18–24 / 58226–58232 | $(4.08 \pm 0.85) \times 10^{-15}$ | $2.17 \pm 0.45$ | 2.4.2. |
| H.E.S.S. | 0.68 - 1.00 TeV | $2.00 \times 10^{26}$ | ≃ 360 | 2018-04-18–24 / 58226–58232 | $(2.09 \pm 0.50) \times 10^{-15}$ | $1.62 \pm 0.39$ | 2.4.2. |
| H.E.S.S. | 1.00 - 1.47 TeV | $2.93 \times 10^{26}$ | ≃ 190 | 2018-04-18–24 / 58226–58232 | $(1.75 \pm 0.38) \times 10^{-15}$ | $1.99 \pm 0.43$ | 2.4.2. |
| H.E.S.S. | 1.47 - 2.15 TeV | $4.30 \times 10^{26}$ | ≃ 190 | 2018-04-18–24 / 58226–58232 | $(9.30 \pm 2.49) \times 10^{-16}$ | $1.53 \pm 0.41$ | 2.4.2. |
| H.E.S.S. | 2.15 - 3.16 TeV | $6.29 \times 10^{26}$ | ≃ 190 | 2018-04-18–24 / 58226–58232 | $(10.56 \pm 2.21) \times 10^{-16}$ | $2.58 \pm 0.54$ | 2.4.2. |
| H.E.S.S. | 3.16 - 4.64 TeV | $9.21 \times 10^{26}$ | ≃ 190 | 2018-04-18–24 / 58226–58232 | $(4.83 \pm 1.48) \times 10^{-16}$ | $1.73 \pm 0.53$ | 2.4.2. |
| H.E.S.S. | 4.64 - 6.81 TeV | $1.36 \times 10^{27}$ | ≃ 190 | 2018-04-18–24 / 58226–58232 | $(3.22 \pm 1.10) \times 10^{-16}$ | $1.69 \pm 0.58$ | 2.4.2. |
| VERITAS | 0.32 - 0.50 TeV | $9.62 \times 10^{25}$ | ≃ 360 | 2018-04-10–21 / 58218–58229 | $(4.75 \pm 1.26) \times 10^{-15}$ | $2.07 \pm 0.55$ | 2.4.2. |
| VERITAS | 0.50 - 0.79 TeV | $1.53 \times 10^{26}$ | ≃ 360 | 2018-04-10–21 / 58218–58229 | $(1.70 \pm 0.60) \times 10^{-15}$ | $1.19 \pm 0.42$ | 2.4.2. |
| VERITAS | 0.79 - 1.26 TeV | $2.42 \times 10^{26}$ | ≃ 360 | 2018-04-10–21 / 58218–58229 | $(2.07 \pm 0.4) \times 10^{-15}$ | $2.35 \pm 0.53$ | 2.4.2. |
| VERITAS | 1.26 - 2.00 TeV | $3.83 \times 10^{26}$ | ≃ 360 | 2018-04-10–21 / 58218–58229 | $(4.19 \pm 2.07) \times 10^{-16}$ | $0.75 \pm 0.37$ | 2.4.2. |
| MAGIC | 0.03 - 0.075 TeV | $1.21 \times 10^{25}$ | ≅ 330 | 2018-04-12–25 / 58220–58233 | $< 3.47 \times 10^{-13}$ | $< 37.84$ | 2.4.2. |
| MAGIC | 0.075 - 0.19 TeV | $3.04 \times 10^{25}$ | ≅ 300 | 2018-04-12–25 / 58220–58233 | $(9.03 \pm 5.96) \times 10^{-15}$ | $2.51 \pm 1.66$ | 2.4.2. |
| MAGIC | 0.19 - 0.48 TeV | $7.64 \times 10^{25}$ | ≅ 240 | 2018-04-12–25 / 58220–58233 | $(2.39 \pm 1.20) \times 10^{-15}$ | $1.68 \pm 0.84$ | 2.4.2. |
| MAGIC | 0.48 - 1.2 TeV | $1.92 \times 10^{26}$ | ≅ 190 | 2018-04-12–25 / 58220–58233 | $(7.58 \pm 3.50) \times 10^{-16}$ | $1.32 \pm 0.61$ | 2.4.2. |

**Notes.** [(a)] Spatial scale of the source emission (FWHM of the instrumental beam or PSF). Besides the EHT observation for other radio observations if the core is not resolved we use the flux peak obtained from the beam size as an upper limit on the core emission. For elliptical beams, we list the average of the axes as a representative angular scale. At higher energies, the scale typically corresponds to the PSF within a given band or the diameter of the region used for data extraction.

**Table D.1.** Fractional variability of the VHE $\gamma$-ray and X-ray bands during the observational campaign in March 2018. For comparison, we also show $F_{\rm var}$ for the same bands in 2017. The calculation of $F_{\rm var}$ of the total flux in 2018 could not be performed because the calculated sample variance was smaller than the measurement uncertainties.

| | $F_{\rm var}$ in band | | |
|---|---|---|---|
| Year | VHE $\gamma$-rays | *Swift*-XRT total (2–10 keV) | *Swift*-XRT core + HST-1 + jet (2–10 keV) |
| 2017 | 0.34 ± 0.41 | 0.06 ± 0.05 | 0.39 ± 0.05 |
| 2018 | 0.17 ± 0.25 | - | 0.26 ± 0.06 |

## Appendix D: Multi-wavelength light curve variability

We estimate the degree of variability in different wavebands by calculating the fractional variability $F_{\rm var}$ from Eq. 10 in Vaughan et al. (2003):

$$F_{\rm var} = \sqrt{\frac{S^2 - \overline{\sigma_{\rm err}^2}}{\overline{x}^2}} \tag{D.1}$$

where $S^2 = \frac{1}{N-1}\sum_{i=1}^{N}(x_i - \overline{x})^2$ is the sample variance where $\overline{x}$ is the arithmetic mean of the flux measurements $x_i$ and $\sigma_{\rm err}^2 = \frac{1}{N}\sum_{i=1}^{N}\sigma_{{\rm err},i}^2$ the mean squared measurement uncertainty. The uncertainties of $F_{\rm var}$ are calculated following Poutanen et al. (2008) as:

$$\Delta F_{\rm var} = \sqrt{F_{\rm var}^2 + err\left(\sigma_{\rm NXS}^2\right)} - F_{\rm var} \tag{D.2}$$

with

$$err\left(\sigma_{\rm NXS}^2\right) = \sqrt{\left(\sqrt{\frac{2}{N}}\frac{\overline{\sigma_{\rm err}^2}}{\overline{x}^2}\right)^2 + \left(\sqrt{\frac{\overline{\sigma_{\rm err}^2}}{N}}\frac{2F_{\rm var}}{\overline{x}^2}\right)^2}. \tag{D.3}$$

We present the calculated values of $F_{\rm var}$ for the 2018 observational campaign in Table D.1 and also show the same values of $F_{\rm var}$ during the 2017 campaign (M87 MWL2017). As described in Sect. 2.4.2 the observed VHE flare has a time variability scale $O$(1 day). Therefore, following the Nyquist-Shannon sampling theorem (Whittaker 1915), flux points closer than 0.5 days are combined using their weighted average value. To be able to compare the fractional variabilities of different bands it is necessary to only select simultaneous data. Therefore, we can only compare the fractional variability of the VHE $\gamma$-ray instruments and *Swift*-XRT listed in Table D.1.